\documentclass[12pt]{article}
\pdfoutput=1

\usepackage{draft}
\usepackage{putex}

 \usepackage{dsfont}
\usepackage[all]{xy}

\usepackage{stackrel,amssymb}

\usepackage{graphicx}
\usepackage{caption}
\usepackage{amsmath,bm}
\usepackage{array}
\usepackage{subcaption}
\usepackage{epstopdf}
\usepackage{enumerate}
\usepackage{cite}
\usepackage{tensor}
\usepackage{slashed}
\usepackage[utf8]{inputenc}
\usepackage{rotating}
\usepackage{bigfoot}
\usepackage[
      colorlinks=true,
      linkcolor=blue,
      urlcolor=blue,
      filecolor=black,
      citecolor=red,
      ]{hyperref}

\usepackage{tikz-cd}
\usepackage{tikz}
\usetikzlibrary{snakes}
\usetikzlibrary{arrows}
\usetikzlibrary{positioning}
\usetikzlibrary{decorations.pathmorphing}
\usetikzlibrary{ decorations.markings}
\tikzset{snake it/.style={decorate, decoration=snake}}
\usetikzlibrary{shapes.misc}
\tikzset{cross/.style={cross out, draw=black, minimum size=2*(#1-\pgflinewidth), inner sep=0pt, outer sep=0pt},
cross/.default={1pt}}
\usetikzlibrary{shapes.geometric}
\definecolor{bleudefrance}{rgb}{0.19, 0.55, 0.91}
\definecolor{candyapplered}{rgb}{1.0, 0.03, 0.0}

\def\XXint#1#2#3{{\setbox0=\hbox{$#1{#2#3}{\int}$ }
		\vcenter{\hbox{$#2#3$ }}\kern-.6\wd0}}

\usepackage{adjustbox}
\usepackage{multirow}

\numberwithin{equation}{section}

\newcommand{\cI}{{\cal I}}

\def\<{\langle}
\def\>{\rangle}

\def\pa{\partial}
\def\ve{\varepsilon}
\def\ep{\epsilon}

\newcommand{\gym}{g}
\newcommand{\gymd}{\tilde{g}}
\newcommand{\lambdad}{\tilde{\lambda}}
\newcommand{\ap}{\alpha'}
\newcommand{\zbar}{\overline{z}}
\newcommand{\es}[2] {\begin{equation} \label{#1} \begin{split} #2 \end{split} \end{equation}}

\newcommand{\abs}[1]{\left\lvert #1 \right\rvert}

\makeatletter
\DeclareFontFamily{OMX}{MnSymbolE}{}
\DeclareSymbolFont{MnLargeSymbols}{OMX}{MnSymbolE}{m}{n}
\SetSymbolFont{MnLargeSymbols}{bold}{OMX}{MnSymbolE}{b}{n}
\DeclareFontShape{OMX}{MnSymbolE}{m}{n}{
    <-6>  MnSymbolE5
   <6-7>  MnSymbolE6
   <7-8>  MnSymbolE7
   <8-9>  MnSymbolE8
   <9-10> MnSymbolE9
  <10-12> MnSymbolE10
  <12->   MnSymbolE12
}{}
\DeclareFontShape{OMX}{MnSymbolE}{b}{n}{
    <-6>  MnSymbolE-Bold5
   <6-7>  MnSymbolE-Bold6
   <7-8>  MnSymbolE-Bold7
   <8-9>  MnSymbolE-Bold8
   <9-10> MnSymbolE-Bold9
  <10-12> MnSymbolE-Bold10
  <12->   MnSymbolE-Bold12
}{}

\let\llangle\@undefined
\let\rrangle\@undefined
\DeclareMathDelimiter{\llangle}{\mathopen}%
                     {MnLargeSymbols}{'164}{MnLargeSymbols}{'164}
\DeclareMathDelimiter{\rrangle}{\mathclose}%
                     {MnLargeSymbols}{'171}{MnLargeSymbols}{'171}

\makeatother

\newcommand{\leftrarrows}{\mathrel{\raise.75ex\hbox{\oalign{%
				$\scriptstyle\leftarrow$\cr
				\vrule width0pt height.5ex$\hfil\scriptstyle\relbar$\cr}}}}
\newcommand{\lrightarrows}{\mathrel{\raise.75ex\hbox{\oalign{%
				$\scriptstyle\relbar$\hfil\cr
				$\scriptstyle\vrule width0pt height.5ex\smash\rightarrow$\cr}}}}
\newcommand{\Rrelbar}{\mathrel{\raise.75ex\hbox{\oalign{%
				$\scriptstyle\relbar$\cr
				\vrule width0pt height.5ex$\scriptstyle\relbar$}}}}

\makeatletter
\def\leftrightarrowsfill@{\arrowfill@\leftrarrows\Rrelbar\lrightarrows}
\newcommand{\xleftrightarrows}[2][]{\ext@arrow 3399\leftrightarrowsfill@{#1}{#2}}
\makeatother

\begin{document}

\preprint{PUPT-2641}

\title{
 Scattering From $(p,q)$-Strings in $\text{AdS}_5\times \text{S}^5 $ 
}

\authors{Silviu S.~Pufu,\worksat{\PU, \PCTS} Victor A.~Rodriguez,\worksat{\PU} and Yifan Wang\worksat{\NYU}}

\institution{PU}{Joseph Henry Laboratories, Princeton University, Princeton, NJ 08544, USA}
 
\institution{PCTS}{Princeton Center for Theoretical Science, Princeton University, Princeton, NJ 08544, USA}
	
\institution{NYU}{Center for Cosmology and Particle Physics, New York University, New York, NY 10003, USA}

\abstract{Motivated by understanding the scattering of gravitons and their superpartners 
from extended $(p, q)$-strings in type IIB string theory via AdS/CFT, we study an integrated two-point function of stress tensor multiplet operators in the presence of a half-BPS line defect in ${\cal N} = 4$ $SU(N)$ super-Yang-Mills theory.  We determine this integrated correlator at the five lowest non-trivial orders in $1/\sqrt{N}$ at fixed Yang-Mills coupling and $\theta$ angle.  Our calculations are performed explicitly when the line defect is a Wilson line, in which case we find a finite number of perturbative contributions at each order in $1/\sqrt{N}$, as well as instanton contributions.  Using $SL(2, \mZ)$ transformations, our results can also be applied to Wilson-'t Hooft line defects dual to extended $(p, q)$-strings in the bulk.  We analyze features of these integrated correlators in the weak coupling expansion by comparing with open-closed amplitudes of type IIB string theory on $\text{AdS}_5\times \text{S}^5$, as well as in its flat space limit.  We predict new higher-derivative interaction vertices on the D1-brane and, more generally, on $(p,q)$-strings.
}

\date{May 2023}

\maketitle

\tableofcontents

\pagebreak

\section{Introduction and Summary}
 
It is a tantalizing idea that the 4d $\cN=4$ super-Yang-Mills (SYM) theory can be used to elucidate, quantitatively, aspects of quantum gravity and string theory.  One such aspect studied recently \cite{Binder:2019jwn,Chester:2019jas,Chester:2020dja,Chester:2020vyz} is the small momentum expansion of the graviton S-matrix of type IIB string theory at finite string coupling $g_s$.\footnote{See \cite{Chester:2018dga,Binder:2018yvd,Binder:2019mpb,Binder:2020ckj} for similar progress on the M-theory and type IIA S-matrices at small momentum using the Aharony-Bergman-Jafferis-Maldacena (ABJM) \cite{Aharony:2008ug} and Aharony-Bergman-Jafferis (ABJ) \cite{Aharony:2008gk} gauge theories.}  The regime of finite $g_s$ is a regime where conventional string worldsheet perturbation theory is of little use, and previous progress required the use of supersymmetry and conjectured string dualities~\cite{Green:1997tv,Green:1997as,Green:1998by,Green:1999pu,Wang:2015jna,Wang:2015aua}, such as the $SL(2, \mZ)$ duality of type IIB string theory discussed below.  In \cite{Binder:2019jwn,Chester:2019jas,Chester:2020dja,Chester:2020vyz}, the first few terms in the small momentum expansion of the S-matrix were rederived from a non-standard large-$N$ limit of the ${\cal N} =4$ $SU(N)$ SYM theory: 
 \es{VeryStrongCoupling}{
  N \to \infty \,, \qquad \tau = \frac{\theta}{2\pi} + \frac{4 \pi i}{g^2} \ \  \text{fixed} \,,
  }
with $\theta\sim \theta+2\pi$ being the theta angle and $g$ the Yang-Mills coupling.   The limit of large $N$ and fixed $\tau$ was referred to in \cite{Binder:2019jwn,Chester:2019jas,Chester:2020dja,Chester:2020vyz} as a ``very strong coupling'' limit,  because in this limit the 't Hooft coupling $\lambda = g^2 N$  becomes large, in contrast with the more conventional 't Hooft limit taken in large $N$ gauge theories where $\lambda$ is kept fixed while $N$ is taken to infinity.   

Via the anti-de Sitter / Conformal Field Theory (AdS/CFT) duality \cite{Maldacena:1997re,Gubser:1998bc,Witten:1998qj}, the ${\cal N} = 4$ SYM theory in the limit \eqref{VeryStrongCoupling} is still dual to type IIB string theory on ${\rm AdS_5} \times {\rm S}^5$, 
and the $1/N$ expansion translates into the low energy expansion of the superstring theory, 
which at leading order  is described by type IIB supergravity on the same background.  Computations in the very strong coupling limit, at the first few orders in $1/N$, were made possible due to powerful analytic bootstrap constraints for four-point holographic correlators \cite{Rastelli:2016nze,Rastelli:2017udc}, particularly the analytic properties of Witten diagrams in Mellin space \cite{Penedones:2010ue,Fitzpatrick:2011hu,Fitzpatrick:2011jn,Fitzpatrick:2011dm}.  These methods were combined with supersymmetric localization results for the sphere partition function of the mass-deformed ${\cal N} = 4$ SYM theory \cite{Pestun:2007rz,Russo:2013kea}, which made possible the exact evaluation of certain ${\cal N} = 4$ SYM integrated correlators \cite{Binder:2018yvd,Binder:2019jwn,Chester:2020dja}.  

\begin{figure}[t!]
\begin{center} \begin{minipage}[c]{1in} \includegraphics [width=\textwidth] {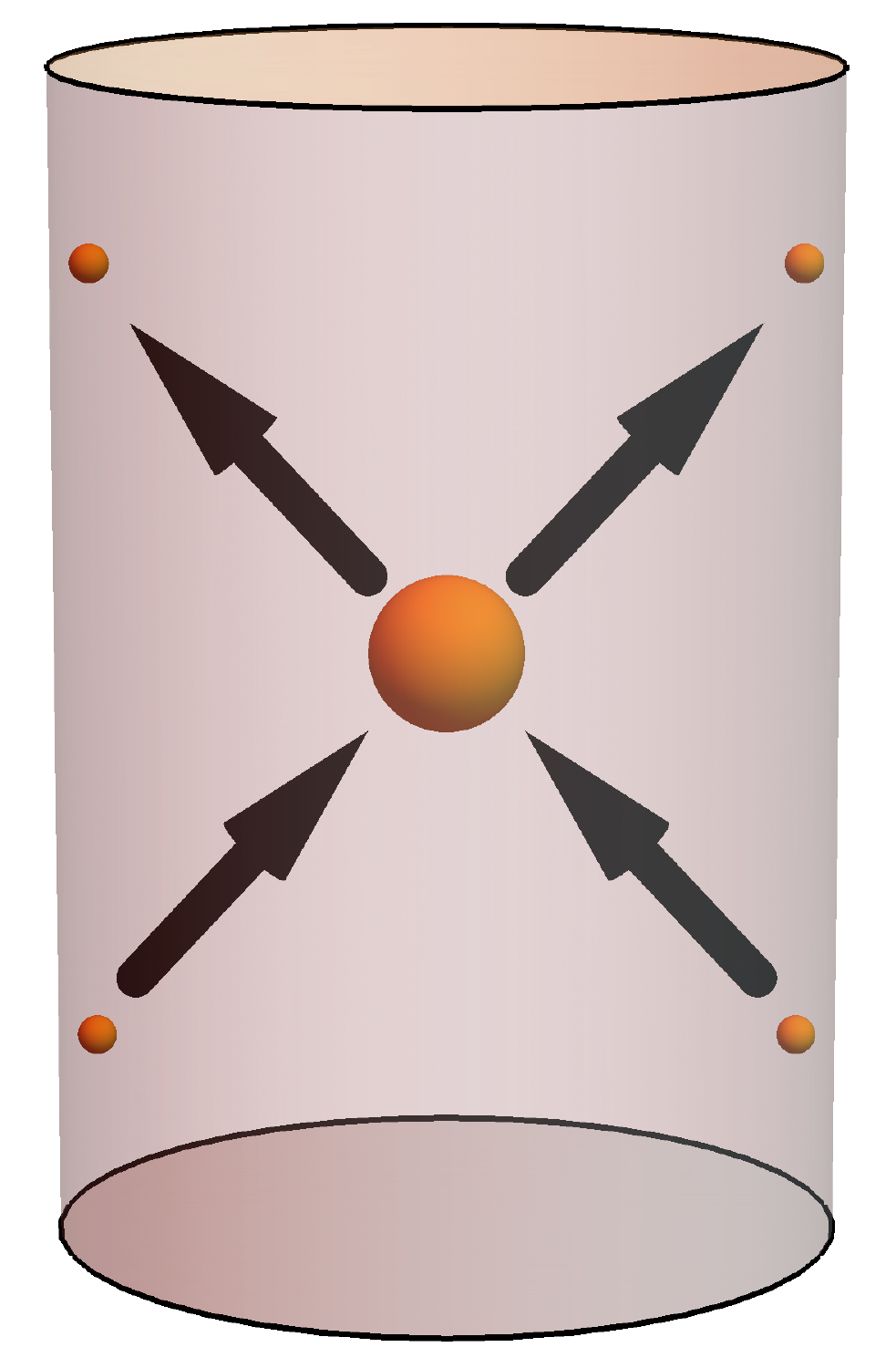}  \end{minipage}
     \hspace{0.3in}
    {\Large $\xrightarrow[\text{limit}]{\text{flat space}}$}
    \hspace{0.3in}
    \begin{minipage}[c]{1in} \includegraphics [width=\textwidth] {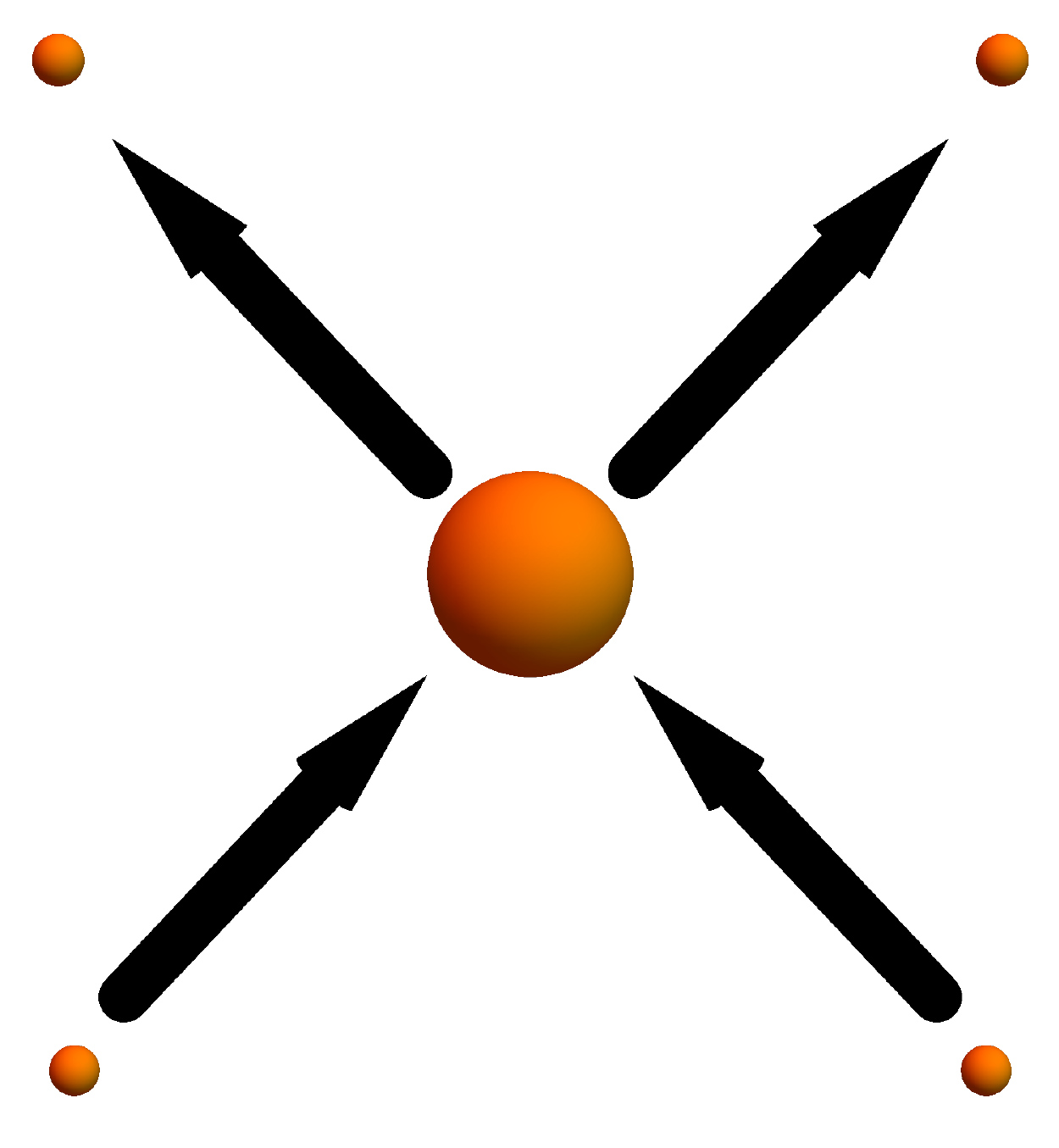} \end{minipage}
\caption{A pictorial representation of a $2 \to 2$ scattering process of closed string states and its flat space limit.  On the left, the ${\rm AdS_5}$ space is viewed as a cylinder.  Using local operators on the boundary, one creates wave packets that then interact in the bulk.  On the right, wave packets come in from infinity and interact in flat space.}
\label{FlatSpaceFigure}
\end{center}
\end{figure}

Dating back to the ideas of \cite{Polchinski:1999ry,Susskind:1998vk,Giddings:1999jq}, the correlators of single-trace $\frac{1}{2}$-BPS local operators in ${\cal N} = 4$ SYM theory can be interpreted as scattering amplitudes on AdS space.  The AdS/CFT dictionary \cite{Maldacena:1997re,Gubser:1998bc,Witten:1998qj}
 \es{Dictionary}{
  \frac{L}{\ell_s}=\lambda^{\frac 14} \,,\qquad \tau=\chi+ \frac{i}{g_s}\,,
 }
where $L$ is the radius of ${\rm AdS_5}$, $\ell_s = \sqrt{\alpha'}$ the string length, and $\chi$ the expectation value of the axion, implies that, in the very strong coupling limit, the radius of AdS becomes very large in string units, and therefore the AdS scattering amplitudes reduce to scattering amplitudes in flat space (when the kinematic variables are properly scaled) \cite{Polchinski:1999ry,Susskind:1998vk,Giddings:1999jq}.   For a pictorial representation, see Figure~\ref{FlatSpaceFigure}.   The precise relation between the scattering amplitudes in flat space and the CFT correlators uses Mellin space and was made precise in~\cite{Penedones:2010ue,Fitzpatrick:2011hu,Fitzpatrick:2011jn,Fitzpatrick:2011dm}.   Importantly, single-trace $1\over 2$-BPS local operators in the ${\cal N} = 4$ SYM theory reduce, in the flat space limit, to massless closed string modes, i.e.~gravitons and their superpartners, moving in five out of the ten spacetime dimensions.

The goal of this paper is to take this program one step further and consider what ${\cal N} = 4$ SYM theory can teach us about scattering of closed strings from extended objects such as long strings or more general D-branes in type IIB string theory.  From the ${\cal N} = 4$ SYM point of view, the simplest setup is given by the correlator between a $1\over 2$-BPS Wilson loop in the fundamental representation of $SU(N)$ and two local operators.
If we view global ${\rm AdS_5}$ as a cylinder with $\mR \times S^3$ boundary, one can consider a $1\over 2$-BPS Wilson loop running along $\mR$ in opposite directions at two antipodal points on $S^3$.\footnote{This is equivalent to the conformal line defect extending along a straight line on $\mR^4$ by a Weyl transformation.}  In the bulk, the fundamental Wilson line extends to a fundamental string worldsheet \cite{Maldacena:1998im,Rey:1998ik} stretching across the ${\rm AdS_5}$ space. This setup generalizes straightforwardly to the case where the Wilson loop is replaced by a ${1\over 2}$-BPS fundamental 't Hooft loop,\footnote{The half-BPS 't Hooft loops are classified by their magnetic fluxes \cite{Kapustin:2005py}. Here, the fundamental 't Hooft loop is labeled by the weight vector that corresponds to the fundamental representation ${\bm N}$ of $SU(N)$  and is a non-genuine line operator (see Footnote~\ref{footnote:globalstructure}).} which corresponds to a D1-brane extending in the bulk,  or by a  more general Wilson-'t Hooft loop of dyonic charges $(p,q)$ \cite{Kapustin:2005py} (see the discussion around \eqref{SL2dyonic}) that corresponds to a $(p,q)$-string \cite{Harvey:1995rn,Schwarz:1995dk,Witten:1995im,Callan:1995xx} in the bulk.\footnote{The F1-string (Wilson loop) and the D1-brane ('t Hooft loop) correspond to $(p,q)=(1,0)$ and $(p,q)=(0,1)$ respectively.}

\begin{figure}[thbp]
\begin{center} \begin{minipage}[c]{1in} \includegraphics [width=\textwidth] {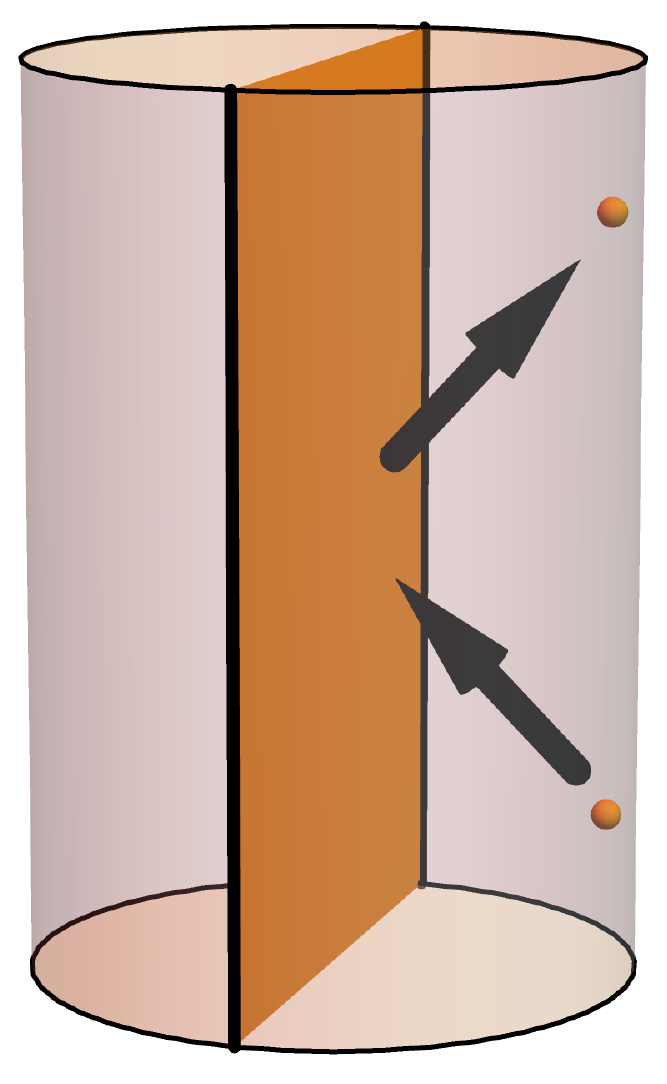}  \end{minipage}
     \hspace{0.4in}
    {\Large $\xrightarrow[\text{limit}]{\text{flat space}}$}
    \hspace{0.4in}
    \begin{minipage}[c]{1in} \includegraphics [width=0.71\textwidth] {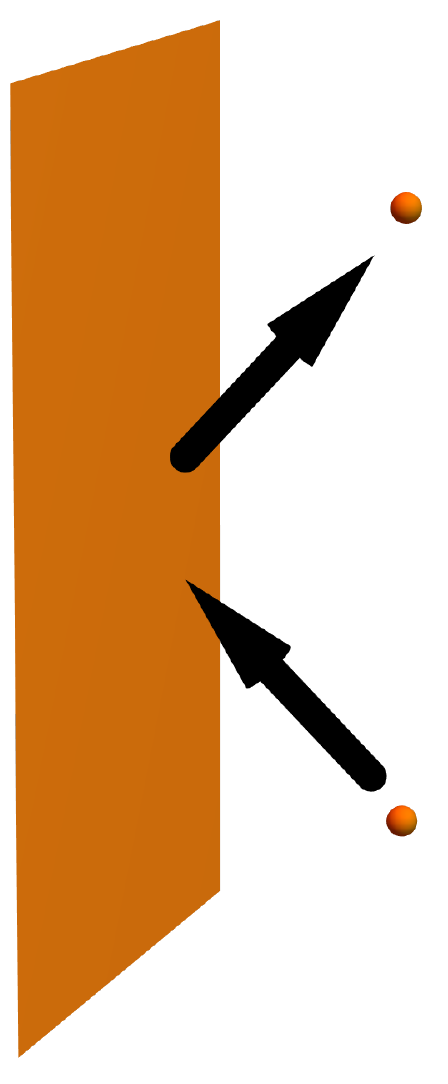} \end{minipage}
\caption{A pictorial representation of a $1 \to 1$ scattering process from an extended $(p,q)$-string (e.g. a long string or a D1-brane) and its flat space limit.  On the left, the ${\rm AdS_5}$ space is viewed as a cylinder.   The $(p,q)$-string worldsheet is depicted in orange and it stretches from one side of AdS to the other, ending onto the contour of the corresponding Wilson-'t Hooft line on the boundary. Using local operators on the boundary, one creates a wave packet that then interacts with the $(p,q)$-string and scatters back off to the boundary.   On the right, a wave packet comes in from infinity and interacts with the $(p,q)$-string in flat space.}
\label{FlatSpaceBraneFigure}
\end{center}
\end{figure}

In the presence of the Wilson-'t Hooft loop, we can then consider single trace $1\over 2$-BPS local operators inserted on the boundary \cite{Berenstein:1998ij,Semenoff:2001xp,Pestun:2002mr}. These local operators can be used to create wave packets that interact with the extended string in the bulk. Even though the precise details have not been worked out, we expect that this setup gives the scattering amplitude of a massless string mode scattering from an extended $(p, q)$-string  in asymptotically flat space.  See Figure~\ref{FlatSpaceBraneFigure}.

The ${\cal N} = 4$ SYM theory is expected to enjoy an exact electromagnetic duality, exhibiting covariance under $SL(2,\mZ)$ transformations of the coupling
\ie 
    \tau \to {a\tau +b \over c\tau +d}\,,
 \label{tauTransf}
 \fe 
with $a,b,c,d\in \mZ$, and $ad-bc=1$. Via the AdS/CFT correspondence, this maps to the $SL(2,\mZ)$ duality in type IIB string theory.  The $SL(2,\mZ)$ duality, in particular the S-duality transformation that inverts the coupling $\tau \to -{1\over \tau}$, ties together the perturbative and non-perturbative contributions to the CFT observables, leading to nontrivial predictions for their modular properties. For example, the correlation functions of local operators that are $SL(2,\mZ)$ invariant, such as the half-BPS primary operators, produce nontrivial modular functions such as Eisenstein series or generalized Eisenstein series \cite{Chester:2019jas,Chester:2020vyz,Green:2020eyj,Dorigoni:2021rdo,Dorigoni:2021guq,Dorigoni:2022zcr,Dorigoni:2021bvj,Paul:2022piq}.  After the flat space limit, these $SL(2,\mZ)$-invariant correlators yield $SL(2,\mZ)$-invariant interactions in the closed string effective action.

In contrast to the BPS local operators, the  BPS line operators transform nontrivially under $SL(2,\mZ)$. In particular, the Wilson and 't Hooft loops transform into each other under S-duality, and more generally into  Wilson-'t Hooft (dyonic) loops that come from the worldlines of probe dyons with electric and magnetic charges labelled by a pair of coprime integers $(p,q)$.\footnote{\label{footnote:globalstructure}It is well known from \cite{Aharony:2013hda} that 
under $SL(2,\mZ)$, not only the coupling $\tau$ of the $\cN=4$ SYM transforms,
but also the global forms of the gauge group and associated discrete theta angles change accordingly, constituting an intricate duality web (see \cite{Aharony:2013hda} for details). For example, the $SU(N)$ SYM is S-dual to the $(PSU(N))_0$ SYM where the subscript on the gauge group labels the discrete $\mZ_N$ theta angle for $PSU(N)$. While the $SU(N)$ theory is invariant under the $T$ transformation, the distinct $(PSU(N))_n$ theories with $n\in \mZ_N$  are permuted under $T$. Nonetheless the topology of $S^4$ does not distinguish among these discrete choices. For the same reason, the Wilson-'t Hooft loops labelled by coprime integers $(p,q)$ in the $SU(N)$ SYM are genuine line operators only when $q\equiv 0 \,({\rm mod}\, N)$. More generally, they are attached to open topological surface operators $U_q(\Sigma)$ at the boundary of the surface $\Sigma$ and $U_q$ generates the $\mathbb{Z}_N$ one-form symmetry in the $SU(N)$ SYM \cite{Gaiotto:2014kfa}. Since the additional local operator insertions are insensitive to one-form symmetries, we will not make this distinction (between genuine and non-genuine line operators) in the rest of the paper.}
 Explicitly, the $SL(2,\mZ)$ duality (see \eqref{tauTransf}) acts on the dyonic charges by
 \ie 
 (p,q) \to (p,q)\begin{pmatrix}
  a &  -c \\ -b & d 
 \end{pmatrix}\,. 
 \label{SL2dyonic}
 \fe 
Therefore, the correlator of a Wilson-'t Hooft loop and two $1\over 2$-BPS single-trace local operators will not be $SL(2, \mZ)$-invariant.  Analogously, the scattering amplitude of a closed string mode from a $(p,q)$-string will not be $SL(2, \mZ)$-invariant either. Consequently, not much is known non-perturbatively about the interactions that contribute to this scattering amplitude, and the ${\cal N} = 4$ SYM correlator seems to be the most promising way of determining at least some of them.

The nontrivial $SL(2,\mZ)$ action on the line defects also means that once we understand the non-perturbative effects in one duality frame (for instance, for the Wilson loop), we will immediately have access to the non-perturbative scattering amplitudes of closed strings from general $(p,q)$-strings with $p$ and $q$ coprime extended over the the same submanifold on ${\rm AdS_5}\times {\rm S^5}$.\footnote{\label{footnote:howtheyend}The discrete global data that labels the SYM translates to choices of boundary conditions for the NSNS and the RR 2-form gauge fields $(B_2,C_2)$ coupled together by a mixed Chern-Simons action in the bulk \cite{Aharony:1998qu,Witten:1998wy,Bergman:2022otk}. This boundary condition specifies which bulk $(p,q)$-strings can end on the boundary. When the SYM line defect is non-genuine (see Footnote~\ref{footnote:globalstructure}), the corresponding $(p,q)$-string worldsheet does not end on the boundary of ${\rm AdS_5}$.  Instead, the non-genuine line defect is attached to a piece of the string worldsheet on the boundary, which becomes the topological operator for the one-form symmetry. As noted in Footnote~\ref{footnote:globalstructure},we suppress these subtle distinctions in the main text since they do not affect the observables we study here.}

Concretely, we will consider the $S^4$ partition function of the ${\cal N} = 4$ SYM theory deformed by an ${\cal N} = 2$-preserving mass parameter $m$ (referred to as the ${\cal N} = 2^*$ theory) and study the quantity
 \es{eq:ZWLtwoderiv}{
  {\cal I}_{\mathbb{W}}(\tau, \bar \tau) = \partial_m^2 \log \langle \mathbb{W} \rangle \Big|_{m=0} \,, 
  }
where $\langle \mathbb{W} \rangle$ is the expectation value of a circular $1\over 2$-BPS fundamental Wilson loop along a great circle of $S^4$.  This quantity can be interpreted as a certain two-point function $\langle \mathbb{W} {\cal O}(\vec{x}_1) {\cal O}(\vec{x}_2) \rangle$ of $1\over 2$-BPS local operators in the presence of a Wilson loop $\mathbb{W}$, further integrated in $\vec{x}_1$ and $\vec{x}_2$ with a supersymmetry-preserving measure.  In the very strong coupling limit, \eqref{eq:ZWLtwoderiv} has an expansion in $1/\sqrt{N}$ of the form
 \es{Expansion}{
  {\cal I}_{\mathbb{W}}(\tau, \bar \tau) = {\cal I}_{\mathbb{W},-1/2}(\tau, \bar \tau) \sqrt{N} 
   + {\cal I}_{\mathbb{W},0}(\tau, \bar \tau)
    + \frac{{\cal I}_{\mathbb{W},1/2}(\tau, \bar \tau)}{\sqrt{N}} + \frac{{\cal I}_{\mathbb{W},1}(\tau, \bar \tau)}{N} 
     + \cdots \,.
 }
At each order in $1/\sqrt{N}$, the integrated correlator \eqref{Expansion} receives contributions from a number of Witten diagrams that grows with the order of the expansion.  Nevertheless, at low enough orders in the $1/\sqrt{N}$ expansion, we expect that the unintegrated two-point function $\langle \mathbb{W} {\cal O}(\vec{x}_1) {\cal O}(\vec{x}_2) \rangle$ is completely determined by \eqref{Expansion} as well as analytic bootstrap conditions, such as crossing symmetry, supersymmetry, and analyticity of holographic correlators.  It is at these orders that one can then draw a direct connection between \eqref{Expansion} and the flat space scattering amplitude of a closed string from a long string (or from a D$1$-brane, in the S-dual frame, and from more general $(p,q)$-strings).  We will provide a more detailed explanation in Section~\ref{sec:unintegrated}.

In Section~\ref{INTEGRATED}, we will evaluate explicitly all the coefficients ${\cal I}_{\mathbb{W},n}$, with $n = -1/2, \ldots, 3/2$.  As we will show, the first coefficient that receives non-perturbative contributions is ${\cal I}_{\mathbb{W},1}$, and, as expected, it is not modular invariant.  Even though we do not establish the precise connection between ${\cal I}_{\mathbb{W},1}$ and the flat space scattering of a graviton from a long string or a general $(p,q)$-string, we conjecture that both the long string and the $(p,q)$-string (with $q\neq 0$) effective actions include a supersymmetric completion of a six-derivative term of the schematic form $D^2 R^2$, where $D$ denotes tangential derivatives  and $R$ is the ten-dimensional Riemann tensor pulled back to the two-dimensional worldvolume. In the string case, the coefficient of $D^2 R^2$ is given by ${\cal I}_{\mathbb{W},1}(\tau, \bar \tau)$, which when $\theta = 0$, takes the form
 \es{FnOfCoupling}{
  {\cal I}_{\mathbb{W},1} = \frac{\frac{3}{8} + \frac{3 \zeta(3)}{2}}{4 \pi g_s} + \frac{\pi^3 g_s^3}{180}  
   - \frac{6 g_s}{\pi} \sum_{k=1}^\infty K_2 (2 \pi k / g_s) \sum_{p|k} \frac{1}{p^2} \,.
 }
For the D1-brane case, we should simply replace $g_s \to 1/g_s$ in \eqref{FnOfCoupling}.  More details will be given in Section~\ref{sec:tL}.

The rest of this paper is organized as follows.  In Section~\ref{INTEGRATED}, we begin with a systematic evaluation of \eqref{eq:ZWLtwoderiv} in the very strong coupling limit.  For each of the terms in \eqref{Expansion}, we will evaluate both the perturbative contribution as well as the non-perturbative contribution coming from the Nekrasov partition function \cite{Nekrasov:2002qd,Nekrasov:2003rj}.  These computations will be performed with the help of the $1/N$ expansion of the $n$-body resolvents, which can be computed efficiently using the recursion relations of \cite{Eynard:2008we,Eynard:2004mh}. 
In Section~\ref{sec:tL}, we discuss the integrated correlator  for general $1\over 2$-BPS Wilson-'t Hooft lines obtained from $SL(2,\mZ)$ transformations. In particular,
we will analyze the strong coupling limit $\gym \to \infty$ of ${\cal I}_{\mathbb{W}}(\tau, \bar \tau)$ in order to extract the weak coupling expansion of the analogous integrated correlator in the presence of a $1\over 2$-BPS 't Hooft line ${\cal I}_{\mathbb{T}}(\tilde \tau, \bar{\tilde \tau})$, defined in (\ref{ZTLtwoderiv}), order by order in the $1/N$ expansion. We will find that nontrivial cancellations between contributions from perturbative and non-perturbative terms in $\gym$ in (\ref{eq:ZWLtwoderiv}) result in a well-defined $1/N$ expansion in terms of the dual 't Hooft coupling $\lambdad = \tilde g^2 N$, providing a strong consistency check on our calculation of (\ref{eq:ZWLtwoderiv}). 
In Section~\ref{sec:unintegrated}, we describe the road map for obtaining the full (un-integrated) two-point correlator in the presence of a general $1\over 2$-BPS Wilson-'t Hooft line to the first few orders in the large $N$ limit. 
In Section~\ref{sec:EFT}, we provide a qualitative understanding of our results for ${\cal I}_{\mathbb{W}}(\tau, \bar \tau)$ and ${\cal I}_{\mathbb{T}}(\tilde \tau, \bar{\tilde \tau})$ (also for more general Wilson-'t Hooft lines) in terms of effective field theory in AdS through Witten diagrams, as well as a qualitative comparison against the structure of the small momentum expansion of scattering amplitudes of gravitons off a D1-brane (and more general $(p,q)$-strings) in type IIB string theory \cite{Hashimoto:1996bf}. 
We conclude in Section~\ref{sec:discussion} with a discussion of our results and future research directions.  Various technical details are relegated to the Appendices.

\section{Integrated Two-point Function with Wilson Loop from Localization} 
\label{INTEGRATED}

Let us begin by describing the calculation of the second mass derivative ${\cal I}_{\mathbb{W}}(\tau, \bar \tau)$ introduced in \eqref{eq:ZWLtwoderiv} in the very strong coupling limit.   Since this computation requires a technical understanding of Hermitian matrix models, let us first present the final result, and the reader interested in the string theory implications of this result can then skip the rest of this section.  We find that the first few coefficients ${\cal I}_{\mathbb{W},p}(\tau, \bar \tau)$ in the $1/N$ expansion of ${\cal I}_{\mathbb{W}}(\tau, \bar \tau)$ (see \eqref{Expansion}) are: 
 \es{IExplicit}{
   {\cal I}_{\mathbb{W},-1/2} &= g\,, \\
   {\cal I}_{\mathbb{W},0} &= \frac{1}{2} - \frac{\pi^2}{3} \,, \\
   {\cal I}_{\mathbb{W},1/2} &= \frac{3}{8g} - \frac{g^3}{32} \,, \\
   {\cal I}_{\mathbb{W},1} &= \frac{3 ( 1+ 4 \zeta(3))}{8 g^2} + \frac{g^6}{11520}  
   - \frac{3 g^2}{2\pi^2}  \sum_{k=1}^\infty \cos (k \theta) K_2 (8 \pi^2 k / g^2) \sigma_{-2}(k) \,, \\
   {\cal I}_{\mathbb{W},3/2} &= \frac{3(21+ 64 \zeta(3))}{128g^3}  - \frac{g}{256}
     + \frac{7 g^5}{10240} - \frac{g^9}{1935360} \\
     &{}+\sum_{k=1}^\infty \biggl[\cos (k \theta) \frac{9 g \sigma_{-2}(k)}{4 \pi^2}  \left(  K_2 (8 \pi^2 k / g^2) + \frac{5 g^2 K_3 (8 \pi^2 k / g^2)}{4 \pi^2 k } \right) \\
     &{}- \sin (k \theta) \frac{3g  \left( 3 k \sigma_{-3}(k) - 4 \sigma_{-2}(k) + \sigma_{-1}(k)  \right)}{2 \pi}  \left(  K_2 (8 \pi^2 k / g^2) + \frac{15 g^2 K_3 (8 \pi^2 k / g^2)}{16 \pi^2 k }  
       \right)  \biggr]  \,, \\
    &\vdots   
 }
where $\tau = \frac{\theta}{2 \pi} + \frac{4 \pi i }{g^2}$ as before, and  the divisor function $\sigma_a$ is defined by $\sigma_a(k) \equiv \sum_{p | k} p^a$.

Let us now describe the derivation of this result.   As mentioned above, the reader interested in the interpretation of \eqref{IExplicit} can skip the rest of this section, while the reader interested in the details of the matrix model computations should continue reading.

Without a Wilson loop insertion, similar computations were performed in \cite{Binder:2019jwn, Chester:2019jas, Chester:2020dja, Chester:2020vyz} for mass derivatives of the partition function without the Wilson loop insertion.  The addition of the Wilson loop, however, poses significant new challenges that requires us to develop a more systematic approach, especially for the study of non-perturbative contributions.

\subsection{Setting up the matrix model}

The starting point in the derivation of \eqref{IExplicit} is that the expectation value $\langle \mathbb{W} \rangle$, whose second mass derivative is evaluated at $m=0$ in \eqref{eq:ZWLtwoderiv}, can be written as a ratio of unnormalized partition functions 
 \es{WRatio}{ 
  \langle \mathbb{W} \rangle = \frac{Z_\mathbb{W}(m, \tau, \bar \tau)}{Z(m, \tau, \bar \tau)} \,,
 }
where both the partition function $Z$ and the partition function $Z_\mathbb{W}$ with a Wilson loop insertion can be computed using supersymmetric localization \cite{Pestun:2007rz}.

For the $U(N)$ SYM, both $Z(m, \tau, \bar \tau)$ and $Z_\mathbb{W}(m, \tau, \bar \tau)$ can be written in terms of expectation values in a Gaussian matrix model.  The Gaussian ensemble is defined by the partition function
 \es{PartGaussian}{
  Z_G = \frac{1}{N!} \int \left( \prod_i da_i \right) \,  \left( \prod_{i<j} a_{ij}^2 \right) e^{-\frac{8\pi^2}{g^2} \sum_i a_i^2} \,, 
 }
where the indices $i, j$ run from $1$ to $N$, and  $a_{ij} \equiv a_i - a_j$.  The normalized expectation values in this Gaussian ensemble of some function $f(a_i)$ is
 \es{fExp}{
  \llangle f(a_i) \rrangle = \frac{1}{Z_G} \frac{1}{N!} \int \left( \prod_i da_i \right) \,  \left( \prod_{i<j} a_{ij}^2 \right) e^{-\frac{8\pi^2}{g^2} \sum_i a_i^2} f(a_i) \,.
 }
With this notation, we can write $Z$ and $Z_\mathbb{W}$ as
 \es{ZPestun}{
  Z(m, \tau, \bar \tau) &= \left \llangle Z_\text{pert}(m, a_{ij}) \abs{Z_{\rm inst} (m,\tau,a_{ij})}^2 \right \rrangle \,, \\
   Z_\mathbb{W}(m, \tau, \bar \tau) &= \left \llangle Z_\text{pert}(m, a_{ij}) \abs{Z_{\rm inst} (m,\tau,a_{ij})}^2 \mathbb{W}(a_i) \right \rrangle \,,
 }
where, in the localization computation of \cite{Pestun:2007rz}, $Z_\text{pert}(m, a_{ij})$ arises from the one-loop determinant of fluctuations around the localizing saddle, $Z_{\rm inst} (m,\tau,a_{ij})$ arises as a contribution from instantons localized at a pole of $S^4$, and $\mathbb{W}(a_i)$ is the classical contribution corresponding to the $1\over 2$-BPS fundamental Wilson loop insertion.  Explicitly, 
 \es{ZpertDef}{
  Z_\text{pert}(m, a_{ij}) = \prod_{i, j=1}^N \frac{ H(a_{ij})}{H(a_{ij}+m) } \,, \qquad
   \mathbb{W} (a_i) = \frac{1}{N} \sum_i e^{2\pi a_i} \,,
 }
and $Z_{\rm inst} (m,\tau,a_{ij})$ has a more complicated expression that will be described in more detail later.  In \eqref{ZpertDef}, the function $H$ is given by 
 \es{HDef}{
  H(m) \equiv e^{-(1 + \gamma) m^2} G(1 + i m) G(1-im) \,,
 }
with $G$ being the Barnes $G$-function and $\gamma$ the Euler-Mascheroni constant.  Note that when $m=0$, we have $Z_\text{pert}(0, a_{ij}) = Z_\text{inst}(0, \tau, a_{ij}) = 1$.
 
The formulas \eqref{PartGaussian}--\eqref{ZpertDef} are valid for the SYM theory with gauge group $U(N)$.   However, in all these formulas, the contributions from the $SU(N)$ subgroup of $U(N)$ and those from the $U(1)$ center factorize.  Moreover, as can be seen from taking the ratio of the two expressions in  \eqref{ZPestun}, in the $U(1)$ theory $\langle \mathbb{W} \rangle$ is independent of $m$.  These two facts imply that ${\cal I}_{\mathbb{W}}(\tau, \bar \tau)$ in \eqref{eq:ZWLtwoderiv} is identical for the $U(N)$ and $SU(N)$ theories, even though $Z$ and $Z_\mathbb{W}$ are not.  Thus, even though we are ultimately interested in the $SU(N)$ ${\cal N} = 4$ SYM theory, we will be working in the $U(N)$ case, because many of the formulas below are simpler in this case.

The quantities $Z_\text{pert}$ and $Z_\text{inst}$ are even functions of $m$, which means that all non-zero contributions to $\cI_{\mathbb{W}}(\tau, \bar \tau)$ come from terms where both derivatives with respect to $m$ act either on  $Z_\text{pert}$, $Z_\text{inst}$, or $Z_\text{inst}^*$.  Thus, ${\cal I}_{\mathbb{W}}(\tau, \bar \tau)$ in \eqref{eq:ZWLtwoderiv} can be split into a perturbative contribution and an instanton contribution as
 \es{eq:ZWLtwoderivSum}{
  {\cal I}_{\mathbb{W}}(\tau, \bar \tau) = {\cal I}_{\mathbb{W},\text{pert}} (\tau, \bar \tau) + {\cal I}_{\mathbb{W},\text{inst}}(\tau, \bar \tau) \,.
 }
For the perturbative contribution, we have
 \es{IPertDef}{
  {\cal I}_{\mathbb{W},\text{pert}} (\tau, \bar \tau)  = \frac{\left \llangle \mathbb{W}(a_i) Z_\text{pert}''(0, a_{ij})\right \rrangle
   -  \left\llangle  Z_\text{pert}''(0, a_{ij}) \right \rrangle \left \llangle \mathbb{W}(a_i) \right \rrangle  }{\left \llangle \mathbb{W}(a_i) \right \rrangle }  \,,
 } 
where the double primes denote derivatives with respect to $m$.  The instanton contribution is 
 \es{IInstDef}{
  {\cal I}_{\mathbb{W},\text{inst}} (\tau, \bar \tau)  = \frac{\left \llangle \mathbb{W}(a_i) Z_\text{inst}''(0, \tau, a_{ij})\right \rrangle
   -  \left\llangle  Z_\text{inst}''(0, a_{ij}) \right \rrangle \left \llangle \mathbb{W}(a_i) \right \rrangle  }{\left \llangle \mathbb{W}(a_i) \right \rrangle }  + \text{c.c.} \,.
 }
We will now evaluate these two contributions in turn, after first introducing some useful notation in the next subsection.

\subsection{$n$-body resolvents and $\langle \mathbb{W} \rangle$ at $m=0$}

We will rewrite \eqref{IPertDef} and \eqref{IInstDef} in terms of $n$-body resolvents, which in turn can be efficiently evaluated in a $\frac{1}{N^2}$ expansion using topological recursion \cite{Eynard:2004mh}.  The $n$-body resolvent $W^n(y_1,\ldots,y_n)$ is defined as the expectation value
 \es{eq:defresolv}{
W^n(y_1,\ldots,y_n) \equiv N^{n-2} \Big\llangle \sum_{i_1}\frac{1}{y_1-a_{i_1}} \cdots \sum_{i_n}\frac{1}{y_n-a_{i_n}} \Big\rrangle_{\rm c} \,,
 }
where the subscript ``$c$'' on the angle brackets means that we are extracting only the connected contribution.\footnote{The connected contribution means that we treat each sum in \eqref{eq:defresolv} as an operator and perform Wick contractions only between different operators.}   Furthermore, it admits a large $N$ expansion of the form\footnote{This is often referred to as the ``genus" expansion of the $n$-body resolvent.}
 \es{ResolventExpansion}{
W^n(y_1,\ldots,y_n) = \sum_{m=0}^{\infty} \frac{1}{N^{2m}} W^{n}_{m}(y_1,\ldots,y_n) \,,
 }
where the functions $W^{n}_{m}(y_1,\ldots,y_n)$ can be computed using Eynard's recursion relations \cite{Eynard:2004mh,Eynard:2008we}, which we review in Appendix~\ref{EYNARD}\@.  The powers of $N$ in \eqref{eq:defresolv} are such that the genus-$m$ $n$-body resolvent $W^n_m$ is, up to an overall power of $a = \frac{2 \pi}{\sqrt{\lambda}}$, a function of just $ay_i$:
 \es{ScaleWnm}{
  W^n_m(y_1, \ldots, y_n) = a^n F^n_m (a y_1, \ldots, a y_n) \,, \qquad a \equiv \frac{2 \pi}{\sqrt{\lambda}} \,.
 }
For instance, the first few one-body resolvents are
 \es{Onebody}{
  W_0^1(y) &= 2a^2 y \left(1 - \sqrt{1 - (ay)^{-2}}  \right)   \,, \\
  W_1^1(y) &= a \frac{1}{(a y)^{5} \left( 1 - (a y)^{-2} \right)^{5/2}} \,, \\
  W_2^1(y) &= a \frac{21 (1 + 4 (a y)^2) }{1024 (a y)^{11} \left( 1 - (a y)^{-2} \right)^{11/2}} \,,
 }
and the leading two-body resolvent is
 \es{Twobody}{
  W_0^2(y, z) = \frac{a^2 yz -1 - a^2 yz \sqrt{1 - (ay)^{-2}} \sqrt{1 - (az)^{-2}}}
   {2 (y- z)^2 a^2 y z \sqrt{1 - (ay)^{-2}} \sqrt{1 - (az)^{-2}}} \,.
 } 

Connected expectation values of exponentials, such as that of the Wilson loop operator insertion $\mathbb{W}(a_i)$, can be obtained from the $n$-body resolvent by performing an inverse Laplace transform:
 \es{InverseLaplace}{
  N^{n-2} \Big\llangle \sum_{i_1} e^{z_1 a_{i_1}} \cdots \sum_{i_n} e^{z_n a_{i_n}} \Big\rrangle_{\rm c} = \cL^{-1}\left[ W^{n} \right](z_1,\ldots, z_n) = \int \left(\prod_{i=1}^n \frac{dy_i}{2\pi i}e^{z_i\,y_i} \right) W^n(y_1,\ldots,y_n) \,,
 }
where the integral over $y_i$ is along a vertical contour $(\gamma_i-i\infty, \gamma_i+i\infty)$, with $\gamma_i\in \bR$ being such that the integration contour resides to the right of all singularities of the integrand. 

As an application, we can calculate the expectation value of the $1\over 2$-BPS Wilson loop at zero mass:
\ie
  \langle W \rangle \big|_{m=0} = \left \llangle \mathbb{W}(a_i) \right \rrangle = \cL^{-1} \left[ W^1 \right](2\pi) = \sum_{m=0}^\infty \frac{1}{N^{2m}} \cL^{-1} \left[ W^1_m \right](2\pi).
\label{eq:ZW0}
\fe
Term by term, one can use the inverse Laplace transforms in Appendix~\ref{LAPLACE} (see Eq.~\eqref{LT}), obtaining
\es{WilsonExp}{
  \langle W \rangle \big|_{m=0} = \frac{2 I_1(\sqrt{\lambda})}{\sqrt{\lambda}}
   + \frac{\lambda I_2(\sqrt{\lambda})}{48 N^2}
    + \frac{36 \lambda^2 I_4(\sqrt{\lambda})+ 5 \lambda^{\frac 52} I_5(\sqrt{\lambda})}{46080 N^4} + O(N^{-6}) \,.
}

At leading order in $1/N$ and large $\lambda$, this expression can be approximated as 
 \es{WLeading}{
  \langle W \rangle_{m=0} \approx \sqrt{\frac{2}{\pi}} \frac{e^{\sqrt{\lambda}}}{\lambda^{3/4}} \,.
  }
The exponent $\sqrt{\lambda}$ is matched onto the area of a classical string worldsheet in ${\rm AdS_5}$, and the prefactor multiplying $e^{\sqrt{\lambda}}$ comes from the one-loop fluctuation determinant around this string worldsheet.  The entire expression \eqref{WilsonExp} can be written as the classical contribution \eqref{WLeading} times a function of $\lambda$ and $N$ that can also be expanded in powers of $1/\sqrt{N}$ at fixed $g$:
 \es{Wm0Again}{
   \langle W \rangle \big|_{m=0} =\sqrt{\frac{2}{\pi}} \frac{e^{\sqrt{\lambda}}}{\lambda^{3/4}}
    \left[1 + \frac{-36 + g^4}{96 g \sqrt{N}}
     + \frac{-2160 - 360 g^4 + g^8}{18432 g^2 N} + O(N^{-3/2})   \right]\,.
 }
This expression will be useful later.

\subsection{Perturbative contributions}

In this section, we compute the perturbative contribution \eqref{IPertDef} to the integrated correlator $\cI_{\mathbb{W}}(\tau,\bar\tau)$ using the resolvent technology introduced in the previous section.
One of the main ingredients is the identity  \cite{Russo:2013kea}
 \es{ZpertDer}{
   Z_\text{pert}''(0, a_{ij}) 
    = - \sum_{i, j}\int_0^\infty d\omega\,  \frac{\omega \left[ e^{2 i \omega a_{ij}} + e^{-2 i \omega a_{ij}}  - 2 \right]}{\sinh^2 \omega} \,,
 }
which can be used to write  $Z_\text{pert}''(0, a_{ij}) $ in terms of expectation values of exponentials as in \eqref{InverseLaplace}.
After plugging this expression into \eqref{IPertDef} and using \eqref{InverseLaplace}, we obtain:
 \es{eq:MMpert}{
{\cal I}_{\mathbb{W},\text{pert}} &= - \int_0^\infty d\omega \frac{2\omega}{\sinh^2\omega} \frac{1}{\cL^{-1}[W^{1}](2\pi)} \left[ \frac{1}{N^2}\cL^{-1}[W^3](2i\omega, -2i\omega, 2\pi) \right.\\
&{}+ \left. \cL^{-1}[W^2](2i\omega, 2\pi)\cL^{-1}[W^1](-2i\omega) + \cL^{-1}[W^2](-2i\omega, 2\pi)\cL^{-1}[W^1](2i\omega) \vphantom{\frac{1}{1}}\right] \,.
 }

Expanding each resolvent in powers of $1/N^2$ as in \eqref{ResolventExpansion}, using the explicit formulas such as \eqref{Onebody} and \eqref{Twobody}, as well as the Laplace transforms in \eqref{LT} and \eqref{eq:LinvW2m0}, we can find explicit expressions for (\ref{eq:MMpert}) to any desired order in $1/N$. In particular, with the definition
 \es{IPertExpansion}{
  {\cal I}_{\mathbb{W},\text{pert}}(\lambda, N) = \sum_{m=0}^\infty \frac{1}{N^{2m}} {\cal I}_{\mathbb{W},\text{pert}, m}(\lambda) \,,
 }
one obtains
 \es{eq:MMpertLO}{
{\cal I}_{\mathbb{W},\text{pert}, 0}  =\, & \frac{2\pi\sqrt{\lambda}}{I_1(\sqrt{\lambda})} \int_0^\infty d\omega \frac{\omega}{(\omega^2 + \pi^2) \sinh^2 \omega} J_1\left(\frac{\omega}{\pi}\sqrt{\lambda}\right) \\
&{}\times\left[ \pi I_0\left(\sqrt{\lambda}\right)J_1\left(\frac{\omega}{\pi}\sqrt{\lambda}\right) - \omega I_1\left(\sqrt{\lambda}\right)J_0\left(\frac{\omega}{\pi}\sqrt{\lambda}\right) \right]. \vphantom{\sum_0^0}
 }
This result precisely agrees\footnote{Note that \cite{Russo:2013kea} computed the \emph{unnormalized} Wilson loop expectation value, i.e.~without the factor of $\frac{\sqrt{\lambda}}{2I_1(\sqrt{\lambda})}=\frac{1}{\cL^{-1}[W^1_0](2\pi)}$.} with that found in \cite{Russo:2013kea}.  The analogous results for ${\cal I}_{\mathbb{W},\text{pert}, 1}$, ${\cal I}_{\mathbb{W},\text{pert}, 2}$, and ${\cal I}_{\mathbb{W},\text{pert}, 3}$ are given in Appendix~\ref{PERTURBATIVE}.

Using the method presented in Appendix~\ref{PERTURBATIVE}, these expressions can be further expanded at large~$\lambda$:
 \es{Expansions}{
   {\cal I}_{\mathbb{W},\text{pert}, 0}  &=\sqrt{\lambda} + \left( \frac{1}{2} - \frac{\pi^2}{3} \right) +  \frac{3}{8} \frac{1}{\sqrt{\lambda}} +  \left( \frac{3 \zeta(3)}{2} + \frac{3}{8} \right)\frac{1}{\lambda} 
 + \left(\frac{3 \zeta(3)}{2} + \frac{63}{128} \right) \frac{1}{\lambda^{\frac 32}} \\
 &{}+ \frac{9 \left( 10 \zeta(5) + 7 \zeta(3) + 3 \right)}{32} \frac{1}{\lambda^2}
 + \frac{9\left(640 \zeta(5) + 384 \zeta(3) + 211 \right) }{1024}\frac{1}{\lambda^{\frac 52}} +  \cdots \,, \\
    {\cal I}_{\mathbb{W},\text{pert}, 1}& = -\frac{\lambda^{\frac 32}}{32} - \frac{\sqrt{\lambda}}{256}
    - \frac{4 \zeta(3) + 1}{128}
     + \frac{9\left( 48 \zeta(3) - 7\right)}{4096} \frac{1}{\sqrt{\lambda}}
     - \frac{3 \left(10 \zeta(5) + 7 \zeta(3) + 3 \right) }{256} \frac{1}{\lambda} \\
     &{}-\frac{15 \left(-960 \zeta(5) + 384 \zeta(3) + 211 \right) }{32768}
      \frac{1}{\lambda^{\frac 32}} + \cdots\,, \\
   {\cal I}_{\mathbb{W},\text{pert}, 2}& = \frac{\lambda^3}{11520} + \frac{7 \lambda^{5/2}}{10240}
    - \frac{\lambda^2}{480} + \frac{803 \lambda^{3/2}}{245760} 
    +\frac{9 (-67 + 20 \zeta(3)) \lambda}{163840} + \cdots \,, \\
     {\cal I}_{\mathbb{W},\text{pert}, 3}& = - \frac{\lambda^{\frac 92}}{1935360} 
      + \frac{271 \lambda^4}{3870720} 
       - \frac{4723 \lambda^{\frac 72}}{20643840} 
        - \frac{(237 + 70 \pi^2)\lambda^3}{483840} +  \cdots  \,.
 }

The expansion \eqref{Expansions} is a double expansion in $1/N$ and $1/\lambda$, where first $N$ is taken to be large at fixed $\lambda$, and then $\lambda$ is taken to be large.  The same expansion can be reorganized as a double expansion in $1/N$ and $g$, where first $N$ is taken to be large at fixed $g$, and afterwards each term is expanded at small $g$.  In this limit, \eqref{Expansions} gives
 \es{IPertLargeN}{
  {\cal I}_{\mathbb{W},\text{pert}} &= g \sqrt{N} 
   + \left( \frac 12 - \frac{\pi^2}{3} \right) 
       + \left( \frac{3 }{8 g} - \frac{g^3}{32} \right)  \frac{1}{\sqrt{N}}
     + \left( \frac{3 (1 + 4 \zeta(3))}{8 g^2} + \frac{g^6}{11520} \right) \frac{1}{N} \\
    &{}+ \left( \frac{3(21+ 64 \zeta(3))}{128g^3}  - \frac{g}{256}
     + \frac{7 g^5}{10240} - \frac{g^9}{1935360} \right) \frac{1}{N^{\frac 32}} + O(N^{-2})  \,.
 }
Even though the computations presented above do not prove rigorously that at each order in $1/N$ the power series expansion in $g$ terminates after the terms written in \eqref{IPertLargeN}, we will provide an argument for this fact after we compute the instanton contributions.  As we will see, the instanton contributions start at order $1/N$, so the first three terms in \eqref{IPertLargeN} provide the complete integrated correlator at their respective orders.

\subsection{Instanton contributions}

Let us now turn to the evaluation of ${\cal I}_{\mathbb{W},\text{inst}}(\tau, \bar \tau)$ in \eqref{IInstDef} which captures the instanton contributions for the integrated correlator.  As explained in \cite{Chester:2019jas}, the second derivative $Z_\text{inst}''(0, \tau, a_{ij}) $ can be expanded into contributions from $k \geq 1$ instantons,\footnote{The explicit expression for $Z_{\rm inst}$ was derived in \cite{Pestun:2007rz} based on the result of \cite{Nekrasov:2002qd} but we will only need the simplified expression \eqref{ZinstFormula} for its second mass derivative at $m=0$ derived in \cite{Chester:2019jas}.}
 \es{ZinstFormula}{
  Z_\text{inst}''(0, \tau, a_{ij}) = \sum_{k=1}^\infty e^{2 \pi i k \tau} \sum_{pq = k, 0< p \leq q} 
   I_{p \times q}(a_{ij})   \,,
 }
where the $k$-instanton contribution is a sum of quantities $I_{p \times q}$, where $p$ and $q$ are divisors of $k$, with $pq=k$.  This is a major simplification of the Nekrasov partition function \cite{Nekrasov:2002qd}, which, in the $k$-instanton sector, takes the form of a sum over Young diagrams.  The simplification in \eqref{ZinstFormula} happens because only rectangular Young diagrams contribute for the second mass derivative \cite{Chester:2019jas}.  The quantity $I_{p\times q}$ can further be written as a one-dimensional integral over an auxiliary variable $z$ \cite{Chester:2019jas}:
 \es{Ipq}{
  I_{p\times q} &= \int \frac{dz}{2 \pi}\Biggl[ \left(c_{p, q} + B(z) \right) e^{A(z)}  - c_{p, q} \Biggr] \,,
 }
where the constant $c_{p, q}$ and the functions $A$ and $B$ are 
 \es{GotA}{
  c_{p, q} &\equiv \frac{4}{1 +\delta_{pq}}  \left( \frac{1}{p^2} + \frac{1}{q^2} \right)\,, \\
  A(z) &\equiv \sum_{m=0}^{p-1} \sum_{n=0}^{q-1} \sum_{j=1}^N
   \log \frac{(z - a_j + (m + n ) i)^2}{(z - a_j + (m + n) i)^2 + 1}
     \,, \\
   B(z) &\equiv  \sum_{j=1}^N \frac{2 i (p+q)(q-p)^2 / pq }{(z - a_j + (p + q - 1)i )
    (z - a_j + ( q - 1)i ) (z - a_j + (p - 1)i ) }
    \,.
 }
All that remains to do is to plug \eqref{ZinstFormula} into \eqref{IInstDef}, and evaluate the resulting expression order by order in $1/N$.  This is a complicated task, so let us begin with the simpler case $p=q=1$, where $B(z)=0$ and the expression for $A(z)$ also simplifies:
 \es{ASimp}{
   p=q=1: \qquad A(z) =  \sum_{j=1}^N
   \log \frac{(z - a_j )^2}{(z - a_j )^2 + 1}
     \,, \qquad B(z) = 0 \,.
 }

 \subsubsection{One-instanton contribution}

The $p=q=1$ contribution to \eqref{IInstDef} involves the quantity (we suppress the $a_i$ dependence below)
  \es{ToEvalGen}{
 \frac{\llangle \mathbb{W} I_{1 \times 1}\rrangle -\llangle \mathbb{W} \rrangle \llangle I_{1 \times 1} \rrangle}{\llangle \mathbb{W} \rrangle} 
  = \frac{2}{\pi} \int dz \, 
    \frac{ \left \llangle \mathbb{W} e^A  \right \rrangle - \llangle \mathbb{W} \rrangle \llangle e^A \rrangle }{\llangle \mathbb{W} \rrangle} \,.
  }
In order to evaluate this expression at large $N$, we should expand the integrand in connected correlators as follows: 
  \es{WeA}{
    \left \llangle \mathbb{W} e^A  \right \rrangle - \llangle \mathbb{W} \rrangle \llangle e^A \rrangle
     &= \left( \llangle \mathbb{W} A \rrangle_c + \frac 12 \llangle \mathbb{W} A A \rrangle_c 
       + \frac 16 \llangle \mathbb{W} A A A \rrangle_c + \cdots \right)  \\
       &{}\times \exp \left[ \llangle A \rrangle 
   + \frac 12 \llangle A A \rrangle_c + \frac{1}{6} \llangle AAA \rrangle_c +  \cdots \right]  \,.
  }
 
When $N$ is large and $g$ is kept fixed, the 't Hooft coupling $\lambda = g^2 N$ is large, and thus the typical values of the $a_i$, of order $\sqrt{\lambda}$, will also be large.  Since in the expression for $A(z)$, the variable $a_i$ appears in the combination $a_i - z$, we should also consider values of $z$ that are large and of order $\sqrt{\lambda}$.  Thus, we should expand $A(z)$ in the regime where both $z$ and $a_j$ are large and of the same order.  In this limit, we have
  \es{ExponentExpansion}{
   A(z)
    =  -  R_2(z) + \frac 12 R_4(z) + \cdots   \,,
  }
where we defined
 \es{RDef}{
  R_n(z) \equiv  \sum_j \frac{1}{(z - a_j)^n} \,.
 }

The contribution to the first two non-trivial orders in the $1/N$ expansion (which will end up being $1/N$ and $1/N^{3/2}$) are given by keeping only the first terms in the prefactor and exponent of \eqref{WeA} after plugging in \eqref{ExponentExpansion}:
  \es{WeALeading}{
   \frac{\left \llangle \mathbb{W} e^A  \right \rrangle - \llangle \mathbb{W} \rrangle \llangle e^A \rrangle}{\llangle \mathbb{W} \rrangle}
     \approx -  \frac{\llangle \mathbb{W} R_2 \rrangle_c}{\llangle \mathbb{W} \rrangle} e^{-\llangle R_2 \rrangle} \,.
  }
It follows from \eqref{RDef} and the definition \eqref{eq:defresolv} that $\llangle R_2 \rrangle$ can be written as a derivative of the one-body resolvent and that $\llangle \mathbb{W} R_2 \rrangle_c$ can be written as an inverse Laplace transform of the two-body resolvent with respect to only one of the variables:
  \es{WR2}{
  \llangle R_2(z) \rrangle &= - N \partial_z W^1(z) \,, \\
  \llangle \mathbb{W} R_2(z) \rrangle_c &= - \frac{1}{N} \int \frac{dy}{2 \pi i} \partial_z W^2(y, z) e^{2 \pi y}  \,.
  }
 Thus, for \eqref{ToEvalGen}, we obtain
  \es{WI1Exp2}{
      \frac{\llangle \mathbb{W} I_{1 \times 1} \rrangle - \llangle \mathbb{W}\rrangle \llangle I_{1 \times 1} \rrangle}{ \llangle \mathbb{W} \rrangle}
    \approx \frac{2}{N \pi} \int dz \int \frac{dy}{2 \pi i} \frac{\partial_z W^2(y, z)}{\llangle \mathbb{W} \rrangle} e^{2 \pi y}  \, 
    e^{N \partial_z W^1(z)}  \,.
  }

One can perform a systematic analysis of which terms to keep in the large $N$ limit.  It turns out that for up to order $1/N^{3/2}$ in the large $N$ expansion, we can approximate $W^2$ up to second order in $1/N^2$ and keep only the leading term in $W^1$:
 \es{WI1Exp2Approx}{
      \frac{\llangle \mathbb{W} I_{1 \times 1} \rrangle - \llangle \mathbb{W}\rrangle \llangle I_{1 \times 1} \rrangle}{\llangle \mathbb{W} \rrangle}
    \approx \frac{2}{N \pi} \int dz \int \frac{dy}{2 \pi i} \frac{ \partial_z W^2_0(y, z)
     + \frac{1}{N^2} \partial_z W^2_1(y, z)}{\llangle \mathbb{W} \rrangle} e^{2 \pi y}  \, 
    e^{N \partial_z W^1_0(z)}  \,.
  }

The remaining challenge is to first evaluate the inverse Laplace transform, and then the integral over $z$.  The Laplace transform of $W^2_0$ is not available in closed form, but an asymptotic form at large $\lambda$ (or small $a = 2 \pi /\sqrt{\lambda}$) and fixed $a z$ suffices:
 \es{TwoBodyLapApprox}{
  \int \frac{dy}{2 \pi i} W^2_0(y, z) e^{2 \pi y}
   &= \frac{e^{ \frac{2 \pi}{a}} a^{\frac 32} \sgn z}{\sqrt{ a^2 z^2 - 1}}
    \left[ \frac{1}{ 4 \pi( a z - 1)} - \frac{3 (3 + a z)}{64 \pi^2 (a z - 1)^2} a + \cdots \right] \,.
 } 
The first term will contribute at order $1/N$ in the final answer and the second term at order $1/N^{3/2}$.  The inverse Laplace transform of $W^2_1$ can be written in closed form, but its approximate form at small $a$ is also sufficient
 \es{TwoBodyCorrLapApprox}{
 \int \frac{dy}{2 \pi i} W^2_1(y, z) e^{2 \pi y}
   &= \frac{e^{ \frac{2 \pi}{a}}  \sgn z}{a^{\frac 32} \sqrt{ a^2 z^2 - 1}} \left[ \frac{\pi^2}{48 (a z - 1)} + \cdots \right] \,,
 } 
and it contributes only at order $1/N^{3/2}$ in the final answer.

Plugging \eqref{TwoBodyLapApprox} and \eqref{TwoBodyCorrLapApprox} into \eqref{WI1Exp2Approx}, evaluating the $z$ integral, and expanding at large $N$ and fixed $g$, one then finds 
 \es{instFinal}{
  \frac{\llangle \mathbb{W} I_{1 \times 1}\rrangle -\llangle \mathbb{W} \rrangle \llangle I_{1 \times 1} \rrangle}{\llangle \mathbb{W} \rrangle}
   &= - \frac{ 3 g^2 e^{\frac{8 \pi^2}{g^2}} K_2 ( \frac{8 \pi^2}{ g^2})}{4 \pi^2} \frac{1}{N} \\
    &{}+ \frac{9 g e^{\frac{8 \pi^2}{g^2}} 
     \left( 4 \pi^2 K_2 ( \frac{8 \pi^2}{ g^2}) + 5 g^2 K_3 ( \frac{8 \pi^2}{ g^2}) \right) }
     {32 \pi^4} \frac{1}{N^{\frac 32}} + O(N^{-2})  \,.
 }

\subsubsection{$k$-instanton contributions}

For general instanton contributions to \eqref{IInstDef}, we need to evaluate
  \es{WIpqExp}{
   \frac{\llangle \mathbb{W} I_{p\times q} \rrangle 
    - \llangle \mathbb{W}\rrangle \llangle I_{p \times q} \rrangle}{\llangle \mathbb{W} \rrangle}
    =  \int \frac{dz}{2 \pi}
     \left[ c_{p, q}  \frac{ \left \llangle \mathbb{W} e^A  \right \rrangle - \llangle \mathbb{W} \rrangle \llangle e^A \rrangle }{\llangle \mathbb{W} \rrangle} 
   +  \frac{ \left \llangle \mathbb{W}Be^A   \right \rrangle - \llangle \mathbb{W} \rrangle \llangle  Be^A \rrangle }{\llangle \mathbb{W} \rrangle} \right]  
  }
instead of \eqref{ToEvalGen}.   For the first term in the square bracket, we can use the identity \eqref{WeA}, while for the second term we can use 
  \es{WBeA}{
    \left \llangle \mathbb{W}Be^A   \right \rrangle - \llangle \mathbb{W} \rrangle \llangle  Be^A \rrangle 
     = \biggl( \llangle \mathbb{W}B\rrangle_c
     + \llangle \mathbb{W}A\rrangle_c  \llangle B\rrangle
      + \llangle \mathbb{W}AB \rrangle_c + \cdots \biggr)e^{\llangle A \rrangle
     + \frac 12 \llangle AA \rrangle_c + \cdots}
  }

To obtain the analog of \eqref{WeALeading}, we should expand the functions $A$ and $B$ in terms of the $R_n$ defined in \eqref{RDef}:
 \es{ABExpansions}{
  A &= - k R_2 + i k (p + q - 2) R_3 
   + \frac{k ( 2 p^2 + 2 q^2  + 3 p q - 6 p - 6 q + 6}{2} R_4 + \cdots \,, \\
   B &= \frac{2 i (p+q)(p-q)^2}{k} R_3 +
    \frac{2 (p-q)^2 (p+q)(2p + 2q - 3)}{k }R_4 +  \cdots \,.
 }
In order to obtain the contribution at orders $1/N$ and $1/N^{3/2}$, we should keep only the leading terms in \eqref{WeA} and \eqref{WBeA} (both in the exponent and in the prefactor), and then further approximate the $A$ appearing in the exponent with $-kR_2$ and keep all the terms involving $R_2$ and $R_3$ in the prefactor:
 \es{WIpqLeading}{
  \frac{\llangle \mathbb{W} I_{p \times q}\rrangle -\llangle \mathbb{W} \rrangle \llangle I_{p \times q} \rrangle}
   {\llangle \mathbb{W} \rrangle} 
   &= \int \frac{dz}{2 \pi} \biggl[
     c_{p, q}  \frac{-k\llangle \mathbb{W} R_2 \rrangle_c
     + i k (p + q - 2)    \llangle \mathbb{W} R_3 \rrangle_c }{\llangle \mathbb{W} \rrangle} \\
     &{}+ \frac{2 i (p+q)(p-q)^2}{k} \frac{\llangle \mathbb{W}R_3 \rrangle_c}{\llangle \mathbb{W} \rrangle} 
     +\cdots \biggr]
     \times e^{-k \llangle R_2 \rrangle
      + \cdots} \,.
 }
Now, in addition to \eqref{WR2}, we also need the fact that $ \llangle \mathbb{W} R_3(z) \rrangle_c = -\frac 12 \partial_z \llangle \mathbb{W} R_2(z) \rrangle_c$, so we can write 
 \es{WIpqLeadingAgain}{
  \frac{\llangle \mathbb{W} I_{p \times q}\rrangle -\llangle \mathbb{W} \rrangle \llangle I_{p \times q} \rrangle}
   {\llangle \mathbb{W} \rrangle} 
   &= \frac{k}{2N} \int \frac{dz}{2 \pi} \int \frac{dy}{2 \pi i} \biggl[
     c_{p, q}  \frac{2 \partial_z W^2(y, z) 
     + i  (p + q - 2)   \partial_z^2 W^2(y, z) }{\llangle \mathbb{W} \rrangle} \\
     &{}+ \frac{2 i (p+q)(p-q)^2}{k^2} \frac{ \partial_z^2 W^2(y, z)}{\llangle \mathbb{W} \rrangle} 
     +\cdots \biggr]
     e^{2 \pi y}  e^{k N \partial_z W^1} \,.
 }

The next step is to expand each of the $W^n$'s on the RHS in $1/N^2$ and keep only the first few terms.  For obtaining the final answer up to order $1/N^{3/2}$, we need
 \es{WIpqLeadingAgain2}{
&  \frac{\llangle \mathbb{W} I_{p \times q}\rrangle -\llangle \mathbb{W} \rrangle \llangle I_{p \times q} \rrangle}
   {\llangle \mathbb{W} \rrangle} 
   \\
   &= \frac{k}{2N} \int \frac{dz}{2 \pi} \int \frac{dy}{2 \pi i} \biggl[
     c_{p, q}  \frac{2 \partial_z W_0^2(y, z) + \frac{2}{N^2}  \partial_z W_1^2(y, z) 
     + i  (p + q - 2)   \partial_z^2 W_0^2(y, z) }{\llangle \mathbb{W} \rrangle} \\
     &{}+ \frac{2 i (p+q)(p-q)^2}{k^2} \frac{ \partial_z^2 W_0^2(y, z)}{\llangle \mathbb{W} \rrangle} 
     +\cdots \biggr]
     e^{2 \pi y}  e^{k N \partial_z W^1_0} \,.
 }
Using again \eqref{TwoBodyLapApprox} and \eqref{TwoBodyCorrLapApprox}  to do the $y$ integrals, and then evaluating the $z$ integral and expanding at large $N$ and fixed $g$, one finds the final answer
 \es{instFinal2}{
  \frac{\llangle \mathbb{W} I_{p \times q}\rrangle -\llangle \mathbb{W} \rrangle \llangle I_{p \times q} \rrangle}{\llangle \mathbb{W} \rrangle}
   &= - \frac{ 3 g^2 (p^2 + q^2) e^{\frac{8 k \pi^2}{g^2}} K_2 ( \frac{8 \pi^2 k}{ g^2})}{4 \pi^2 k^2 (1 + \delta_{p, q}) } \frac{1}{N} \\
    &\hspace{-1.7in}{}+ \biggl[ e^{\frac{8 k \pi^2}{g^2}} K_2 ( \frac{8 \pi^2 k}{ g^2})
     \left( \frac{3 i g (p-q)^2 (p+q)}{4 \pi k^2}
      + \frac{ 3 g (p^2 + q^2) (3 + 4 \pi i (p + q - 2))}{8 \pi^2 k^2 (1 + \delta_{p, q})} \right) \\
      &\hspace{-1.7in}{}+e^{\frac{8 k \pi^2}{g^2}} K_3 ( \frac{8 \pi^2 k}{ g^2})
     \left( \frac{45  i g^3 (p-q)^2 (p+q)}{64 \pi^3 k^3}
      + \frac{ 45 g^3 (p^2 + q^2) (1 +  \pi i (p + q - 2))}{32 \pi^4 k^3 (1 + \delta_{p, q})} \right)  \biggr]  \frac{1}{N^{\frac 32}}\\
      &\hspace{-1.7in}{}+ O(N^{-2})  \,.
 }

As per \eqref{IInstDef}, the full contribution to ${\cal I}_{\mathbb{W},\text{inst}}$ is
 \es{IInstSum0}{
  {\cal I}_{\mathbb{W},\text{inst}} = 
   \sum_{k=1}^\infty e^{2 \pi i k \tau} \sum_{pq=k, 0<p\leq q} \left[ \frac{\llangle \mathbb{W} I_{p \times q}\rrangle -\llangle \mathbb{W} \rrangle \llangle I_{p \times q} \rrangle}{\llangle \mathbb{W} \rrangle} \right] + \text{c.c.} \,.
 }
Plugging \eqref{instFinal2} into \eqref{IInstSum0}, and rewriting all the sums over $p$ and $k$ in terms of the divisor functions $\sigma_a(k) = \sum_{p|k} p^a$, we find
 \es{IInstSum}{
  {\cal I}_{\mathbb{W},\text{inst}} &=  
  \sum_{k=1}^\infty \biggl[  - \frac{3 g^2}{2\pi^2}   \cos (k \theta) K_2 (8 \pi^2 k / g^2) \sigma_{-2}(k) \frac{1}{N} \\
     &{}+ \frac{1}{N^{3/2}} \biggl[\cos (k \theta) \frac{9 g \sigma_{-2}(k)}{4 \pi^2}  \left(  K_2 (8 \pi^2 k / g^2) + \frac{5 g^2 K_3 (8 \pi^2 k / g^2)}{4 \pi^2 k } \right) \\
     &{}- \sin (k \theta) \frac{3g  \left( 3 k \sigma_{-3}(k) - 4 \sigma_{-2}(k) + \sigma_{-1}(k)  \right)}{2 \pi}  \left(  K_2 (8 \pi^2 k / g^2) + \frac{15 g^2 K_3 (8 \pi^2 k / g^2)}{16 \pi^2 k }  
       \right)  \biggr] \\
       &{}+ \cdots \biggr] \,.
 }
Combining this expression with the perturbative result \eqref{IPertLargeN}, we obtain \eqref{IExplicit}.

\section{$SL(2,\mZ)$ duality and Wilson-'t Hooft Loop}
\label{sec:tL}

\subsection{Integrated two-point function with 't Hooft loop}

As mentioned in the Introduction, the $\cN=4$ SYM theory enjoys an $SL(2,\bZ)$ electromagnetic duality. In particular, under S-duality, the  Wilson line $\mathbb{W}$ in the $SU(N)$ SYM theory is mapped to the 't Hooft line $\mathbb{T}$ in the $PSU(N)_0$ SYM theory, where the subscript of $PSU(N)$ denotes the $\mZ_N$ discrete theta angle \cite{Aharony:2013hda} (see also Footnote~\ref{footnote:globalstructure}), and the two-point function of local 1/2-BPS operators in the presence of the Wilson loop is expected to obey
\ie \label{Sdual2pt}
\left\langle \mathbb{W} \cO(\vec{x}_1) \cO(\vec{x}_2) \right\rangle^{SU(N)}_{\tau,\bar{\tau}} = \left\langle \mathbb{T} \cO(\vec{x}_1) \cO(\vec{x}_2) \right\rangle^{PSU(N)_0}_{-{1/\tau}, -{1/ \bar{\tau}}}\,.
\fe
Here, we have used the fact that the canonically normalized $1\over 2$-BPS operators are $SL(2,\mZ)$ invariant \cite{Intriligator:1998ig,Intriligator:1999ff}.
Under the T-transformation, the $SU(N)$ Wilson loops are invariant,
\ie \label{Tdual2pt0}
\left\langle \mathbb{W} \cO(\vec{x}_1) \cO(\vec{x}_2) \right\rangle^{SU(N)}_{\tau,\bar{\tau}} = \left\langle \mathbb{W} \cO(\vec{x}_1) \cO(\vec{x}_2) \right\rangle^{SU(N)}_{\tau+1,\bar{\tau}+1}\,,
\fe
while the 't Hooft loops in the $PSU(N)_n$ SYM with discrete theta angle $n\in \mZ_N$ transform into Wilson-'t Hooft loops \cite{Aharony:2013hda} as a consequence of the Witten effect \cite{Witten:1978mh},
\ie \label{Tdual2pt}
\left\langle \mathbb{W}_k\mathbb{T} \cO(\vec{x}_1) \cO(\vec{x}_2) \right\rangle^{PSU(N)_k}_{\tau,\bar{\tau}}
=
\left\langle \mathbb{W}_{k-1}\mathbb{T} \cO(\vec{x}_1) \cO(\vec{x}_2) \right\rangle^{PSU(N)_{k-1}}_{\tau+1,\bar{\tau}+1}\,,
\fe
where $\mathbb{W}_{k}\mathbb{T}$ denotes the charge $(p,q)=(k,1)$ dyonic loop. The more general $(p,q)$ dyonic loops (with coprime $p$ and $q$) can be obtained by combining the $S$ and $T$ transformations (see \cite{Aharony:2013hda} for details). 

Here we will mostly focus on the Wilson line in the fundamental representation of $SU(N)$ and correspondingly the dual  fundamental 't Hooft line. We comment on the generalization to charge $(p,q)$ dyonic loops towards the end of the section.
For reasons explained in  Footnote~\ref{footnote:globalstructure}, on $S^4$, \eqref{Sdual2pt}, \eqref{Tdual2pt0}, and \eqref{Tdual2pt} can all be interpreted as relations between defect correlators in the same $SU(N)$ SYM.\footnote{It would be interesting to study the SYM on manifolds that are sensitive to such discrete data, such as on $\mathbb{RP}^4$ \cite{Wang:2020jgh,Caetano:2022mus}.}

For the case of the 1/2-BPS 't Hooft line, the RHS of (\ref{Sdual2pt}), integrated over $\vec{x}_1$ and $\vec{x}_2$ in an appropriate sense, may be computed in the very strong coupling limit of large $N$ and fixed $\tilde \tau \equiv -{1\over \tau}$ through supersymmetric localization yielding a corresponding matrix model computation of the form (\ref{eq:ZWLtwoderiv}) \cite{Gomis:2009ir, Gomis:2009xg,Gomis:2010kv,Gomis:2011pf}. In this paper, we will not attempt this direct matrix model calculation using the 't Hooft loop description. Instead, we will take  the non-perturbative results presented in Section \ref{INTEGRATED} for the LHS of (\ref{Sdual2pt}) and implement the S transformation which determines the RHS of \eqref{Sdual2pt} immediately. In particular, the strong gauge coupling limit of the Wilson loop correlator in \eqref{Sdual2pt} translates to the weak coupling expansion for the 't Hooft loop.
As we will see, the fact that the RHS of (\ref{Sdual2pt}) should have a well-defined expansion in the dual 't Hooft coupling $\lambdad=\gymd^2 N$, with $\gymd=\frac{4 \pi}{\gym}$ at large $N$  provides a strong consistency check of the results of Section \ref{INTEGRATED}.

Just as in the case of the Wilson loop in \eqref{eq:ZWLtwoderiv}, let us denote the second mass derivative of the $S^4$ partition function in the presence of a $1\over 2$-BPS 't Hooft loop by
 \es{ZTLtwoderiv}{
   {\cal I}_{\mathbb{T}}(\tilde \tau, \bar{\tilde \tau}) = \partial_m^2 \log \langle \mathbb{T} \rangle \big|_{m=0} \,,
 }
and expand this quantity in $1/N$ by analogy with \eqref{Expansion}
 \es{ExpansiontHooft}{
   {\cal I}_{\mathbb{T}}(\tilde \tau, \bar{\tilde \tau}) =
     {\cal I}_{\mathbb{T},-1/2}(\tilde\tau, \bar{\tilde \tau}) \sqrt{N} 
   +{\cal I}_{\mathbb{T},0}(\tilde\tau, \bar{\tilde \tau})
    +\frac{\tilde{\cal I}_{\mathbb{T},1/2}(\tilde \tau, \bar{\tilde \tau})}{\sqrt{N}} + \frac{ {\cal I}_{\mathbb{T},1}(\tilde\tau, \bar{\tilde \tau})}{N} 
     + \cdots\,.
 }
Of course, at each order in the $1/N$ expansion, we can simply plug in $\tau = - 1/ \tilde \tau$ in the expressions in \eqref{IExplicit} given at the beginning of the previous section:
 \es{ItildeIRelation}{
 {\cal I}_{\mathbb{T},k} (\tilde \tau, \bar{\tilde \tau})
   =  {\cal I}_{\mathbb{W},k} (-1/\tilde \tau, -1/\bar{\tilde \tau}) \,.
 }

While the expansion \eqref{ExpansiontHooft} is in the very strong coupling limit, one can also consider the resummation of the perturbative terms into the 't Hooft expansion in this duality frame.  In \eqref{IPertExpansion}, we saw that the 't Hooft expansion of the Wilson loop correlator is in even powers of $1/N$, starting at $N^0$, with each term being a function of the 't Hooft coupling $\lambda$.  In the case of the 't Hooft loop correlator $\langle \mathbb{T} {\cal O} {\cal O} \rangle$, we expect an expansion of the perturbative part $\tilde {\cal I}_{\mathbb{T},\text{pert}}$ of \eqref{ExpansiontHooft} of the following form
 \es{PerttHooft}{
 {\cal I}_{\mathbb{T},\text{pert}} = N \sum_{m=0}^\infty \frac{1}{N^m}   {\cal I}_{\mathbb{T},\text{pert}, m}(\tilde \lambda) \,.
 }
As we will see in Section~\ref{sec:EFT}, this expression is in agreement with the effective field theory in AdS\@.

\subsection{Very strong coupling expansion}
\label{sec:verystrongtH}

Writing $\tilde \tau = \frac{\tilde \theta}{2 \pi} + \frac{4 \pi i}{\tilde g}$, we can easily convert the expressions \eqref{ItildeIRelation} into functions of $\tilde g$ and $\tilde \theta$.  At the first few orders in $1/N$ this is easily done, as follows.

\paragraph{Order $N^{1/2}$.} In particular, at order $N^{1/2}$, we have
 \es{tildeIm12}{
  {\cal I}_{\mathbb{T},-1/2} = \sqrt{\frac{16 \pi^2}{\tilde g^2} + \frac{\tilde g^2 \tilde{\theta}^2}{4 \pi^2}} \,.
 }
At small $\tilde{g}$, this expression gives, approximately, $4 \pi \sqrt{N} / \tilde g = 4 \pi N / \sqrt{\tilde \lambda}$, which matches the form \eqref{PerttHooft}.

\paragraph{Order $N^{0}$.} At order $N^0$, given that ${\cal I}_{\mathbb{W},0}$ in \eqref{IExplicit} is independent of $\tau$ and $\bar \tau$, the relation \eqref{ItildeIRelation} implies that ${\cal I}_{\mathbb{T},0}$ is also independent of $\tilde g$ and $\tilde \theta$:
 \es{tildeI0}{
  {\cal I}_{\mathbb{T},0}= \frac{1}{2} - \frac{\pi^2}{3} \,.
 }

\paragraph{Order $N^{-1/2}$.}  Next, at order $N^{-1/2}$, we have
 \es{I12}{
 {\cal I}_{\mathbb{T},1/2} =   \frac{-\frac{8 \pi^4}{\tilde g^4} + \frac 38 - \frac{\tilde \theta^2}{4} - \frac{\tilde g^4 \tilde \theta^4}{512 \pi^4}}{\sqrt{\frac{16 \pi^2}{\tilde g^2} + \frac{\tilde g^2 \tilde{\theta}^2}{4 \pi^2}}} \,.
 }
At small $\tilde g$, this gives $- 2 \pi^3 / (\tilde g^3 N^{1/2}) = - 2 \pi^3 N / \tilde \lambda^{3/2} $, which again matches the form \eqref{PerttHooft}.

\paragraph{Order $N^{-1}$.} The calculation becomes more intricate starting at order $1/N$, where there is an interesting interplay between the perturbative terms and the instanton contributions from \eqref{IExplicit}. For simplicity, let us restrict our discussion to $\tilde \theta = \theta = 0$, where from \eqref{IExplicit} and \eqref{ItildeIRelation} we have 
  \es{tildeI1}{
  {\cal I}_{\mathbb{T},1} \big|_{\tilde \theta = 0}  &= \frac{16 \pi^6}{45 \tilde g^6} + \frac{3 \tilde g^2 ( 1+ 4 \zeta(3))}{128 \pi^2}  
   - \frac{24 }{\tilde g^2}  \sum_{k=1}^\infty  K_2 \left( k \tilde g^2 / 2 \right) \sigma_{-2}(k) \,.
  }
If we considered only the first two terms in this equation, we would conclude that at small $\tilde g$ the $\langle \mathbb{T} {\cal O} {\cal O} \rangle$ correlation function behaves as $\sim 1/(N \tilde g^6) = N^2/ \tilde \lambda^3$, which is in contradiction with the expected planar limit expansion in \eqref{PerttHooft}.  Thus, it must be that the behavior of the last term in \eqref{ItildeIRelation} at small $\tilde g$ is also $\sim 1/ \tilde g^6$, with a coefficient that precisely cancels the contribution from the first term.  

Indeed, the small $\tilde g$ asymptotics of the last term in \eqref{tildeI1} can be found by passing to the Mellin representation of the Bessel $K$ functions:
 \es{BesselKMellin}{
   K_t(2 \pi  y)= \frac{1}{8 \pi i } \int_{\C-i\infty}^{\C+i\infty}   \frac{ds}{ (\pi y)^s}  \Gamma \left( \frac{s + t}{2} \right) \Gamma \left( \frac{s-t}{2} \right)  \,,
 }
with $\C >{\rm Re}(t)$. This, together with the identity
 \es{sigmaSum}{
   \sum_{k=1}^\infty \frac{\sigma_a(k)}{k^b}=\zeta(b)\zeta(b-a)\,,
 }
gives the general formula
 \es{BesselSum}{
   \sum_{k=1}^\infty \frac{\sigma_{-\B}(k)}{k^\A} K_t(2 \pi k y)
      =
 \frac{1}{ 8 \pi i} \int_{\C-i\infty}^{\C+i\infty} \frac{ds}{(\pi y)^s} \Gamma \left(\frac{s+t}{2}\right) \Gamma\left(\frac{s-t}{2}\right)
\zeta(\A+s)\zeta(\A+s+\B) \,.
 }
For the case relevant for \eqref{tildeI1}, we have $t = \beta = 2$ and $\alpha = 0$, so 
 \es{BesselSum2}{
   \sum_{k=1}^\infty \sigma_{-2}(k) K_2(2\pi k  y)
      =
 \frac{1}{ 8 \pi i} \int_{\C-i\infty}^{\C+i\infty} \frac{ds}{(\pi y)^s} \Gamma \left(\frac{s+2}{2}\right) \Gamma\left(\frac{s-2}{2}\right)
\zeta(s)\zeta(s+2) \,.
 }
It is useful to define the completed Riemann zeta function as $\Lambda(s)=\pi^{-s} \Gamma(s)\zeta(2s)$, which satisfies the reflection symmetry
\ie 
\Lambda(s)=\Lambda\left({1\over 2}-s\right)\,.
\label{zetareflection}
\fe
Consequently, after implementing the reflection identity \eqref{zetareflection} and a change of variables $s\to -s$, \eqref{BesselSum2} can be rewritten as
 \es{sigmaK2}{
\sum_{k=1}^\infty \sigma_{-2}(k) K_2(2\pi k  y)
      =
 \frac{i}{4} \int_{-\C-i\infty}^{-\C+i\infty} ds\,   \frac{ y^s}{s+2}\Lambda\left(\frac{s+1}{ 2}\right)\Lambda\left(\frac{s-1}{2}\right) \,.
 }
The completed Riemann zeta function $\Lambda(s)$ is meromorphic with a pair of simple poles at $s=0,{1\over 2}$ related by the reflection \eqref{zetareflection} and the residues are ${\rm Res}_{s=0}\Lambda(s)=-{\rm Res}_{s={1\over 2}}\Lambda(s)=-{1\over 2}$.

Closing the contour on the right in \eqref{sigmaK2} (recall $\C>2$) and picking up the five poles at $s=\pm 2, \pm 1,0$, we obtain
 \es{BesselAsymp}{
 \sum_{k=1}^\infty  \sigma_{-2}(k)  K_2(2\pi k   y)=&\,
\frac{\pi ^4}{1080 y^2}-\frac{ \zeta (3)}{4  y}+\frac{\pi ^2}{24}-\frac{\pi^2 y}{36}+\frac{y^2 \zeta (3)}{16 }
\\
-& 
{y\over 2 }\sum_{k=1}^\infty \sigma_{-2}(k) 
G^{3,0}_{1,3} 
\left(
\left.\begin{array}{c}
{5\over 2} \\
 0,1,{3\over2} \\
\end{array} \right| (\pi k /y)^2
\right)
\,,
 }
where the remaining contour integral gives a sum of the Meijer $G$-function $G^{3,0}_{1,3}$. Using the asymptotic formula \cite{Luke:1969:SFA},
\ie 
G^{3,0}_{1,3} 
\left(
\left.\begin{array}{c}
{5\over 2} \\
 0,1,{3\over2} \\
\end{array} \right| w
\right) = {\sqrt{\pi}\over w^{1\over 4}}e^{-2\sqrt{w}} \left ( 1-\frac{25}{16 \sqrt{w}}+\frac{1785}{512 w}-\frac{78435}{8192 w^{3/2}}+\frac{16309755}{524288 w^2} +\dots \right)\,,
\fe
we see clearly that the Meijer $G$-functions capture exponentially suppressed contributions at small $y$.

Quite nicely, after plugging in $y = {\tilde g^2 \over 4\pi}$ and using \eqref{BesselAsymp} in \eqref{tildeI1}, we see that the most singular term at small $\tilde g$ in \eqref{tildeI1} as well as the term with an explicit $\zeta(3)$ dependence are canceled and the final result takes the form:
 \es{tildeI1Again}{
  {\cal I}_{\mathbb{T},1} \big|_{\tilde \theta = 0}  &=  
  \frac{24 \pi \zeta(3)}{\tilde g^4} - \frac{\pi^2}{\tilde g^2}
   + \frac{\pi}{6} + \frac{3 \tilde g^2}{128 \pi^2}
  +{3\tilde g\over 2\pi^{3\over 2}} e^{-{8\pi^2 \over \tilde g^2}}
  \left(1-{25 \tilde g^2\over 64\pi^2}+\cO(\tilde g^4)\right)+\cO\left(e^{-{16\pi^2 \over \tilde g^2}}\right)
   \,.
  }
Remarkably, this expression has only a finite number of perturbative terms.  The leading such term gives a contribution to the $\langle \mathbb{T} {\cal O} {\cal O} \rangle$ correlator proportional to $1/(\tilde g^4 N) \sim N / \tilde \lambda^2$, which matches the expectation in \eqref{PerttHooft}.

\paragraph{Order $N^{-3/2}$.} One can perform a similar analysis at order $1/N^{3/2}$, where, for $\tilde \theta = \theta = 0$ from \eqref{IExplicit}, we find
 \es{Itilde32}{
   {\cal I}_{\mathbb{T},\frac 32} \big|_{\tilde \theta = 0} &= 
   - \frac{ 128 \pi^9}{945 \tilde g^9} + \frac{7 \pi^5}{10 \tilde g^5}  - \frac{ \pi}{64 \tilde g} + \frac{3(21+ 64 \zeta(3)) \tilde g^3}{8192 \pi^3 }
       \\
     &{}+\sum_{k=1}^\infty \biggl[ \frac{9  \sigma_{-2}(k)}{\pi \tilde g}  \left(  K_2 (k \tilde g^2/2) + \frac{20  K_3 (k \tilde g^2/2)}{ k  \tilde g^2} \right)  \biggr] \,.
 }
Again, if we were to ignore the last term, we would conclude that there is a contribution to the $\langle \mathbb{T} {\cal O} {\cal O} \rangle$ correlator proportional to $1/(\tilde g^9 N^{3/2}) \propto N^3 / \lambda^{9/2}$, which would disagree with the expectation \eqref{PerttHooft}.  However, we can also expand the last term in \eqref{Itilde32} at small $\tilde g$ using the formula
  \es{BesselAsymp2}{
 \sum_{k=1}^\infty  \frac{\sigma_{-2}(k)}{k}  K_3(2 \pi k y) &=
   \frac{\pi^{7}}{85050 y^3}-\frac{\pi^5}{1080 y}+\frac{\pi  \zeta (3)}{6}-\frac{\pi^3 y}{48}+\frac{\pi^3 y^2}{90}
    - \frac{\zeta(3) \pi  y^3}{48  } \\
    &{}+
    {\pi y^2\over 2 }\sum_{k=1}^\infty \sigma_{-2}(k) 
G^{3,0}_{1,3} 
\left(
\left.\begin{array}{c}
{7\over 2} \\
 0,1,{3\over 2} \\
\end{array} \right| (\pi k /y)^2
\right)
\,,
 }
which was derived in the same way as \eqref{BesselAsymp}.  Using both \eqref{BesselAsymp} and \eqref{BesselAsymp2}, we can then rewrite
 \es{Itilde32Again}{
   {\cal I}_{\mathbb{T},\frac 32} \big|_{\tilde \theta = 0} &= 
   \frac{ \pi^5}{6 \tilde g^5} + \frac{21 \zeta(3)}{\tilde g^3} - \frac{37 \pi}{64 \tilde g} + \frac{\tilde g}{16} + \frac{63 \tilde g^3}{8192 \pi^3 }
  \\
 &  -{9\tilde g^2\over 16\pi^{5\over 2}} e^{-{8\pi^2 \over \tilde g^2}}
  \left(1-{105\tilde g^2\over 64\pi^2}
   +\cO(\tilde g^4)\right)+\cO\left(e^{-{16\pi^2 \over \tilde g^2}}\right)
     \,.
 }
The leading term gives a contribution to the $\langle \mathbb{T} {\cal O} {\cal O} \rangle$ correlator proportional to $1/(\tilde g^5 N^{3/2}) \sim N / \tilde \lambda^{5/2}$, which again matches the expectation in \eqref{PerttHooft}.

\paragraph{Nonzero $\tilde{\theta}$.} We can also consider the case for small but nonzero $\tilde{\theta}$. The linear term in $\tilde{\theta}$ comes from the term proportional to $\sin(k\theta)K_3(8\pi^2 k/g^2)$ in ${\cal I}_{3\over 2}$. Following the same procedure described above, we find that the leading order term in a small $\tilde{\theta}$ and small $\tilde{g}$ expansion takes the form
\ie
 {\cal I}_{\mathbb{T},\frac 32} \big|_{{\cal O}(\tilde{\theta})} = \frac{\tilde{\theta}}{\tilde{g}^5} \left( 180\zeta(5) - \frac{1440\zeta(3)\zeta(5)}{\pi^2} + \frac{\pi\zeta(3)}{90} \right) + {\cal O}(\tilde{\theta}\tilde{g}^{-4}) \,.
\label{eq:linearthetat1}
\fe
This is in contrast to the case of the integrated two-point correlator in the presence of the Wilson line (\ref{IExplicit}), in which any nontrivial dependence on $\theta$ is further exponentially suppressed by instantons at small $g$. 
Re-expanding (\ref{eq:linearthetat1}) at large $N$ and fixed $\lambdad$, this linear in $\tilde \theta$ contribution to ${{\cal I}}_{\mathbb{T},\text{pert}}$ becomes
\ie
\frac{\tilde{\theta} N}{\lambdad^{5\over 2}} \left( 180\zeta(5) - \frac{1440\zeta(3)\zeta(5)}{\pi^2} + \frac{\pi\zeta(3)}{90} \right),
\label{eq:linearthetat2}
\fe
which is also of the form (\ref{PerttHooft}), providing a mild yet nontrivial consistency check on the term proportional to $\sin(k\theta)$ of the result (\ref{IExplicit}).

\subsection{The S-dual 't Hooft limit}

Collecting the results above and re-expanding at large $N$ and fixed dual 't Hooft coupling $\tilde \lambda$, we can generate the first few terms in the 't Hooft limit when $\tilde \theta = 0$,
 \es{DualTthooftexpansion}{
   {{\cal I}}_{\mathbb{T},\text{pert}} \big|_{\tilde \theta = 0} &= N \left( \frac{4\pi}{\lambdad^{1\over 2}} - \frac{2\pi^3}{\lambdad^{3\over 2}} +\frac{24\pi\zeta(3)}{\lambdad^{2}} + \frac{\pi^5}{6\lambdad^{5\over 2}} + O(\lambdad^{-3}) \right) \\
   &{}+ \left( \left( \frac{1}{2} - \frac{\pi^2}{3} \right) - \frac{\pi^2}{\lambdad} + \frac{21\zeta(3)}{\lambdad^{3\over 2}} + O(\lambdad^{-2}) \right) 
  + \frac{1}{N} \left( \frac{3\lambdad^{1\over 2}}{32\pi} + \frac{\pi}{6} - \frac{37\pi}{64\lambdad^{1\over 2}} + O(\lambdad^{-1}) \right)  \\
   &{}+ \frac{1}{N^2} \left( \frac{3 \lambdad}{128 \pi^2}  + \frac{\lambdad^{\frac 12}}{16} + O(\lambdad^0)\right) 
    + \frac{1}{N^3} \left( \frac{63 \lambdad^{\frac 32}}{8192 \pi^3} + O(\lambdad) \right) + O(N^{-4}) \,,
 }
 as well as the leading order contribution at small $\tilde \theta$ in (\ref{eq:linearthetat2}). 
In the next section, we will present the interpretation of some of these terms from the bulk dual description.

As was already mentioned, it would be interesting to reproduce the structure in \eqref{DualTthooftexpansion} from a direct matrix model calculation of the integrated correlation function of two 1/2-BPS local operators in the presence of the 1/2-BPS 't Hooft loop, starting from the results of \cite{Gomis:2009ir, Gomis:2009xg,Gomis:2010kv,Gomis:2011pf}. We leave this localization computation to future work.

\subsection{Dyonic loops}
The generation of the above analysis to the case of dyonic loops of charge $(p,q)=(p,1)$ is straightforward. 

It follows from \eqref{Sdual2pt} and \eqref{Tdual2pt} that 
\ie 
\cI_{\mathbb{W}_p\mathbb{T}}(\tilde \tau,\bar{\tilde \tau})=\cI_{\mathbb{T}}(\tilde \tau+p,\bar{\tilde \tau}+p)=\cI_{\mathbb{W}}(\tau,\bar \tau)\,,
\label{IWT}
\fe
where the dual couplings are related by an ST$^p$ transformation,
\ie 
\tau= -{1\over p+\tilde \tau}\,.
\fe
Correspondingly, in the very strong coupling limit, the $1/\sqrt{N}$ expansion \eqref{Expansion} of $\cI_{\mathbb{W}}(\tau,\bar \tau)$ carries over immediately for the dyonic loop.

We can also re-expand the expression $\cI_{\mathbb{W}_p\mathbb{T}}(\tilde \tau,\bar{\tilde \tau})$ in the usual 't Hooft large $N$ limit in the current duality frame, by keeping the dual 't Hooft coupling $\tilde\lambda$ fixed.\footnote{It would also be interesting to derive the full expansion of $\cI_{\mathbb{W}_p\mathbb{T}}(\tilde \tau,\bar{\tilde \tau})$ at weak gauge coupling $\tilde g$ that includes the instanton contributions but we will not do it here.} In this limit, the instanton effects are suppressed and we have a similar expansion as in the 't Hooft loop case,\footnote{The $p$ dependence in the 't Hooft limit takes the form a series of positive powers in $p \tilde \lambda  \over N$.}
\ie 
{\cal I}_{\mathbb{W}_p\mathbb{T},\text{pert}} = N \sum_{m=0}^\infty \frac{1}{N^m}   {\cal I}_{\mathbb{W}_p\mathbb{T},\text{pert}, m}(\tilde \lambda)\,.
\label{WTexpand}
\fe 
In particular, the leading terms of ${\cal I}_{\mathbb{W}_p\mathbb{T},\text{pert}, m}(\tilde \lambda)$ for $m=0,1$ 
in the ${1\over\tilde \lambda}$ expansion is independent of $n$ (and thus matches with those for the 't Hooft loop). Furthermore, at nonzero small $\tilde \theta$, the same contribution \eqref{eq:linearthetat2} carries over for the dyonic loop.

As we will see in Section~\ref{sec:EFT}, both of these expansions have natural interpretations in AdS and they predict new couplings on the worldsheet of $(p,q)$-strings that are consistent with the type IIB $SL(2,\mZ)$ duality. 

\subsection{Differential equations}
\label{sec:diffeqns}

Interestingly, the functions ${\cal I}_{\mathbb{W},1}$ and ${\cal I}_{\mathbb{W},3/2}$ in \eqref{IExplicit} obey simple partial differential equations.  When written in terms of $\tau = \frac{\theta}{2 \pi} + \frac{4 \pi i}{g^2} = \tau_1 + i \tau_2$, the function ${\cal I}_1$ obeys the equation\footnote{We thank Pierre Vanhove for pointing this out to us.}
 \es{I1Eq}{
  \left( \Delta_\tau + 3 \tau_2 \partial_{\tau_2} - 3 \right) {\cal I}_1 = 0 \,,
 }
where $\Delta_\tau = \tau_2^2 \left( \partial_{\tau_1}^2 + \partial_{\tau_2}^2 \right)$ is the Laplacian on the upper half plane with the hyperbolic metric.

At the next order, we can split the function ${\cal I}_{\mathbb{W},3/2}$ in \eqref{IExplicit} into three parts
 \es{I32Three}{
  {\cal I}_{\mathbb{W},3/2} = - \frac{g}{256} + {\cal I}_{\mathbb{W},3/2}^{(1)} + {\cal I}_{\mathbb{W},3/2}^{(2)} \,,
 }
where
 \es{I32parts}{
   {\cal I}_{\mathbb{W},3/2}^{(1)} &= \frac{3(21+ 64 \zeta(3))}{128g^3} - \frac{C}{g^3}  + \frac{7 g^5}{10240}  \\
     &{}+\sum_{k=1}^\infty \biggl[\cos (k \theta) \frac{9 g \sigma_{-2}(k)}{4 \pi^2}  - \sin (k \theta) \frac{3g  \left( 3 k \sigma_{-3}(k) - 4 \sigma_{-2}(k) + \sigma_{-1}(k)  \right)}{2 \pi}    \biggr]  K_2 (8 \pi^2 k / g^2)   \,, \\
    {\cal I}_{\mathbb{W},3/2}^{(2)} &= \frac{C}{g^3} - \frac{g^9}{1935360} \\
     &{}+\sum_{k=1}^\infty \biggl[\cos (k \theta) \frac{9 g \sigma_{-2}(k)}{4 \pi^2}   \frac{5 g^2 K_3 (8 \pi^2 k / g^2)}{4 \pi^2 k }  \\
     &{}- \sin (k \theta) \frac{3g  \left( 3 k \sigma_{-3}(k) - 4 \sigma_{-2}(k) + \sigma_{-1}(k)  \right)}{2 \pi}   \frac{15 g^2 K_3 (8 \pi^2 k / g^2)}{16 \pi^2 k }  
       \biggr]   \,,
 }
where $C$ is an arbitrary constant. The functions ${\cal I}_{\mathbb{W},3/2}^{(1)}$ and ${\cal I}_{\mathbb{W},3/2}^{(2)}$ obey the differential equations 
 \es{DiffEqs}{
  \left( \Delta_\tau + 2 \tau_2 \partial_{\tau_2} - \frac{15}{4} \right) {\cal I}_{\mathbb{W},3/2}^{(1)} &= 0 \,,\\
   \left( \Delta_\tau + 4 \tau_2 \partial_{\tau_2} - \frac{27}{4} \right) {\cal I}_{\mathbb{W},3/2}^{(2)} &= 0 \,.
 }

The differential equations \eqref{I1Eq}--\eqref{DiffEqs} were derived by examining the expressions for ${\cal I}_{\mathbb{W}, 1}$ and ${\cal I}_{\mathbb{W}, 3/2}$.  It would be interesting to explore whether there exists a more abstract derivation of these equations that does not rely on knowing these functions.

\section{Towards the Un-integrated Correlation Function}
\label{sec:unintegrated}

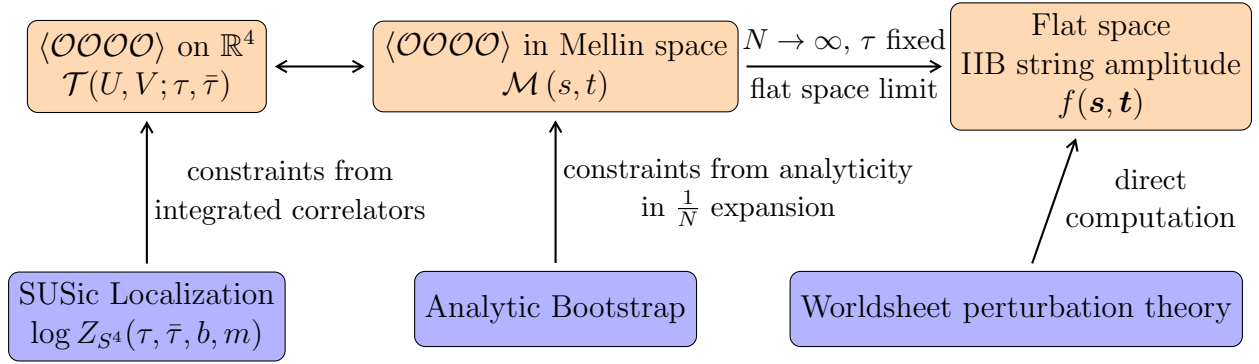
\begin{figure}[h!]
\centering
\usetikzlibrary{arrows.meta, positioning, quotes,shapes}
\begin{tikzpicture}[
node distance = 12mm and 12mm, 
every text node part/.style={align=center},
   box/.style = {draw, rounded corners, 
                 minimum width=22mm, minimum height=10mm, align=center},
            > = {Straight Barb[angle=60:2pt 3]},
   bend angle = 15,
         auto = right,
                        ]
\node (n1)  [box,fill=orange,fill opacity=.3,text opacity = 1]  {$\la\cO\cO\cO\cO \ra$ on $\mR^4$
\\
$\cT(U,V;\tau,\bar\tau)$};
\node (n2)  [box, right=40pt of n1,fill=orange,fill opacity=.3,text opacity = 1]     {$\la\cO\cO\cO\cO \ra$ in Mellin space
\\
$\cM\left(  s,  t\right)$};
\node (n3)  [box, right=80pt of n2,fill=orange,fill opacity=.3,text opacity = 1]     {Flat space \\
 IIB string amplitude
\\
$f(\bm s,\bm t)$};
\node (n4)  [box, below=60pt of n1,fill=blue,fill opacity=.3,text opacity = 1]     {SUSic Localization
\\
$\log Z_{S^4}(\tau,\bar\tau,b,m)$};
\node (n5)  [box, below=60pt of n2,fill=blue,fill opacity=.3,text opacity = 1] {Analytic Bootstrap};
\node (n6)  [box, right=35pt of n5,fill=blue,fill opacity=.3,text opacity = 1]    {Worldsheet perturbation theory};
\draw[<->,thick,shorten >=3pt,shorten <=3pt]        (n1) to [ ]  (n2);
\draw[->,thick,shorten >=3pt,shorten <=3pt]          (n2) -- node [above] {\small $N\to \infty$,
$\tau$ fixed} node [below] {\small flat space  limit} (n3);
\draw[->,thick,shorten >=3pt,shorten <=3pt]         (n4) -- node [right=-0.05cm]{\small constraints from \\ \small integrated correlators}  (n1);
\draw[->,thick,shorten >=3pt,shorten <=3pt]         (n5) 
--node [right=-0.05cm]{\small constraints from 
analyticity \\ \small in ${1\over N}$ expansion}  (n2);
\draw[->,thick,shorten >=3pt,shorten <=3pt]          (n6) --node [right=0cm]{\small direct \\computation}  (n3);
\end{tikzpicture}
\caption{From integrated to un-integrated four-point function of local operators $\cO$.}
\label{fig:4pfsummary}
\end{figure}

Thus far we have focused on the integrated two-point function $\cI_{\mathbb{L}}(\tau,\bar\tau)$ of the stress tensor multiplet in the presence of a half-BPS line defect $\mathbb{L}$.  This integrated correlator can be computed using a matrix model obtained from the supersymmetric localization of the full path integral on $S^4$. By construction, the integrated correlator contains a combination of the OPE data of the ${\cal N} = 4$ SYM theory that is packaged in the un-integrated two-point function. The latter involves a nontrivial function of the conformal cross ratios $\cT_{\mathbb L}(U,V;\tau,\bar\tau)$ which we will review shortly.
A priori, $\cT_{\mathbb L}(U,V;\tau,\bar\tau)$ seems to contain a lot more information than $\cI_\mathbb{L}(\tau,\bar\tau)$ and thus it would be miraculous that the latter determines the former. Nonetheless, a similar scenario in the $\cN=4$ super-Yang-Mills 
suggests that this miracle can happen in the large $N$ limit.  To see how this can happen, let us first recall the successful route from the integrated to the un-integrated four-point functions without defects in the $\cN=4$ SYM, summarized pictorially in Figure~\ref{fig:4pfsummary}.

\subsection{Review of the un-integrated four-point function }
As a consequence of the superconformal Ward identities, the four-point functions of local operators in the $\cN=4$ stress tensor multiplet are determined by a single function $\cT(U,V;\tau,\bar\tau)$ of the conformally invariant cross ratios $U,V$ \cite{Belitsky:2014zha}. In particular, the four-point function of the superconformal primary 
 \ie
 \cO(\vec x,Y) \equiv \cO_{IJ}(\vec x)Y^IY^J\,,
  \label{cODef}
\fe
where $Y^I$ is a null $SO(6)_R$ polarization vector, takes the following form \cite{Dolan:2001tt},
\ie
\la 
\cO(\vec x_1,Y_1)\cO(\vec x_2,Y_2)\cO(\vec x_3,Y_3) \cO(\vec x_4,Y_4)\ra
={1\over \vec x_{12}^4 \vec x_{34}^{4}}
\left(
\cT_{\rm free}(U,V,Y_{ij})
+\Theta(U,V,Y_{ij})\cT(U,V;\tau,\bar\tau)
\right )
\label{4pf}
\fe 
where 
\ie 
U\equiv {\vec x_{12}^2\vec x_{34}^2\over \vec x_{13}^2\vec x_{24}^2}\,,\quad V\equiv {\vec x_{14}^2\vec x_{23
}^2\over \vec x_{13}^2 \vec x_{24}^2}\,,\quad Y_{ij}\equiv Y_i\cdot Y_j\,,
\fe
and $\Theta(U,V,Y_{ij})$ is a fixed known function.  The main take-away of \eqref{4pf} is that the entire content of the separated-point 4-point function of the superconformal primary of the stress tensor multiplet (and also of its descendants) is contained in the function $\cT(U, V; \tau,\bar\tau)$ \cite{Dolan:2001tt}. 

The integrated four-point function first studied in \cite{Binder:2019jwn}, which has a matrix model expression derived from the localization formula for the partition function $Z_{S^4}(\tau,\bar\tau,m)$ of the ${\cal N} = 4$ SYM theory deformed by an $\cN=2^*$-preserving mass $m$, is 
\ie 
G(\tau,\bar\tau)=\left.{ \pa_\tau \pa_{\bar\tau} \pa_m^2 \log Z_{S^4}(\tau,\bar\tau,m) \over  \pa_\tau \pa_{\bar\tau}  \log Z_{S^4}(\tau,\bar\tau,m)}\right|_{m=0}\,.
\label{4pfint}
\fe
The mass parameter $m$ couples to operators that belong to an $\cN=2$ conserved current multiplet which is a submultiplet of the $\cN=4$ stress tensor multiplet, while $\tau,\bar\tau$ couple to operators in other submultiplets known as $\cN=2$ Coulomb branch chiral multiplets. Therefore, the RHS of \eqref{4pfint} is naturally a linear combination of integrals of four-point functions of the $\cN=4$ stress tensor multiplet. The $\cN=4$ superconformal Ward identity implies that $G(\tau,\bar\tau)$ is entirely determined by an integral of $\cT(U,V;\tau,\bar\tau)$. Indeed, as derived in \cite{Binder:2019jwn} and further simplified in \cite{Chester:2020dja}, $G(\tau,\bar\tau)$ takes the form
\ie 
G(\tau,\bar\tau)\equiv -{2\over \pi}\int_0^\infty  dr\int_0^{\pi} d\theta \left. {r^3 \sin^2\theta \over U^2}\cT(U,V;\tau,\bar\tau) \right|_{U=1+r^2-2r \cos\theta,V=r^2}\,.
\label{Gintegralform}
\fe
By the same logic, one can construct another integrated four-point function of the stress tensor multiplet by four mass derivatives,
\ie 
F(\tau,\bar\tau)\equiv  \pa_m^4 \left. \log Z_{S^4}(\tau,\bar\tau,m) \right|_{m=0}\,,
\label{4pfint2}
\fe 
which involves an integral of $\cT(U,V;\tau,\bar\tau)$ with a different measure
\ie 
F(\tau,\bar\tau)=&-{32c^2\over \pi}
\int_0^\infty dr \int_0^\pi d\theta 
\left. r^3\sin^2\theta {1+U+V\over U^2}\bar D_{1,1,1,1}(U,V)\cT(U,V;\tau,\bar\tau) \right|_{U=1+r^2-2r \cos\theta,V=r^2}
\\
&-48\zeta(3)c\,,
\label{Fintegralform}
\fe
where $c={N^2-1\over 4}$ is the central charge of the $\cN=4$ $SU(N)$ SYM and $\bar D_{1,1,1,1}(U,V)$ is a fixed function (see   \cite{Chester:2020dja} for details).

In the large $N$ limit, it is useful to consider the Mellin transform of the correlator
 $\cT(U,V;\tau,\bar\tau)$~\cite{Zhou:2017zaw},
\ie 
\cT(U,V;\tau,\bar\tau)=\int_{-i\infty}^{i\infty} {ds\, dt\over (4\pi i)^2}\,  \cM(s,t) U^{s\over 2}V^{{u\over 2}-2}\Gamma\left(4-s\over 2\right)^2\Gamma\left(4-t\over 2\right)^2\Gamma\left(4-u\over 2\right)^2 \,,
\fe 
which defines the reduced Mellin amplitude $\cM(s,t)$ and the Mellin space variables $s,t,u$ are related by  $s+t+u=4$.
The key idea is that $\cM(s,t)$ 
has a simple analytic structure in the large $N$ limit coming from the bulk Witten diagrams. After taking into account the crossing symmetry in Mellin space,  $\cM(s,t)=\cM(t,s)=\cM(s,u)$, the analytic structure leads to the following ansatz 
\ie 
\cM(s,t)=&{1\over c}{\A \over (s-2)(t-2)(u-2)} +{\B \over c^{7\over 4}}
+{\cM_{1-\rm loop}(s,t)\over c^2}
+{\C_1 (s^2+t^2+u^2)+\C_2\over c^{9\over 4}}
\\&+{\D_1  stu+\D_2 (s^2+t^2+u^2) + \D_3 \over c^{5\over 2}} + \cO(c^{-{11\over 4}} )
\,,
\label{Mstexp}
\fe
where $\A,\B,\C_i,\D_i$ are functions of $(\tau, \bar \tau)$.  Apart from $\cM_{1-\rm loop}(s,t)$, which is non-analytic in $s$ and $t$, 
the  rest of the Mellin amplitude $\cM(s,t)$ is analytic in $s$ and $t$ to the order given in \eqref{Mstexp}.  The term of order $1/c$ corresponds to an exchange amplitude with two-derivative interaction vertices, and the subleading analytic terms come from contact higher-derivative interactions.  

The integrated correlators \eqref{4pfint} and \eqref{4pfint2} each give one constraint on $\cM(s,t)$ at each order in the ${1\over c}$ (equivalently $1\over N$)  expansion.
Consequently, they determine $\A,\B,\C_1,\C_2$. For higher order terms, the localization constraints from \eqref{4pfint} and \eqref{4pfint2} are not sufficient. 

Another set of constraints come from the relation between the Mellin amplitude $\cM(s,t)$ which describes the graviton four-point  scattering in type IIB string theory
for ${\rm AdS}_5\times {\rm S}^5$ and the flat space IIB amplitude $\cA({\bm s},{\bm t})$.  As a consequence of type IIB supersymmetry, $\cA({\bm s},{\bm t})$ takes the  form 
\ie 
\cA({\bm s},{\bm t})=\cA_{\rm SG\,tree}({\bm s},{\bm t}) f({\bm s},{\bm t}) \,,
\fe
where $\cA_{\rm SG\,tree}({\bm s},{\bm t})$ is the type IIB supergravity tree amplitude, and $\bm{s},\bm{t}$ are the usual Mandelstam variables in 10d flat space. Here $f({\bm s},{\bm t})$ is the flat space analog of the reduced Mellin amplitude $\cM(s,t)$. The two are related by the   flat space limit formula \cite{Chester:2018aca,Binder:2019jwn,Alday:2018pdi},
\ie 
f({\bm s},{\bm t})={\bm{ stu}\over 2^{11} \pi^2g_s^2 \ell_s^8} \lim_{L/\ell_s\to \infty}  L^{14}\int_{\kappa-i\infty}^{\kappa+i\infty} {d\A\over 2\pi i} e^\A \A^6 \cM\left ({L^2\over \A}{\bm s},{L^2\over \A}{\bm t} \right)\,,
\label{flatspacelimit}
\fe
with $\bm{u}\equiv-{\bm s}-{\bm t}$.
This formula implies that at each order in $1\over c$, only the leading term in $\cM(s,t)$ at large $s,t$ is related to $f({\bm s},{\bm t})$. Furthermore, the ${1\over c}$ expansion of $\cM(s,t)$ translates into the derivative expansion of $f({\bm s},{\bm t})$. 
The derivative expansion of $f({\bm s},{\bm t})$ keeps track of higher derivative interactions in the IIB effective action,
\ie 
{f({\bm s},{\bm t})\over \bm{stu}} &= {1\over \bm{stu}}+f_{R^4}(\tau,\bar\tau)\ell_s^6+f_{1-\rm loop}({\bm s},{\bm t})\ell_s^8
+f_{D^4R^4}(\tau,\bar\tau) (\bm{s}^2+\bm{t}^2+\bm{u}^2)\ell_s^{10}
\\
&{}+f_{D^6R^4}(\tau,\bar\tau)  \bm{stu} \ell_s^{12} +\cdots\,.
\label{fstexp}
\fe
Here, we have included the leading terms from four-point vertices of the types $f_{R^4}(\tau,\bar\tau)R^4$, $f_{D^4R^4}(\tau,\bar\tau)D^4R^4$,  and $f_{D^6R^4}(\tau,\bar\tau)D^6R^4$. The $(\tau,\bar\tau)$ dependence receives both perturbative higher genus contributions and non-perturbative contributions from multiple D-instantons. 
Existing type IIB string perturbation theory has been performed up to genus three (up to the derivative order of $D^6R^4$ \cite{Gomez:2013sla}), while recent progress in the D-instanton contributions is still mostly limited to the case of a single D-instanton \cite{Agmon:2022vdj}. 
Despite having no explicit derivations, the coefficient functions in \eqref{fstexp} have been determined exactly.
This is because 
these leading higher derivative interactions $R^4,D^4R^4$, and $D^6R^4$ are ${1\over 2}$-, ${1\over 4}$-, and ${1\over 8}$-BPS, respectively, and their coefficients satisfy Laplace-type differential equations in $(\tau,\bar\tau)$ as a consequence of the supersymmetry \cite{Green:1998by,Wang:2015jna}. Together with the $SL(2,\mZ)$ duality in type IIB string theory and the perturbative contributions, these differential equations determine $f_{R^4}(\tau,\bar\tau)$, $f_{D^4R^4}(\tau,\bar\tau)$, and $f_{D^6R^4}(\tau,\bar\tau)$ uniquely.
Using the relation \eqref{flatspacelimit}, this determines immediately the coefficients $\A,\B,\C_1,\D_1$. Indeed, the terms in the Mellin amplitude \eqref{Mstexp} multiplying $\B,\C_1,\D_1$ come from contact interactions of the type $R^4$, $D^4R^4$, and $D^6R^4$, respectively, on ${\rm AdS}_5\times {\rm S}^5$.  On  ${\rm AdS}_5\times {\rm S}^5$, the $R^4$, $D^4R^4$, and $D^6R^4$ interaction vertices also receive corrections due to the nontrivial curvature of the background, and these corrections produce the subleading terms in large $s,t$ at each order in ${1\over c}$ in \eqref{Mstexp} (see \cite{Alday:2018pdi,Binder:2019jwn} for more details).

Together, the localization constraints from the two integrated correlators   \eqref{4pfint} and \eqref{4pfint2}, the analytic and crossing constraints on the large $N$ Mellin amplitude \eqref{Mstexp}, and the relation to flat space type IIB string amplitude \eqref{flatspacelimit} determine completely the four-point function of the stress tensor multiplet up to the order $c^{-{5\over 2}}$, where coefficients in \eqref{Mstexp} involve nontrivial modular invariant functions from (generalized) non-holomorphic Eisenstein series.  See Figure~\ref{fig:4pfsummary} for a summary of the various connections and limits. 
Moreover, for lower orders up to $c^{-{9\over 4}}$, the localization constraints together with crossing and analyticity of the Mellin amplitude are strong enough to fix the $R^4$ and $D^4R^4$ contributions from IIB string theory on ${\rm AdS}_5\times {\rm S}^5$, thus providing a field theory derivation of the well-known Eisenstein series appearing to this order in flat space closed-string scattering amplitudes. We emphasize that in this field theory computation, the D-instantons are mapped to Yang-Mills instantons, and the multi-instanton contributions are completely transparent, evading the many subtleties in a direct type IIB string worldsheet calculation that are still to be understood. Furthermore, having obtained the un-integrated four-point function gives us access to a wealth of CFT data in the OPE channel, including the anomalous dimensions of non-BPS operators as exact functions of $(\tau, \bar \tau)$ in the large $N$ limit.

\subsection{Towards the un-integrated defect two-point function}
\label{sec:unint2pf}

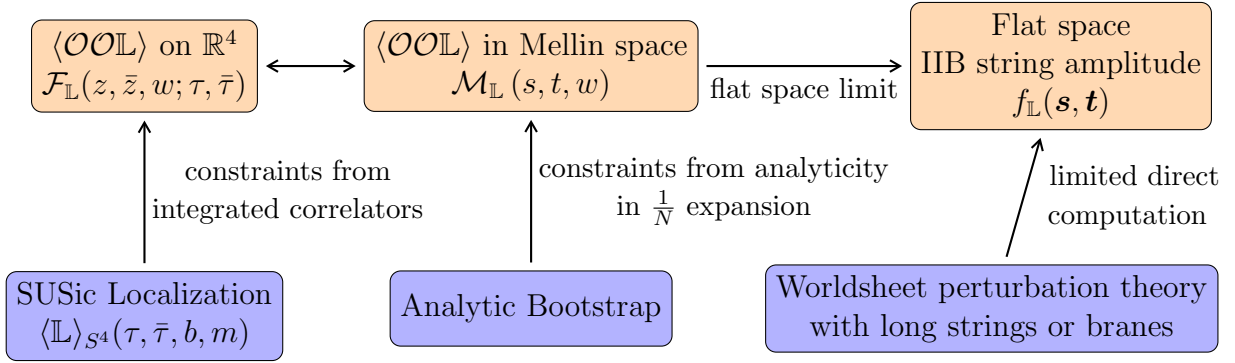
\begin{figure}[h!]
\centering
\usetikzlibrary{arrows.meta, positioning, quotes,shapes}
\begin{tikzpicture}[
node distance = 12mm and 12mm, 
every text node part/.style={align=center},
   box/.style = {draw, rounded corners, 
                 minimum width=22mm, minimum height=10mm, align=center},
            > = {Straight Barb[angle=60:2pt 3]},
   bend angle = 15,
         auto = right,
                        ]
\node (n1)  [box,fill=orange,fill opacity=.3,text opacity = 1]  {$\la  \cO\cO  {\mathbb L}\ra$ on $\mR^4$
\\
$\cF_{\mathbb L}(z,\bar z,w;\tau,\bar\tau)$};
\node (n2)  [box, right=40pt of n1,fill=orange,fill opacity=.3,text opacity = 1]     {$\la   \cO \cO {\mathbb L}\ra$ in Mellin space
\\
$\cM_{\mathbb L}\left(  s,  t,w\right)$};
\node (n3)  [box, right=80pt of n2,fill=orange,fill opacity=.3,text opacity = 1]     {Flat space \\
 IIB string amplitude
\\
$f_{\mathbb L}(\bm s,\bm t)$};
\node (n4)  [box, below=60pt of n1,fill=blue,fill opacity=.3,text opacity = 1]     {SUSic Localization
\\
$\la {\mathbb L} \ra_{S^4}(\tau,\bar\tau,b,m)$};
\node (n5)  [box, below=60pt of n2,fill=blue,fill opacity=.3,text opacity = 1] {Analytic Bootstrap};
\node (n6)  [box, right=35pt of n5,fill=blue,fill opacity=.3,text opacity = 1]    {Worldsheet perturbation theory\\ with long strings or branes};
\draw[<->,thick,shorten >=3pt,shorten <=3pt]        (n1) to [ ]  (n2);
\draw[->,thick,shorten >=3pt,shorten <=3pt]          (n2) -- node [above] { } node [below] {\small flat space  limit} (n3);
\draw[->,thick,shorten >=3pt,shorten <=3pt]         (n4) -- node [right=-0.05cm]{\small constraints from \\\small integrated correlators}  (n1);
\draw[->,thick,shorten >=3pt,shorten <=3pt]         (n5) 
--node [right=-0.05cm]{\small constraints from 
analyticity \\ \small in ${1\over N}$ expansion}  (n2);
\draw[->,thick,shorten >=3pt,shorten <=3pt]          (n6) --node [right=0cm]{\small\, limited direct \\\small computation}  (n3);
\end{tikzpicture}
\caption{From integrated to un-integrated two-point function of the local operators $\cO$ in the presence of a line operator ${\mathbb L}$.}
\label{fig:2pfsummary}
\end{figure}

Given the successful attempt in the case of four-point function of BPS operators (summarized in Figure~\ref{fig:4pfsummary}), here we would like to initiate a parallel program (see Figure~\ref{fig:2pfsummary}) for two-point functions of the stress tensor multiplet with a half-BPS line defect in the $\cN=4$ SYM.

A general half-BPS superconformal line preserves the $\mf{osp}(4^*|4)\subset \mf{psu}(2,2|4)$ subalgebra. In particular, the $\mf{so}(6)_R$ symmetry in the bulk is broken to a $\mf{usp}(4)_R=\mf{so}(5)_R$ subalgebra, which can be specified as the subalgebra fixing a real $\mf{so}(6)_R$ vector $Y$ of unit norm. We will label the half-BPS line as $\mathbb{L}(Y)$ to keep track of its R-symmetry polarization. We split the coordinates on the Euclidean spacetime as $\vec x=(x^\perp,x^4)$ and
take the half-BPS line defect $\mathbb{L}(Y)$ to lie along the $x^4$ direction at $x^\perp=0$. In addition to the $\mf{so}(5)_R$ symmetry, the line preserves the longitudinal conformal symmetry $\mf{so}(1,2)$ and the transverse $\mf{so}(3)_{\rm rot}$ rotation symmetry in $x^\perp$. 

Before introducing the defect two-point function, let us first review the one-point function. The form of the one-point function of a general bulk conformal primary operator $\phi(\vec x)$ of scaling dimension $\Delta_\phi$ is fixed by conformal symmetry up to an overall constant $a_{\phi, \mathbb{L}}$:
\ie 
{\la  \phi(\vec  x)\mathbb{L}(Y_3)\ra\over \la \mathbb{L}(Y_3)\ra}={a_{\phi, \mathbb{L}} \over |x_\perp|^{\Delta_\phi} }\,.
\fe
The one-point function of any conformal primary in the same superconformal multiplet is proportional to that of the superconformal primary \cite{Liendo:2016ymz}.  In particular, the one-point functions of all operators in the stress tensor multiplet are determined by that of the superconformal primary $\cO(\vec x,Y)$ introduced in \eqref{cODef}, which, given that $\cO(\vec x,Y)$ has scaling dimension $2$, takes the form below
\ie
{\la \cO(\vec  x,Y_1) \mathbb{L}(Y_3)\ra\over \la \mathbb{L}(Y_3)\ra } =
 \frac{a_{{\cal O}, \mathbb{L}} Y_{13}^2}{ |x_\perp|^2}\,,
\label{O1pf}
\fe

In the following, we will normalize the bulk operator to have the canonical two-point functions
\ie 
\la \cO(\vec x_1,Y_1)\cO(\vec x_2,Y_2) \ra = {Y_{12}^2\over |\vec x_{12}|^4}
\label{2pfnorm}
\fe
in the absence of any defects. Then the one-point coefficient in \eqref{O1pf} is unambiguously fixed.\footnote{  This is related to the stress energy tensor (with canonical normalization) one-point function coefficient $h_\mathbb{L}$, namely in $\la T_{44}(x)\mathbb{L}(Y_3)\ra= {h_\mathbb{L}  x_\perp^{-4}} \la \mathbb{L}(Y_3)\ra$, by the $\cN=2$ superconformal Ward identity \cite{Gomis:2008qa,Fiol:2015spa}, 
\ie 
a_{\cO,\mathbb{L} }= {3\over 8}h_\mathbb{L}\,.
\fe
}

Let us now discuss the two-point function of $\cO$ in the presence of the $1\over 2$-BPS line defect.  The defect two-point function $\la \cO(\vec x_1,Y_1)\cO(\vec x_2,Y_2) \mathbb{L}(Y_3)\ra $ is a much richer observable that is akin to the bulk four-point functions in the absence of defects.  To elucidate the kinematic structure,
 it is useful to introduce the defect conformal cross-ratios
  \es{CrossRatios}{
   {\rm U} = \frac{x_{1}^\perp \cdot x_{2}^\perp}{\abs{x_{1}^\perp} \abs{x_{2}^\perp}} \,, \qquad {\rm V} = \frac{\vec x_{12}^2}{\abs{x_{1}^\perp} \abs{x_{2}^\perp}} \,,
  }
which are invariant under the $\mf{so}(1,2)\times \mf{so}(3)$ subalgebra of the (Euclidean) conformal symmetry preserved by the defect.  Following \cite{Barrat:2020vch}, we can use the residual symmetry to fix $\vec x_1=(1,0,0,0)$ and $\vec x_2=(x,y,0,0)$.  The conformal cross-ratios are then parametrized by
 \es{UVExplicit}{
  {\rm U} = \frac{z + \bar z}{2 \sqrt{z \bar z}} \,, \qquad 
   {\rm V} = \frac{(1-z)(1-\bar z)}{\sqrt{ z \bar z}} \,, \qquad z=x+iy\,,\quad \bar 
 z=x-iy\,.
 }
 We also define the R-symmetry invariant $w$ by
 \ie 
 {2 w\over (1-w)^2}=-{ Y_{13} Y_{23}\over Y_{12}  }\,.
 \fe
Similar to the case of four-point function, $\cN=4$ superconformal symmetry dictates the two-point function of the half-BPS primary $\cO(\vec x,Y)$ with a half-BPS line operator $\mathbb{L}(Y)$ must take the form \cite{Liendo:2016ymz,Barrat:2020vch}
\ie  
{\la\cO(\vec x_1,Y_1)\cO(\vec x_2,Y_2) \mathbb{L}(Y_3)\ra
\over 
\mathbb{L}(Y_3)}
={Y_{13}^2 Y_{23}^2 \over  x_{1\perp}^2 x_{2\perp}^2
 }\cF_\mathbb{L}(z,\bar z,w)\,,
\label{2pfL}
\fe 
where $\cF_\mathbb{L}(z,\bar z,w)$ receives contributions from three R-symmetry channels 
 \ie 
 \cF_\mathbb{L}(z,\bar z,w) = \Omega^2 F_0(z,\bar z)+\Omega^1 F_1(z,\bar z)
 +  F_2(z,\bar z)\,,
\quad 
 \Omega\equiv {\sqrt{z\bar z}\over (1-z)(1-\bar z)} {(1-w)^2\over 4w}\,,
 \label{2pFRchannels}
 \fe
 and satisfies the differential constraints
\ie 
\left( \pa_z +{1\over 2} \pa_w\right )\left.\cF_\mathbb{L} (z,\bar z,w)\right |_{z=w}=0\,,\quad 
\left( \pa_{\bar z} +{1\over 2} \pa_w\right )\left.\cF_\mathbb{L} (z,\bar z,w)\right |_{\bar z=w}=0\,.
\label{Ward}
\fe
Note that due to the different form of the differential constraints (from the Ward identities), the defect two-point function $\cF_\mathbb{L}(z,\bar z,w)$ does not exhibit the simple factorization structure  as in \eqref{4pf} for the four-point function. Also note that the $(\tau,\bar\tau)$ dependence of $\cF_\mathbb{L}(z,\bar z,w)$ is suppressed in order to simplify the notation.

Previous studies \cite{Buchbinder:2012vr,Barrat:2020vch,Barrat:2021yvp,Barrat:2022psm} of the un-integrated defect two-point function $\cF_\mathbb{L}(z,\bar z,w)$ focus on the case where $\mathbb{L}$ is the half-BPS fundamental Wilson loop considered in the planar limit $N\to \infty$ with the 't Hooft coupling $\lambda$ fixed. At leading orders in the weak  and strong coupling limits of $\lambda$, the two-point function $\cF_\mathbb{L}(z,\bar z,w)$ was determined in 
\cite{Buchbinder:2012vr}.  The defect two-point function with the Wilson loop in these limits was further analyzed in relation to the bootstrap equations and the bulk-defect OPE data in 
\cite{Barrat:2020vch,Barrat:2021yvp,Barrat:2022psm}. Here, we would like to go beyond these limits to consider more general line defects in the SYM theory and also to keep track of the nontrivial dependence on the coupling constant $\tau$ beyond the perturbative regime. We propose to achieve this by a program as outlined in Figure~\ref{fig:2pfsummary}. Below, we will discuss each of the ingredients that go into this procedure.

There are two natural integrated two-point functions of the stress tensor multiplet in the presence of the half-BPS line defect $\mathbb{L}$. They are obtained from certain second derivatives of the expectation value $\la \mathbb{L} \ra(\tau,m)$ of the line defect $\mathbb{L}$ along a great circle on $S^4$ in the $\cN=2^*$ theory, further evaluated at $m=0$. Note that  $\la \mathbb{L} \ra(\tau,m)$ is an even function of $m$ as a consequence of the Weyl reflection symmetry from the $SU(2)_F$ subgroup of the $SO(6)_R$ R-symmetry. Therefore the nontrivial second derivatives of $\la \mathbb{L} \ra(\tau,m)$ in $\tau,m$ at the CFT point are
\ie 
\cE_{\mathbb{L}}(\tau,\bar\tau)=  
\pa_\tau \pa_{\bar\tau} \left.\log \la \mathbb{L} \ra  \right|_{m=0}\,,
\label{eq:IL2tauderivs}
\fe
and
\ie 
\cI_{\mathbb{L}}(\tau,\bar\tau)=\left.  \pa^2_m\log \la \mathbb{L} \ra\right |_{m=0}\,.
\label{eq:IL2massderivs}
\fe
As in the case of the fourth derivatives of the four-point function, these second derivatives package certain linear combinations of integrated defect two-point functions of operators in the stress tensor multiplet, which are all related to $\cF_\mathbb{L}(z,\bar z,w)$ in \eqref{2pfL} by superconformal Ward identities.

The first integrated two-point function $\cE_{\mathbb{L}}(\tau,\bar\tau)$ in \eqref{eq:IL2tauderivs} is defined for general 4d $\cN=2$ SCFTs with a half-BPS line operator $\mathbb{L}$ and marginal couplings $(\tau,\bar\tau)$. It has been studied previously for Wilson lines in $\cN=2$ conformal gauge theories \cite{Gomis:2008qa,Giombi:2009ds,Giombi:2012ep,Beccaria:2020ykg}. In Appendix~\ref{app:otherintegrated}, we discuss more details of this integrated correlator and its immediate generalizations. Here we summarize its relation to the un-integrated correlator \eqref{2pfL} for half-BPS line operators $\mathbb{L}$ in the $\cN=4$ SYM theory, in which case the marginal operator that couples to $(\tau,\bar\tau)$ is an $\cN=4$ superconformal descendant of the primary operator $\cO$ in the stress tensor multiplet. 

The $\cN=4$ SYM theory contains a 2d topological subsector in its full operator algebra of both local and extended operators \cite{Drukker:2007yx,Pestun:2009nn,Giombi:2009ds,Giombi:2009ek,Wang:2020seq}. In particular it determines $\cE_{\mathbb{L}}(\tau,\bar\tau)$ completely as we review in Appendix~\ref{app:otherintegrated}.
At the level of the un-integrated defect two-point function \eqref{2pfL},
this topological subsector corresponds to imposing the following relations between spacetime and R-symmetry cross-ratios \cite{Barrat:2020vch},
\ie 
z=\bar z=w\,,
\label{topconstraint}
\fe
in which case the correlator is constrained by the Ward identity \eqref{Ward} to be a $(\tau,\bar\tau)$-dependent constant ($c^+_\mathbb{L}(\tau,\bar\tau),c^-_\mathbb{L}(\tau,\bar\tau)$ below), 
\ie 
\label{topF}
\cF_\mathbb{L}(w,w,w)=\begin{cases}
{1\over 16}F_0(w,w)+ {1\over 4}F_1(w,w)+F_2(w,w) =c^+_\mathbb{L}(\tau,\bar\tau) & w>0\,,
\\
{1\over 16}F_0(w,w)- {1\over 4}F_1(w,w)+F_2(w,w)=c^-_\mathbb{L}(\tau,\bar\tau) & w<0\,.
\end{cases}
\fe
This topological correlator determines the integrated correlator \eqref{eq:IL2tauderivs} as follows (see Appendix~\ref{app:otherintegrated} for details),
\ie 
\cE_{\mathbb{L}}(\tau,\bar\tau)={c\over 8\tau_2^2}(c^-_\mathbb{L}(\tau,\bar\tau)-a_{\cO,\mathbb{L}}(\tau,\bar\tau)^2)\,,
\label{Eintfinal}
\fe
where the one-point function coefficient $a_{\cO,\mathbb{L}}$ (see \eqref{O1pf}) is also an observable in the topological sector and given by the first derivative of the line operator expectation value,
\ie 
a_{\cO,\mathbb{L}}={2\sqrt{2}\tau_2\over \sqrt{c}}\pa_\tau \left.\log \la \mathbb{L}\ra \right|_{m=0} \,.
\label{O1pffinal}
\fe
Here $c$ is the SYM central charge.

The second integrated two-point function $\cI_{\mathbb{L}}(\tau,\bar\tau)$, given  in \eqref{eq:IL2massderivs},  has a much richer structure and is the main focus of this work. In particular, as we have seen, $\cI_{\mathbb{L}}(\tau,\bar\tau)$ in the SYM theory receives non-perturbative instanton contributions in $(\tau,\bar\tau)$ which are absent from $\cE_{\mathbb{L}}(\tau,\bar\tau)$. It also has immediate generalizations to $\cN=2$ SCFTs with a half-BPS line operator $\mathbb{L}$. However in that case, the relevant un-integrated correlator would not be for operators in the stress tensor multiplet but rather for those in the flavor symmetry multiplet (also known as the moment map multiplet) that couples to the mass parameter $m$.

In the $\cN=4$ SYM, the mass $m$ couples to the $\cN=2$  $\mf{su}(2)_F$ flavor symmetry multiplet (a submultiplet of the $\cN=4$ stress tensor multiplet) and the integrated insertion from $\pa_m$ takes the following form  
\ie 
\int d^4 x \sqrt{g} \left ({i\over R} J+ K\right)\,,
\fe
where $J$ is the moment map operator proportional to $\cO_{11}+\cO_{22}-\cO_{33}-\cO_{44}$  and $K$ is its $\cN=2$ descendant (see \cite{Bobev:2013cja,Binder:2019jwn} for details). As a consequence of the superconformal Ward identity, this integrated two-point function is entirely determined by the two-point function $\la J(x_1)J(x_2) \mathbb{L}\ra$. Furthermore, the R-symmetry polarization of $J$ is such that a unique R-symmetry channel in \eqref{2pFRchannels} contributes, namely the $F_0(z,\bar z)$ piece. Therefore, similar to \eqref{Gintegralform} and \eqref{Fintegralform}, we can derive an integral relation between the un-integrated correlator and \eqref{eq:IL2massderivs},
\ie 
\cI_{\mathbb{L}}(\tau,\bar\tau)=\int d^2 z \, \rho(z,\bar z) F_0(z,\bar z)\,,
\fe
where $\rho(z,\bar z)$ is a theory-independent integration measure on the space of cross-ratios. Its explicit form and detail derivation will be given in the subsequent work \cite{toappear}. 

There are further generalizations of the defect integrated correlators \eqref{eq:IL2tauderivs} and \eqref{eq:IL2massderivs} from considering supersymmetric squashing deformations on the $S^4$ with parameter $b$ (and $b=1$ correspond to the round $S^4$), similar to the four-point function case considered in \cite{Chester:2020vyz}. It would be interesting to see whether this leads to a new localization constraint on the un-integrated correlator $\cF_\mathbb{L}(z,\bar z,w)$ independent from \eqref{eq:IL2tauderivs} and \eqref{eq:IL2massderivs}.

Let us now explain how to resolve the nontrivial dependence on the cross-ratios.  As for the four-point function, the strategy is to develop a general ansatz for $\cF_\mathbb{L}(z,\bar z,w)$ in an expansion scheme with a small number of cross-ratio-dependent building blocks at each order so that their coefficients which are functions of $(\tau, \bar \tau)$ can be fixed by the localization constraints.

Such an expansion comes naturally in large $N$ theories with a holographic dual, with ${1\over N}$ playing the role of the expansion parameter. This expansion is especially facilitated in the Mellin space as we have reviewed in the previous section for the four-point function in the large $N$ SYM theory. For general defect two-point functions, the corresponding Mellin amplitude was studied in \cite{Goncalves:2018fwx} (see also upcoming work \cite{Aleix}) and we refer to the Mellin transform of $\cF_{\mathbb{L}}(z,\bar z,w)$ here as $\cM_{\mathbb{L}}(s,t,w)$ where $s,t$ are the Mellin variables.
In the large $N$ SYM theory, for defects that are dual to macroscopic strings or branes in ${\rm AdS}_5\times {\rm {\rm S}^5}$, the defect Mellin amplitude admits a natural ${1\over N}$ expansion that encodes the Witten diagrams that contribute to the correlator. Correspondingly, Mellin amplitudes at a fixed ${1\over N}$ order are strongly constrained by analyticity and cross-symmetry, thus  producing a simple ansatz for the full amplitude such that we can hope to determine the unknown coefficient functions with a small number of dynamical inputs. 

Apart from the integrated defect correlators, another source of input that can be used to constrain the  Mellin amplitude $\cM_{\mathbb{L}}(s,t,w)$ is flat space string theory amplitudes. Here the particular relevant amplitude involves two massless closed string modes (e.g. gravitons) scattering off the extended string or brane (e.g. $(p,q)$-strings) that produces the line defect in the SYM theory. While the full Mellin amplitude $\cM_{\mathbb{L}}(s,t,w)$ captures this scattering process on ${\rm AdS}_5\times {\rm S}^5$, we expect, similar to the four-point function case, a natural flat space limit of $\cM_{\mathbb{L}}(s,t,w)$ (from large $N$ and fixed $\tau$) that reproduces the scattering amplitudes in type IIB string theory on flat space with an extended string or brane, and the ${1\over N}$ expansion  translates to the derivative (i.e.~$\alpha'$) expansion in the flat space amplitude.
If such a flat space amplitude is available (as a function of $\alpha'$ and $\tau$), it would produce constraints on $\cM_{\mathbb{L}}(s,t,w)$ that either help to determine the amplitude to higher order in ${1\over N}$ or 
provide a nontrivial consistency check of AdS/CFT beyond the supergravity limit and beyond the scattering among closed string modes only. 

It turns out that surprisingly little is known about such flat space amplitudes (compared to the four-point case without defects) beyond tree level (genus zero), let alone nonperturbative contributions.\footnote{However, see for example \cite{Green:2000ke,Billo:2002hm,Billo:2006jm,Billo:2007sw,Billo:2007py,Billo:2008pg,Billo:2008sp,Billo:2009di,Alexandrov:2021shf,Alexandrov:2021dyl,Alexandrov:2022mmy} that study D-instanton effects in the presence of certain D-branes (not D1-brane or $(p,q)$-strings) from the string worldsheet perspective.} 
As we will see in Section~\ref{sec:flatspacescattering},
the localization formulas for the integrated correlators provide important hints on the structure of higher genus and nonperturbative corrections to the flat space amplitudes, even at finite string coupling.  

In this work, we initiate the program outlined here (see Figure~\ref{fig:2pfsummary}) by providing the explicit integrated correlator \eqref{eq:IL2massderivs} for half-BPS Wilson-'t Hooft line operators as exact functions of $\tau$ up to the fifth order in the $N^{-{1\over 2}}$ expansion. We comment on the connection to the bulk Witten diagram expansion and the flat space scattering amplitude in Section~\ref{sec:EFT}.
We leave the more extensive analysis to a subsequent publication \cite{toappear}.

\section{Effective Field Theory Interpretation}
\label{sec:EFT}

 \subsection{Effective field theory in AdS}

Although we have not computed the full un-integrated correlator~\eqref{2pfL} described in Section~\ref{sec:unintegrated}, the structure of the perturbative contributions to the integrated correlator of the type (\ref{eq:IL2massderivs}) can nevertheless be understood qualitatively in terms of an effective field theory in AdS\@. In this section, we describe some examples of perturbative AdS diagrams that are expected to contribute to the two-point correlation function in the presence of a $1\over 2$-BPS fundamental Wilson line, or of a more general $1\over 2$-BPS Wilson-'t Hooft line in the same $SL(2,\mZ)$ orbit.  We will work mostly in the 't Hooft strong coupling limit, where $N$ is taken to be large with $\lambda = g^2 N $ fixed, and then $\lambda$ is also taken to be large, but we will also comment on the very strong coupling limit.

Under the AdS/CFT dictionary, a $1/2$-BPS Wilson-'t Hooft lines with dyonic charge $(p, q)$ corresponds to a $(p,q)$-string anchored on the boundary of AdS precisely along the path where the Wilson-'t Hooft line resides.  Let us first consider the case $(p, q)=(1, 0)$ corresponding to a long/macroscopic fundamental (F1) string.

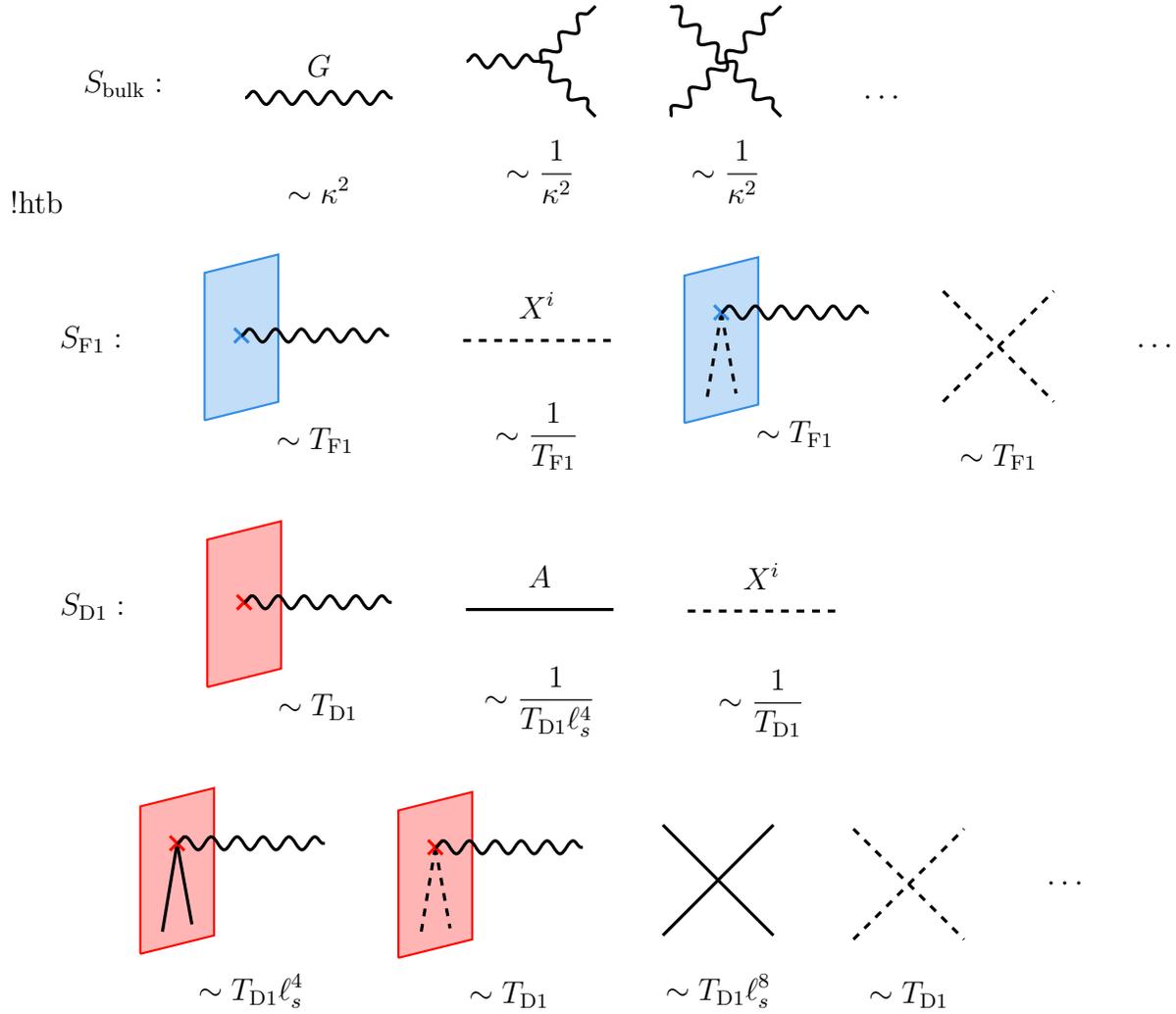
\begin{figure}{!htb}
\centering
\begin{tikzpicture}
\draw[color=white] (0,0) -- (0,-1.75);
\node at (0,0) {$\displaystyle S_{\rm bulk}:$};
\end{tikzpicture} 
~~~~~ 
\begin{tikzpicture}
\node[above] at (0,0.15) {$\displaystyle G$};
\draw [very thick, snake it] (-1,0) -- (1,0);
\node at (0,-1.25) {$\displaystyle \sim \kappa^2$};
\end{tikzpicture} 
~~~~~
\begin{tikzpicture}
\draw [very thick, snake it] (-1,0) -- (0,0);
\draw [very thick, snake it] (0,0) -- (0.75,0.75);
\draw [very thick, snake it] (0,0) -- (0.75,-0.75);
\node at (0,-1.5) {$\displaystyle \sim \frac{1}{\kappa^2}$};
\end{tikzpicture} 
~~~~~
\begin{tikzpicture}
\draw [very thick, snake it] (-0.75,0.75) -- (0.75,-0.75);
\draw [very thick, snake it] (-0.75,-0.75) -- (0.75,0.75);
\node at (0,-1.5) {$\displaystyle \sim \frac{1}{\kappa^2}$};
\end{tikzpicture} 
~~~~~
\begin{tikzpicture}
\draw [color=white] (0,0.75) -- (0,-0.75);
\node at (0,-1.25) {$\displaystyle \vphantom{\sim N^2}$};
\node at (0,0) {$\displaystyle \cdots$};
\end{tikzpicture} 
~~~~~~~~~~~~~~~~~~~~~~~~~~~~~~~
\\ ~~ \\
\begin{tikzpicture}
\draw[color=white] (0,1) -- (0,-1.9);
\node at (0,0) {$\displaystyle S_{\rm F1}:$};
\end{tikzpicture} 
~~~~~ 
\begin{tikzpicture}
\filldraw [color=bleudefrance, thick, fill=bleudefrance!30] (-1.5,-1.5) -- (-1.5,0.5) -- (-0.5,0.75) -- (-0.5,-1.25) -- (-1.5,-1.5);
\node[cross=4pt, color=bleudefrance, very thick] at (-1,-0.35) {};
\draw [very thick, snake it] (-1,-0.35) -- (1,-0.35);
\node at (0,-1.75) {$\displaystyle \sim T_{\rm F1} \vphantom{\frac{1}{1^2}}$};
\end{tikzpicture} 
~~~~~
\begin{tikzpicture}
\node[above] at (0,0.15) {$\displaystyle X^i$};
\draw [very thick, dashed] (-1,0) -- (1,0);
\node at (0,-1.3) {$\displaystyle \sim \frac{1}{T_{\rm F1}} $};
\node at (0,-1.6) {$\vphantom{1}$};
\end{tikzpicture} 
~~~~~ 
\begin{tikzpicture}
\filldraw [color=bleudefrance, thick, fill=bleudefrance!30] (-1.5,-1.5) -- (-1.5,0.5) -- (-0.5,0.75) -- (-0.5,-1.25) -- (-1.5,-1.5);
\draw [very thick, dashed] (-1,0) -- (-1.2,-1.2);
\draw [very thick, dashed] (-1,0) -- (-0.8,-1.1);
\node[cross=4pt, color=bleudefrance, very thick] at (-1,0) {};
\draw [very thick, snake it] (-1,0) -- (1,0);
\node at (0,-2) {$\displaystyle \vphantom{\sim 1}$};
\node at (0,-1.65) {$\displaystyle \sim T_{\rm F1}$};
\end{tikzpicture} 
~~~~~
\begin{tikzpicture}
\draw [very thick, dashed] (-0.75,0.75) -- (0.75,-0.75);
\draw [very thick, dashed] (-0.75,-0.75) -- (0.75,0.75);
\node at (0,-1.5) {$\displaystyle \sim T_{\rm F1} $};
\end{tikzpicture} 
~~~~~
\begin{tikzpicture}
\draw [color=white] (0,0.75) -- (0,-0.75);
\node at (0,-1.5) {$\displaystyle \vphantom{\sim N^2}$};
\node at (0,0) {$\displaystyle \cdots$};
\end{tikzpicture} 
\\ ~~ \\
\begin{tikzpicture}
\draw[color=white] (0,1) -- (0,-1.9);
\node at (0,0) {$\displaystyle S_{\rm D1}:$};
\end{tikzpicture} 
~~~~~ 
\begin{tikzpicture}
\filldraw [color=candyapplered, thick, fill=candyapplered!30] (-1.5,-1.5) -- (-1.5,0.5) -- (-0.5,0.75) -- (-0.5,-1.25) -- (-1.5,-1.5);
\node[cross=4pt, color=candyapplered, very thick] at (-1,-0.35) {};
\draw [very thick, snake it] (-1,-0.35) -- (1,-0.35);
\node at (0,-1.75) {$\displaystyle \sim T_{\rm D1} \vphantom{\frac{1}{1^2}}$};
\end{tikzpicture} 
~~~~~
\begin{tikzpicture}
\node[above] at (0,0.15) {$\displaystyle A$};
\draw [very thick] (-1,0) -- (1,0);
\node at (0,-1.25) {$\displaystyle \sim \frac{1}{T_{\rm D1} \ell_s^4}$};
\end{tikzpicture} 
~~~~~
\begin{tikzpicture}
\node[above] at (0,0.15) {$\displaystyle X^i$};
\draw [very thick, dashed] (-1,0) -- (1,0);
\node at (0,-1.25) {$\displaystyle \sim \frac{1}{T_{\rm D1}}$};
\end{tikzpicture} 
~~~~~
\begin{tikzpicture}
\draw [color=white] (0,0.75) -- (0,-0.75);
\node at (0,-1.25) {$\displaystyle \vphantom{\sim N^2}$};
\node at (0,0) {$\displaystyle \hphantom{\cdots}$};
\end{tikzpicture} 
~~~~~~~~~~~~~~~~~~~~~~
\\ ~~ \\
\begin{tikzpicture}
\filldraw [color=candyapplered, thick, fill=candyapplered!30] (-1.5,-1.5) -- (-1.5,0.5) -- (-0.5,0.75) -- (-0.5,-1.25) -- (-1.5,-1.5);
\draw [very thick] (-1,0) -- (-1.2,-1.2);
\draw [very thick] (-1,0) -- (-0.8,-1.1);
\node[cross=4pt, color=candyapplered, very thick] at (-1,0) {};
\draw [very thick, snake it] (-1,0) -- (1,0);
\node at (0,-2) {$\displaystyle \sim T_{\rm D1} \ell_s^4$};
\end{tikzpicture} 
~~~~~
\begin{tikzpicture}
\filldraw [color=candyapplered, thick, fill=candyapplered!30] (-1.5,-1.5) -- (-1.5,0.5) -- (-0.5,0.75) -- (-0.5,-1.25) -- (-1.5,-1.5);
\draw [very thick, dashed] (-1,0) -- (-1.2,-1.2);
\draw [very thick, dashed] (-1,0) -- (-0.8,-1.1);
\node[cross=4pt, color=candyapplered, very thick] at (-1,0) {};
\draw [very thick, snake it] (-1,0) -- (1,0);
\node at (0,-2) {$\displaystyle \sim T_{\rm D1}$};
\end{tikzpicture} 
~~~~~
\begin{tikzpicture}
\draw [very thick] (-0.75,0.75) -- (0.75,-0.75);
\draw [very thick] (-0.75,-0.75) -- (0.75,0.75);
\node at (0,-1.5) {$\displaystyle \sim T_{\rm D1} \ell_s^8$};
\end{tikzpicture} 
~~~~~
\begin{tikzpicture}
\draw [very thick, dashed] (-0.75,0.75) -- (0.75,-0.75);
\draw [very thick, dashed] (-0.75,-0.75) -- (0.75,0.75);
\node at (0,-1.5) {$\displaystyle \sim T_{\rm D1}$};
\end{tikzpicture} 
~~~~~
\begin{tikzpicture}
\draw [color=white] (0,0.75) -- (0,-0.75);
\node at (0,-1.5) {$\displaystyle \vphantom{\sim N^2}$};
\node at (0,0) {$\displaystyle \cdots$};
\end{tikzpicture} 
\caption{Scaling of interaction vertices and propagators coming from the ten-dimensional type IIB supergravity action (\ref{eq:bulkEFT}), the Nambu-Goto action (\ref{eq:NG}), and the DBI action for a D1-brane (\ref{eq:DBI}), respectively. Here, $T_{\rm F1} = (2\pi\alpha')^{-1} \sim \lambda^{1\over 2}$ and $T_{\rm D1} \sim N \widetilde{\lambda}^{-\frac{1}{2}}$ in the respective 't Hooft limit of large $N$ and fixed 't Hooft coupling $\lambda$ or $\widetilde{\lambda}$.}
\label{fig:Feymanrules}
\end{figure}

Our conventions are such that the bulk effective type IIB supergravity action is normalized as 
\ie
S_{\rm bulk}=\frac{1}{2\kappa^2} \int d^{10}x\, \sqrt{-G} \left( R + \cdots \right) \,, \qquad \frac{1}{\kappa^2} = \frac{4 \pi}{g_s^2 (2 \pi \ell_s)^8} \,, 
\label{eq:bulkEFT}
\fe
where $G=\det G_{\mu\nu}$ is the determinant of the spacetime metric $G_{\mu\nu}$, and $R$ the corresponding Ricci scalar, and where the $\cdots$ include terms involving the other fields in the supergraviton multiplet as well as higher derivative interactions.  Hence, bulk supergraviton propagators scale as $\kappa^{2}$, whereas bulk supergraviton interaction vertices, induced from the two-derivative action, scale as $\kappa^{-2}$. See the first row of Figure~\ref{fig:Feymanrules}.  In AdS diagrams, the 10d gravitational coupling $\kappa^2$ will appear in the dimensionless combination 
 \es{kappaRatio}{
  \frac{\kappa^2}{L^8} \propto \frac{1}{N^2} \,.
 }

{\bf F1-string.}  At the fundamental level, the string theory action is given by the Green-Schwarz action with kappa symmetry \cite{Green:1983wt, Green:1984fu}, which is a supersymmetrization of the Nambu-Goto action, and in particular \cite{Metsaev:1998it,Kallosh:1998ji,Drukker:2000ep} for ${\rm AdS}_5\times {\rm S}^5$.  In our setup, however, one can describe boundary-boundary and bulk-boundary\footnote{Here, ``boundary" refers to the long F1 worldsheet.} interactions using an effective action for the light fields obtained by integrating out massive string modes.  Such an action takes the form of the Green-Schwarz action with additional higher derivative interactions on the string worldsheet.  Schematically, 
\ie
S_{\rm F1} &= - T_{\rm F1} \int d^2 \sigma \,   \sqrt{ -\det G_{\mu\nu}(X)\partial_a X^{\mu}\partial_b X^{\nu} } + \cdots \,, \qquad T_{\rm F1} = \frac{1}{2 \pi \ell_s^2} \,,
\label{eq:NG}
\fe
where $G_{\mu\nu}$ is the target space metric, $X^{\mu}$ with $\mu=0,\ldots, 9$ are the worldsheet fields describing the embedding coordinates of target spacetime, and the determinant is evaluated in the worldsheet coordinates $\sigma^a$ with $a=0,1$. The ellipses in \eqref{eq:NG} stand for the coupling of the fundamental string to the NS-NS two-form $B_{\mu\nu}$, terms involving fermions, as well as higher derivative interactions.   

Expanding the action \eqref{eq:NG} in static gauge where $X^0=\sigma^0$ and $X^1=\sigma^1$ and around a background $G_{\mu\nu}^{(0)}$, $G_{\mu\nu} = G^{(0)}_{\mu\nu} + h_{\mu\nu}$, where $h_{\mu\nu}$ represents a bulk graviton, we can obtain the coupling of a graviton to the F1 worldsheet as well as interaction vertices with boundary degrees of freedom, i.e.~with the scalar fields $X^{i}$ with $i=2,\ldots,9$ living on the F1 worldsheet.  In our conventions, the scaling of interaction vertices and propagators coming from the first term in (\ref{eq:NG})  are given in the second row of Figure~\ref{fig:Feymanrules}.   

In AdS diagrams, the fundamental string tension $T_{\rm F1}$ and $\ell_s$  will appear in the dimensionelss combinations
 \es{TF1Dimless}{
  T_{\rm F1} L^2 \propto \lambda^{\frac 12} \,, \qquad \frac{\ell_s}{L} \propto \lambda^{- \frac 14} \,.
 }

We can then obtain the scaling in $N$ and $\lambda$ of various loop diagrams with propagating gravitons and worldsheet scalars, with $l$ worldsheet loops and $L$ bulk loops:
 \ie
\begin{tikzpicture}[scale=0.75,baseline={([yshift=-.5ex]current bounding box.center)},vertex/.style={anchor=base,
    circle,fill=black!25,minimum size=18pt,inner sep=2pt}]
\draw [color=white, thick] (-2,0) -- (2,0);
\filldraw [color=bleudefrance, thick, fill=bleudefrance!30] (0,-3) -- (0,0.475) -- (-0.75,-0.4) -- (-0.75,-3.875) -- (0,-3);
\draw [color=gray, thick] (0,0) ellipse (1.25 and 0.5);
\draw [color=gray, thick] (-1.25,0) -- (-1.25,-3.5);
\draw [color=gray, thick] (-1.25,-3.5) arc (180:360:1.25 and 0.5);
\draw [color=gray, thick, dashed] (-1.25,-3.5) arc (180:360:1.25 and -0.5);
\draw [color=gray, thick] (1.25,-3.5) -- (1.25,0);
\draw [very thick, dashed] (-0.375,-1.75) ellipse (0.15 and 0.85);
\node[cross=4pt, color=bleudefrance, very thick] at (-0.375,-0.875) {};
\node[cross=4pt, color=bleudefrance, very thick] at (-0.375,-2.625) {};
\filldraw [very thick, fill = black!40] (-0.375,-1.75) ellipse (0.25 and 0.3);
\node[left] at (-0.65,-1.75) {$l$};
\draw [very thick, snake it] (-0.375,-0.875) -- (1,-0.875);
\filldraw (1,-0.875) circle (2pt);
\draw [very thick, snake it] (-0.375,-2.625) -- (1,-2.625);
\filldraw (1,-2.625) circle (2pt);
\end{tikzpicture}
&\sim \, \lambda^{\frac{1-l}{2}},
&~~~~~~
&\begin{tikzpicture}[scale=0.75,baseline={([yshift=-.5ex]current bounding box.center)},vertex/.style={anchor=base,
    circle,fill=black!25,minimum size=18pt,inner sep=2pt}]
\draw [color=white, thick] (-2,0) -- (2,0);
\filldraw [color=bleudefrance, thick, fill=bleudefrance!30] (0,-3) -- (0,0.475) -- (-0.75,-0.4) -- (-0.75,-3.875) -- (0,-3);
\draw [color=gray, thick] (0,0) ellipse (1.25 and 0.5);
\draw [color=gray, thick] (-1.25,0) -- (-1.25,-3.5);
\draw [color=gray, thick] (-1.25,-3.5) arc (180:360:1.25 and 0.5);
\draw [color=gray, thick, dashed] (-1.25,-3.5) arc (180:360:1.25 and -0.5);
\draw [color=gray, thick] (1.25,-3.5) -- (1.25,0);
\node[cross=4pt, color=bleudefrance, very thick] at (-0.375,-1.75) {};
\draw [very thick, snake it] (-0.375,-1.75) -- (0.375,-1.75);
\draw [very thick, snake it] (0.375,-1.75) -- (1,-1);
\filldraw (1,-1) circle (2pt);
\draw [very thick, snake it] (0.375,-1.75) -- (1,-2.5);
\filldraw (1,-2.5) circle (2pt);
\filldraw[very thick, fill = black!40] (0.375,-1.75) circle (7pt);
\node[above] at (0.375,-1.5) {$L$};
\end{tikzpicture}
&\sim \, \frac{\lambda^{\frac 12}}{N^{2L}}, & \\
\begin{tikzpicture}[scale=0.75,baseline={([yshift=-.5ex]current bounding box.center)},vertex/.style={anchor=base,
    circle,fill=black!25,minimum size=18pt,inner sep=2pt}]
\draw [color=white, thick] (-2,0) -- (2,0);
\filldraw [color=bleudefrance, thick, fill=bleudefrance!30] (0,-3) -- (0,0.475) -- (-0.75,-0.4) -- (-0.75,-3.875) -- (0,-3);
\draw [color=gray, thick] (0,0) ellipse (1.25 and 0.5);
\draw [color=gray, thick] (-1.25,0) -- (-1.25,-3.5);
\draw [color=gray, thick] (-1.25,-3.5) arc (180:360:1.25 and 0.5);
\draw [color=gray, thick, dashed] (-1.25,-3.5) arc (180:360:1.25 and -0.5);
\draw [color=gray, thick] (1.25,-3.5) -- (1.25,0);
\node[cross=4pt, color=bleudefrance, very thick] at (-0.375,-0.875) {};
\node[cross=4pt, color=bleudefrance, very thick] at (-0.375,-2.625) {};
\draw [very thick, snake it] (-0.375,-0.875) -- (0.3,-1.75);
\draw [very thick, snake it] (0.3,-1.75) -- (1,-0.875);
\filldraw (1,-0.875) circle (2pt);
\draw [very thick, snake it] (-0.375,-2.625) -- (0.3,-1.75);
\draw [very thick, snake it] (0.3,-1.75) -- (1,-2.625);
\filldraw (1,-2.625) circle (2pt);
\filldraw[very thick, fill = black!40] (0.375,-1.75) circle (7pt);
\node[above] at (0.375,-1.5) {$L$};
\end{tikzpicture}
&\sim\, \frac{\lambda}{N^{2L+2}},
&~~~~~~
&\begin{tikzpicture}[scale=0.75,baseline={([yshift=-.5ex]current bounding box.center)},vertex/.style={anchor=base,
    circle,fill=black!25,minimum size=18pt,inner sep=2pt}]
\draw [color=white, thick] (-2,0) -- (2,0);
\filldraw [color=bleudefrance, thick, fill=bleudefrance!30] (0,-3) -- (0,0.475) -- (-0.75,-0.4) -- (-0.75,-3.875) -- (0,-3);
\draw [color=gray, thick] (0,0) ellipse (1.25 and 0.5);
\draw [color=gray, thick] (-1.25,0) -- (-1.25,-3.5);
\draw [color=gray, thick] (-1.25,-3.5) arc (180:360:1.25 and 0.5);
\draw [color=gray, thick, dashed] (-1.25,-3.5) arc (180:360:1.25 and -0.5);
\draw [color=gray, thick] (1.25,-3.5) -- (1.25,0);
\draw [very thick, dashed] (-0.375,-1.75) ellipse (0.15 and 0.85);
\node[cross=4pt, color=bleudefrance, very thick] at (-0.375,-0.875) {};
\node[cross=4pt, color=bleudefrance, very thick] at (-0.375,-2.625) {};
\filldraw [very thick, fill = black!40] (-0.375,-1.75) ellipse (0.25 and 0.3);
\node[left] at (-0.65,-1.75) {$l$};
\draw [very thick, snake it] (-0.375,-0.875) -- (0.3,-1.75);
\draw [very thick, snake it] (0.3,-1.75) -- (1,-0.875);
\filldraw (1,-0.875) circle (2pt);
\draw [very thick, snake it] (-0.375,-2.625) -- (0.3,-1.75);
\draw [very thick, snake it] (0.3,-1.75) -- (1,-2.625);
\filldraw (1,-2.625) circle (2pt);
\filldraw[very thick, fill = black!40] (0.375,-1.75) circle (7pt);
\node[above] at (0.33,-1.5) {$L$};
\end{tikzpicture}
&\sim\, \frac{\lambda^{\frac{1-l}{2}}}{N^{2L+2}},
& ~~~~ \ldots
\label{eq:loopdiagramsF1}
\fe

In addition, there are higher-derivative contact interactions arising from integrating out massive closed string states at $h$ closed string loop order (coming from worldsheets with $h$ handles), which we denote schematically by $g_s^{2h} D^{2m} R^2$, with $m = 0, 1, 2, \ldots$.  Here, $R$ is the ambient curvature pulled back to the worldsheet and $D$ denotes a derivative in a direction tangential to the string worldsheet.  Such diagrams give
\ie
\begin{tikzpicture}[scale=0.75,baseline={([yshift=-.5ex]current bounding box.center)},vertex/.style={anchor=base,
    circle,fill=black!25,minimum size=18pt,inner sep=2pt}]
\draw [color=white, thick] (-2,0) -- (2,0);
\filldraw [color=bleudefrance, thick, fill=bleudefrance!30] (0,-3) -- (0,0.475) -- (-0.75,-0.4) -- (-0.75,-3.875) -- (0,-3);
\draw [color=gray, thick] (0,0) ellipse (1.25 and 0.5);
\draw [color=gray, thick] (-1.25,0) -- (-1.25,-3.5);
\draw [color=gray, thick] (-1.25,-3.5) arc (180:360:1.25 and 0.5);
\draw [color=gray, thick, dashed] (-1.25,-3.5) arc (180:360:1.25 and -0.5);
\draw [color=gray, thick] (1.25,-3.5) -- (1.25,0);
\node[cross=4pt, color=bleudefrance, very thick] at (-0.375,-1.75) {};
\node[above] at (-1.2,-1.65) {$g_s^{2h} D^{2m} R^2$};
\draw [very thick, snake it] (-0.375,-1.75) -- (1,-1);
\filldraw (1,-1) circle (2pt);
\draw [very thick, snake it ] (-0.375,-1.75) -- (1,-2.5);
\filldraw (1,-2.5) circle (2pt);
\end{tikzpicture}
\sim \, \frac{1}{N^{2h}}\, \lambda^{2h-\frac{m}{2}-\frac{1}{2}} \,.
\label{eq:higherderivF1}
\fe

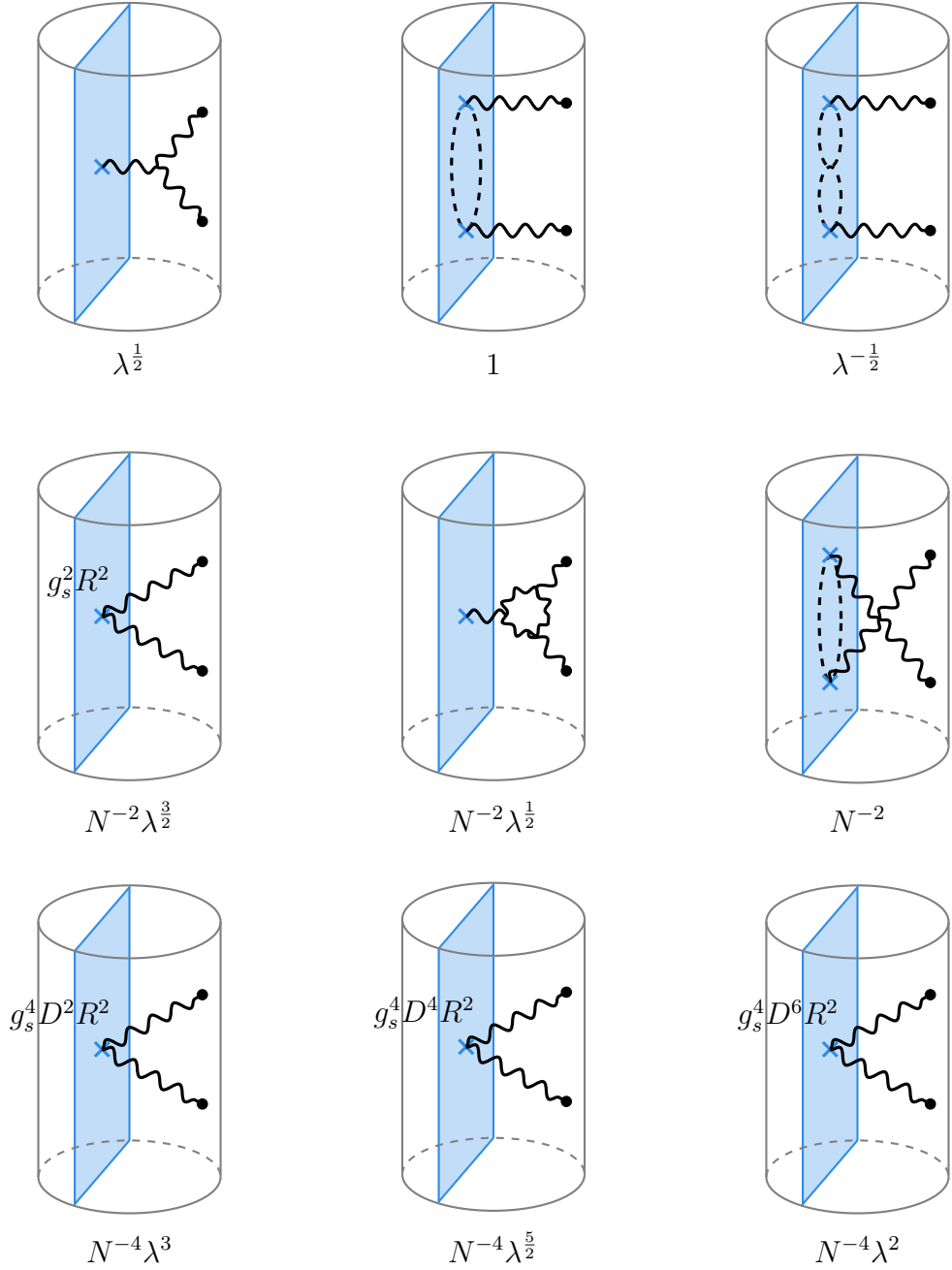
\begin{figure}[!htb]
\centering
\begin{tikzpicture}
\draw [color=white, thick] (-2,0) -- (2,0);
\filldraw [color=bleudefrance, thick, fill=bleudefrance!30] (0,-3) -- (0,0.475) -- (-0.75,-0.4) -- (-0.75,-3.875) -- (0,-3);
\draw [color=gray, thick] (0,0) ellipse (1.25 and 0.5);
\draw [color=gray, thick] (-1.25,0) -- (-1.25,-3.5);
\draw [color=gray, thick] (-1.25,-3.5) arc (180:360:1.25 and 0.5);
\draw [color=gray, thick, dashed] (-1.25,-3.5) arc (180:360:1.25 and -0.5);
\draw [color=gray, thick] (1.25,-3.5) -- (1.25,0);
\node[cross=4pt, color=bleudefrance, very thick] at (-0.375,-1.75) {};
\draw [very thick, snake it] (-0.375,-1.75) -- (0.375,-1.75);
\draw [very thick, snake it] (0.375,-1.75) -- (1,-1);
\filldraw (1,-1) circle (2pt);
\draw [very thick, snake it] (0.375,-1.75) -- (1,-2.5);
\filldraw (1,-2.5) circle (2pt);
\node at (0,-4.5) {$\displaystyle\lambda^{1\over 2} \vphantom{\frac{N}{\widetilde{\lambda}^{1\over 2}}}$};
\end{tikzpicture} 
~~~~~
\begin{tikzpicture}
\draw [color=white, thick] (-2,0) -- (2,0);
\filldraw [color=bleudefrance, thick, fill=bleudefrance!30] (0,-3) -- (0,0.475) -- (-0.75,-0.4) -- (-0.75,-3.875) -- (0,-3);
\draw [color=gray, thick] (0,0) ellipse (1.25 and 0.5);
\draw [color=gray, thick] (-1.25,0) -- (-1.25,-3.5);
\draw [color=gray, thick] (-1.25,-3.5) arc (180:360:1.25 and 0.5);
\draw [color=gray, thick, dashed] (-1.25,-3.5) arc (180:360:1.25 and -0.5);
\draw [color=gray, thick] (1.25,-3.5) -- (1.25,0);
\draw [very thick, dashed] (-0.375,-1.75) ellipse (0.2 and 0.85);
\node[cross=4pt, color=bleudefrance, very thick] at (-0.375,-0.875) {};
\node[cross=4pt, color=bleudefrance, very thick] at (-0.375,-2.625) {};
\draw [very thick, snake it] (-0.375,-0.875) -- (1,-0.875);
\filldraw (1,-0.875) circle (2pt);
\draw [very thick, snake it] (-0.375,-2.625) -- (1,-2.625);
\filldraw (1,-2.625) circle (2pt);
\node at (0,-4.5) {$\displaystyle 1 \vphantom{\frac{N}{\widetilde{\lambda}^{1\over 2}}}$};
\end{tikzpicture}
~~~~~
\begin{tikzpicture}
\draw [color=white, thick] (-2,0) -- (2,0);
\filldraw [color=bleudefrance, thick, fill=bleudefrance!30] (0,-3) -- (0,0.475) -- (-0.75,-0.4) -- (-0.75,-3.875) -- (0,-3);
\draw [color=gray, thick] (0,0) ellipse (1.25 and 0.5);
\draw [color=gray, thick] (-1.25,0) -- (-1.25,-3.5);
\draw [color=gray, thick] (-1.25,-3.5) arc (180:360:1.25 and 0.5);
\draw [color=gray, thick, dashed] (-1.25,-3.5) arc (180:360:1.25 and -0.5);
\draw [color=gray, thick] (1.25,-3.5) -- (1.25,0);
\draw [very thick, dashed] (-0.375,-1.325) ellipse (0.15 and 0.425);
\draw [very thick, dashed] (-0.375,-2.175) ellipse (0.15 and 0.425);
\node[cross=4pt, color=bleudefrance, very thick] at (-0.375,-0.875) {};
\node[cross=4pt, color=bleudefrance, very thick] at (-0.375,-2.625) {};
\draw [very thick, snake it] (-0.375,-0.875) -- (1,-0.875);
\filldraw (1,-0.875) circle (2pt);
\draw [very thick, snake it] (-0.375,-2.625) -- (1,-2.625);
\filldraw (1,-2.625) circle (2pt);
\node at (0,-4.5) {$\displaystyle\lambda^{ - {1\over 2}} \vphantom{\frac{N}{\widetilde{\lambda}^{1\over 2}}}$};
\end{tikzpicture}
\\ ~ \\
\begin{tikzpicture}
\draw [color=white, thick] (-2,0) -- (2,0);
\filldraw [color=bleudefrance, thick, fill=bleudefrance!30] (0,-3) -- (0,0.475) -- (-0.75,-0.4) -- (-0.75,-3.875) -- (0,-3);
\draw [color=gray, thick] (0,0) ellipse (1.25 and 0.5);
\draw [color=gray, thick] (-1.25,0) -- (-1.25,-3.5);
\draw [color=gray, thick] (-1.25,-3.5) arc (180:360:1.25 and 0.5);
\draw [color=gray, thick, dashed] (-1.25,-3.5) arc (180:360:1.25 and -0.5);
\draw [color=gray, thick] (1.25,-3.5) -- (1.25,0);
\node[cross=4pt, color=bleudefrance, very thick] at (-0.375,-1.75) {};
\node[above] at (-0.7,-1.65) {$g_s^2 R^2$};
\draw [very thick, snake it] (-0.375,-1.75) -- (1,-1);
\filldraw (1,-1) circle (2pt);
\draw [very thick, snake it] (-0.375,-1.75) -- (1,-2.5);
\filldraw (1,-2.5) circle (2pt);
\node at (0,-4.5) {$\displaystyle N^{-2} \lambda^{{3\over 2}}$};
\end{tikzpicture}
~~~~~ 
\begin{tikzpicture}
\draw [color=white, thick] (-2,0) -- (2,0);
\filldraw [color=bleudefrance, thick, fill=bleudefrance!30] (0,-3) -- (0,0.475) -- (-0.75,-0.4) -- (-0.75,-3.875) -- (0,-3);
\draw [color=gray, thick] (0,0) ellipse (1.25 and 0.5);
\draw [color=gray, thick] (-1.25,0) -- (-1.25,-3.5);
\draw [color=gray, thick] (-1.25,-3.5) arc (180:360:1.25 and 0.5);
\draw [color=gray, thick, dashed] (-1.25,-3.5) arc (180:360:1.25 and -0.5);
\draw [color=gray, thick] (1.25,-3.5) -- (1.25,0);
\node[cross=4pt, color=bleudefrance, very thick] at (-0.375,-1.75) {};
\draw [very thick, snake it] (-0.375,-1.75) -- (0.375,-1.75);
\draw [very thick, snake it] (0.375,-1.75) -- (1,-1);
\filldraw (1,-1) circle (2pt);
\draw [very thick, snake it] (0.375,-1.75) -- (1,-2.5);
\filldraw (1,-2.5) circle (2pt);
\filldraw[very thick, fill = white, decorate, decoration={snake, segment length=2.4mm, amplitude=0.5mm}] (0.45,-1.75) circle (8.75pt);
\node at (0,-4.5) {$\displaystyle N^{-2} \lambda^{{1\over 2}}$};
\end{tikzpicture}
~~~~~
\begin{tikzpicture}
\draw [color=white, thick] (-2,0) -- (2,0);
\filldraw [color=bleudefrance, thick, fill=bleudefrance!30] (0,-3) -- (0,0.475) -- (-0.75,-0.4) -- (-0.75,-3.875) -- (0,-3);
\draw [color=gray, thick] (0,0) ellipse (1.25 and 0.5);
\draw [color=gray, thick] (-1.25,0) -- (-1.25,-3.5);
\draw [color=gray, thick] (-1.25,-3.5) arc (180:360:1.25 and 0.5);
\draw [color=gray, thick, dashed] (-1.25,-3.5) arc (180:360:1.25 and -0.5);
\draw [color=gray, thick] (1.25,-3.5) -- (1.25,0);
\draw [very thick, dashed] (-0.375,-1.75) ellipse (0.15 and 0.85);
\node[cross=4pt, color=bleudefrance, very thick] at (-0.375,-0.875) {};
\node[cross=4pt, color=bleudefrance, very thick] at (-0.375,-2.625) {};
\draw [very thick, snake it] (-0.375,-0.875) -- (0.3,-1.75);
\draw [very thick, snake it] (0.3,-1.75) -- (1,-0.875);
\filldraw (1,-0.875) circle (2pt);
\draw [very thick, snake it] (-0.375,-2.625) -- (0.3,-1.75);
\draw [very thick, snake it] (0.3,-1.75) -- (1,-2.625);
\filldraw (1,-2.625) circle (2pt);
\node at (0,-4.5) {$\displaystyle N^{-2}$};
\end{tikzpicture}
\\ ~ \\
\begin{tikzpicture}
\draw [color=white, thick] (-2,0) -- (2,0);
\filldraw [color=bleudefrance, thick, fill=bleudefrance!30] (0,-3) -- (0,0.475) -- (-0.75,-0.4) -- (-0.75,-3.875) -- (0,-3);
\draw [color=gray, thick] (0,0) ellipse (1.25 and 0.5);
\draw [color=gray, thick] (-1.25,0) -- (-1.25,-3.5);
\draw [color=gray, thick] (-1.25,-3.5) arc (180:360:1.25 and 0.5);
\draw [color=gray, thick, dashed] (-1.25,-3.5) arc (180:360:1.25 and -0.5);
\draw [color=gray, thick] (1.25,-3.5) -- (1.25,0);
\node[cross=4pt, color=bleudefrance, very thick] at (-0.375,-1.75) {};
\node[above] at (-0.95,-1.65) {$g_s^4 D^2R^2$};
\draw [very thick, snake it] (-0.375,-1.75) -- (1,-1);
\filldraw (1,-1) circle (2pt);
\draw [very thick, snake it] (-0.375,-1.75) -- (1,-2.5);
\filldraw (1,-2.5) circle (2pt);
\node at (0,-4.5) {$\displaystyle N^{-4} \lambda^{3} $};
\end{tikzpicture}
~~~~~ 
\begin{tikzpicture}
\draw [color=white, thick] (-2,0) -- (2,0);
\filldraw [color=bleudefrance, thick, fill=bleudefrance!30] (0,-3) -- (0,0.475) -- (-0.75,-0.4) -- (-0.75,-3.875) -- (0,-3);
\draw [color=gray, thick] (0,0) ellipse (1.25 and 0.5);
\draw [color=gray, thick] (-1.25,0) -- (-1.25,-3.5);
\draw [color=gray, thick] (-1.25,-3.5) arc (180:360:1.25 and 0.5);
\draw [color=gray, thick, dashed] (-1.25,-3.5) arc (180:360:1.25 and -0.5);
\draw [color=gray, thick] (1.25,-3.5) -- (1.25,0);
\node[cross=4pt, color=bleudefrance, very thick] at (-0.375,-1.75) {};
\node[above] at (-0.95,-1.65) {$g_s^4 D^4 R^2$};
\draw [very thick, snake it] (-0.375,-1.75) -- (1,-1);
\filldraw (1,-1) circle (2pt);
\draw [very thick, snake it] (-0.375,-1.75) -- (1,-2.5);
\filldraw (1,-2.5) circle (2pt);
\node at (0,-4.5) {$\displaystyle N^{-4} \lambda^{5\over 2} $};
\end{tikzpicture}
~~~~~
\begin{tikzpicture}
\draw [color=white, thick] (-2,0) -- (2,0);
\filldraw [color=bleudefrance, thick, fill=bleudefrance!30] (0,-3) -- (0,0.475) -- (-0.75,-0.4) -- (-0.75,-3.875) -- (0,-3);
\draw [color=gray, thick] (0,0) ellipse (1.25 and 0.5);
\draw [color=gray, thick] (-1.25,0) -- (-1.25,-3.5);
\draw [color=gray, thick] (-1.25,-3.5) arc (180:360:1.25 and 0.5);
\draw [color=gray, thick, dashed] (-1.25,-3.5) arc (180:360:1.25 and -0.5);
\draw [color=gray, thick] (1.25,-3.5) -- (1.25,0);
\node[cross=4pt, color=bleudefrance, very thick] at (-0.375,-1.75) {};
\node[above] at (-0.95,-1.65) {$g_s^4 D^6 R^2$};
\draw [very thick, snake it] (-0.375,-1.75) -- (1,-1);
\filldraw (1,-1) circle (2pt);
\draw [very thick, snake it] (-0.375,-1.75) -- (1,-2.5);
\filldraw (1,-2.5) circle (2pt);
\node at (0,-4.5) {$\displaystyle N^{-4} \lambda^{2} $};
\end{tikzpicture}
\caption{Sample perturbative diagrams in (a slice of) ${\rm AdS_5}$ that are expected to contribute at the given order in the double expansion in large $N$ and large 't Hooft coupling $\lambda$ (shown below each diagram) to the result of the integrated correlator (\ref{Expansions}) in the presence of a $1\over 2$-BPS Wilson line. The blue sheet stretching across AdS represents the worldsheet of a long fundamental (F1) string, the wavy lines represent 1-particle states in the bulk supergraviton multiplet, and the dashed lines represent 1-particle states living on the two-dimensional F1 worldsheet.}
\label{fig:sampleF1diagrams}
\end{figure}

\begin{figure}[!htb]
\centering
\begin{tikzpicture}
\draw [color=white, thick] (-2,0) -- (2,0);
\filldraw [color=candyapplered, thick, fill=candyapplered!30] (0,-3) -- (0,0.475) -- (-0.75,-0.4) -- (-0.75,-3.875) -- (0,-3);
\draw [color=gray, thick] (0,0) ellipse (1.25 and 0.5);
\draw [color=gray, thick] (-1.25,0) -- (-1.25,-3.5);
\draw [color=gray, thick] (-1.25,-3.5) arc (180:360:1.25 and 0.5);
\draw [color=gray, thick, dashed] (-1.25,-3.5) arc (180:360:1.25 and -0.5);
\draw [color=gray, thick] (1.25,-3.5) -- (1.25,0);
\node[cross=4pt, color=candyapplered, very thick] at (-0.375,-1.75) {};
\draw [very thick, snake it] (-0.375,-1.75) -- (0.375,-1.75);
\draw [very thick, snake it] (0.375,-1.75) -- (1,-1);
\filldraw (1,-1) circle (2pt);
\draw [very thick, snake it] (0.375,-1.75) -- (1,-2.5);
\filldraw (1,-2.5) circle (2pt);
\node at (0,-4.5) {$\displaystyle N \, \widetilde{\lambda}^{-{1\over 2}}$};
\end{tikzpicture} 
~~~~~
\begin{tikzpicture}
\draw [color=white, thick] (-2,0) -- (2,0);
\filldraw [color=candyapplered, thick, fill=candyapplered!30] (0,-3) -- (0,0.475) -- (-0.75,-0.4) -- (-0.75,-3.875) -- (0,-3);
\draw [color=gray, thick] (0,0) ellipse (1.25 and 0.5);
\draw [color=gray, thick] (-1.25,0) -- (-1.25,-3.5);
\draw [color=gray, thick] (-1.25,-3.5) arc (180:360:1.25 and 0.5);
\draw [color=gray, thick, dashed] (-1.25,-3.5) arc (180:360:1.25 and -0.5);
\draw [color=gray, thick] (1.25,-3.5) -- (1.25,0);
\node[cross=4pt, color=candyapplered, very thick] at (-0.375,-1.75) {};
\node[above] at (-0.375,-1.65) {$R^2$};
\draw [very thick, snake it] (-0.375,-1.75) -- (1,-1);
\filldraw (1,-1) circle (2pt);
\draw [very thick, snake it] (-0.375,-1.75) -- (1,-2.5);
\filldraw (1,-2.5) circle (2pt);
\node at (0,-4.5) {$\displaystyle N \, \widetilde{\lambda}^{-{3\over 2}}$};
\end{tikzpicture} 
~~~~~
\begin{tikzpicture}
\draw [color=white, thick] (-2,0) -- (2,0);
\filldraw [color=candyapplered, thick, fill=candyapplered!30] (0,-3) -- (0,0.475) -- (-0.75,-0.4) -- (-0.75,-3.875) -- (0,-3);
\draw [color=gray, thick] (0,0) ellipse (1.25 and 0.5);
\draw [color=gray, thick] (-1.25,0) -- (-1.25,-3.5);
\draw [color=gray, thick] (-1.25,-3.5) arc (180:360:1.25 and 0.5);
\draw [color=gray, thick, dashed] (-1.25,-3.5) arc (180:360:1.25 and -0.5);
\draw [color=gray, thick] (1.25,-3.5) -- (1.25,0);
\node[cross=4pt, color=candyapplered, very thick] at (-0.375,-1.75) {};
\node[above] at (-0.65,-1.65) {$D^2R^2$};
\draw [very thick, snake it] (-0.375,-1.75) -- (1,-1);
\filldraw (1,-1) circle (2pt);
\draw [very thick, snake it] (-0.375,-1.75) -- (1,-2.5);
\filldraw (1,-2.5) circle (2pt);
\node at (0,-4.5) {$\displaystyle N \, \widetilde{\lambda}^{-2} \vphantom{N \, \widetilde{\lambda}^{-{3\over 2}}}$};
\end{tikzpicture} 
\\ ~ \\
\begin{tikzpicture}
\draw [color=white, thick] (-2,0) -- (2,0);
\filldraw [color=candyapplered, thick, fill=candyapplered!30] (0,-3) -- (0,0.475) -- (-0.75,-0.4) -- (-0.75,-3.875) -- (0,-3);
\draw [color=gray, thick] (0,0) ellipse (1.25 and 0.5);
\draw [color=gray, thick] (-1.25,0) -- (-1.25,-3.5);
\draw [color=gray, thick] (-1.25,-3.5) arc (180:360:1.25 and 0.5);
\draw [color=gray, thick, dashed] (-1.25,-3.5) arc (180:360:1.25 and -0.5);
\draw [color=gray, thick] (1.25,-3.5) -- (1.25,0);
\draw [very thick, dashed] (-0.375,-1.75) ellipse (0.2 and 0.85);
\node[cross=4pt, color=candyapplered, very thick] at (-0.375,-0.875) {};
\node[cross=4pt, color=candyapplered, very thick] at (-0.375,-2.625) {};
\draw [very thick, snake it] (-0.375,-0.875) -- (1,-0.875);
\filldraw (1,-0.875) circle (2pt);
\draw [very thick, snake it] (-0.375,-2.625) -- (1,-2.625);
\filldraw (1,-2.625) circle (2pt);
\node at (0,-4.5) {$\displaystyle 1 \vphantom{\widetilde{\lambda}^{-{3\over 2}}}$};
\end{tikzpicture}
~~~~~ 
\begin{tikzpicture}
\draw [color=white, thick] (-2,0) -- (2,0);
\filldraw [color=candyapplered, thick, fill=candyapplered!30] (0,-3) -- (0,0.475) -- (-0.75,-0.4) -- (-0.75,-3.875) -- (0,-3);
\draw [color=gray, thick] (0,0) ellipse (1.25 and 0.5);
\draw [color=gray, thick] (-1.25,0) -- (-1.25,-3.5);
\draw [color=gray, thick] (-1.25,-3.5) arc (180:360:1.25 and 0.5);
\draw [color=gray, thick, dashed] (-1.25,-3.5) arc (180:360:1.25 and -0.5);
\draw [color=gray, thick] (1.25,-3.5) -- (1.25,0);
\node[cross=4pt, color=candyapplered, very thick] at (-0.375,-0.875) {};
\node[cross=4pt, color=candyapplered, very thick] at (-0.375,-2.625) {};
\draw [very thick, snake it] (-0.375,-0.875) -- (0.3,-1.2);
\draw [very thick, snake it] (0.3,-1.2) -- (1,-0.875);
\draw [very thick, snake it] (0.3,-1.2) -- (0.3,-2.3);
\filldraw (1,-0.875) circle (2pt);
\draw [very thick, snake it] (-0.375,-2.625) -- (0.3,-2.3);
\draw [very thick, snake it] (0.3,-2.3) -- (1,-2.625);
\filldraw (1,-2.625) circle (2pt);
\node at (0,-4.5) {$\displaystyle \widetilde{\lambda}^{-1}$};
\end{tikzpicture}
~~~~~
\begin{tikzpicture}
\draw [color=white, thick] (-2,0) -- (2,0);
\filldraw [color=candyapplered, thick, fill=candyapplered!30] (0,-3) -- (0,0.475) -- (-0.75,-0.4) -- (-0.75,-3.875) -- (0,-3);
\draw [color=gray, thick] (0,0) ellipse (1.25 and 0.5);
\draw [color=gray, thick] (-1.25,0) -- (-1.25,-3.5);
\draw [color=gray, thick] (-1.25,-3.5) arc (180:360:1.25 and 0.5);
\draw [color=gray, thick, dashed] (-1.25,-3.5) arc (180:360:1.25 and -0.5);
\draw [color=gray, thick] (1.25,-3.5) -- (1.25,0);
\node[cross=4pt, color=candyapplered, very thick] at (-0.375,-1.75) {};
\node[above] at (-0.85,-1.65) {$g_s D^4R^2$};
\draw [very thick, snake it] (-0.375,-1.75) -- (1,-1);
\filldraw (1,-1) circle (2pt);
\draw [very thick, snake it] (-0.375,-1.75) -- (1,-2.5);
\filldraw (1,-2.5) circle (2pt);
\node at (0,-4.5) {$\displaystyle \widetilde{\lambda}^{-{3\over 2}}$};
\end{tikzpicture}
\\ ~ \\
\begin{tikzpicture}
\draw [color=white, thick] (-2,0) -- (2,0);
\filldraw [color=candyapplered, thick, fill=candyapplered!30] (0,-3) -- (0,0.475) -- (-0.75,-0.4) -- (-0.75,-3.875) -- (0,-3);
\draw [color=gray, thick] (0,0) ellipse (1.25 and 0.5);
\draw [color=gray, thick] (-1.25,0) -- (-1.25,-3.5);
\draw [color=gray, thick] (-1.25,-3.5) arc (180:360:1.25 and 0.5);
\draw [color=gray, thick, dashed] (-1.25,-3.5) arc (180:360:1.25 and -0.5);
\draw [color=gray, thick] (1.25,-3.5) -- (1.25,0);
\node[cross=4pt, color=candyapplered, very thick] at (-0.375,-1.75) {};
\node[above] at (-0.7,-1.65) {$g_s^2 R^2$};
\draw [very thick, snake it] (-0.375,-1.75) -- (1,-1);
\filldraw (1,-1) circle (2pt);
\draw [very thick, snake it] (-0.375,-1.75) -- (1,-2.5);
\filldraw (1,-2.5) circle (2pt);
\node at (0,-4.5) {$\displaystyle N^{-1} \widetilde{\lambda}^{{1\over 2}}$};
\end{tikzpicture}
~~~~~ 
\begin{tikzpicture}
\draw [color=white, thick] (-2,0) -- (2,0);
\filldraw [color=candyapplered, thick, fill=candyapplered!30] (0,-3) -- (0,0.475) -- (-0.75,-0.4) -- (-0.75,-3.875) -- (0,-3);
\draw [color=gray, thick] (0,0) ellipse (1.25 and 0.5);
\draw [color=gray, thick] (-1.25,0) -- (-1.25,-3.5);
\draw [color=gray, thick] (-1.25,-3.5) arc (180:360:1.25 and 0.5);
\draw [color=gray, thick, dashed] (-1.25,-3.5) arc (180:360:1.25 and -0.5);
\draw [color=gray, thick] (1.25,-3.5) -- (1.25,0);
\node[cross=4pt, color=candyapplered, very thick] at (-0.375,-1.75) {};
\node[above] at (-0.95,-1.65) {$g_s^2 D^2R^2$};
\draw [very thick, snake it] (-0.375,-1.75) -- (1,-1);
\filldraw (1,-1) circle (2pt);
\draw [very thick, snake it] (-0.375,-1.75) -- (1,-2.5);
\filldraw (1,-2.5) circle (2pt);
\node at (0,-4.5) {$\displaystyle N^{-1} $};
\end{tikzpicture}
~~~~~
\begin{tikzpicture}
\draw [color=white, thick] (-2,0) -- (2,0);
\filldraw [color=candyapplered, thick, fill=candyapplered!30] (0,-3) -- (0,0.475) -- (-0.75,-0.4) -- (-0.75,-3.875) -- (0,-3);
\draw [color=gray, thick] (0,0) ellipse (1.25 and 0.5);
\draw [color=gray, thick] (-1.25,0) -- (-1.25,-3.5);
\draw [color=gray, thick] (-1.25,-3.5) arc (180:360:1.25 and 0.5);
\draw [color=gray, thick, dashed] (-1.25,-3.5) arc (180:360:1.25 and -0.5);
\draw [color=gray, thick] (1.25,-3.5) -- (1.25,0);
\node[cross=4pt, color=candyapplered, very thick] at (-0.375,-1.75) {};
\node[above] at (-0.95,-1.65) {$g_s^2 D^4R^2$};
\draw [very thick, snake it] (-0.375,-1.75) -- (1,-1);
\filldraw (1,-1) circle (2pt);
\draw [very thick, snake it] (-0.375,-1.75) -- (1,-2.5);
\filldraw (1,-2.5) circle (2pt);
\node at (0,-4.5) {$\displaystyle N^{-1} \widetilde{\lambda}^{-{1\over 2}}$};
\end{tikzpicture}
\caption{Sample perturbative diagrams in (a slice of) ${\rm AdS_5}$ that are expected to contribute at the given order in the double expansion in large $N$ and large dual 't Hooft coupling $\widetilde{\lambda}$ (shown below each diagram) to the result of the integrated correlator $\cI_{\mathbb{T}}(\tau,\bar\tau)$ in (\ref{DualTthooftexpansion}) in the presence of a $1\over 2$-BPS 't Hooft line. The red sheet stretching across AdS represents the worldsheet of a D1-brane, the wavy lines represent 1-particle states in the bulk supergraviton multiplet, and the solid lines represent 1-particle states in the gauge supermultiplet living on the two-dimensional D1 worldsheet.}
\label{fig:sampleAdSdiagramsD1}
\end{figure}
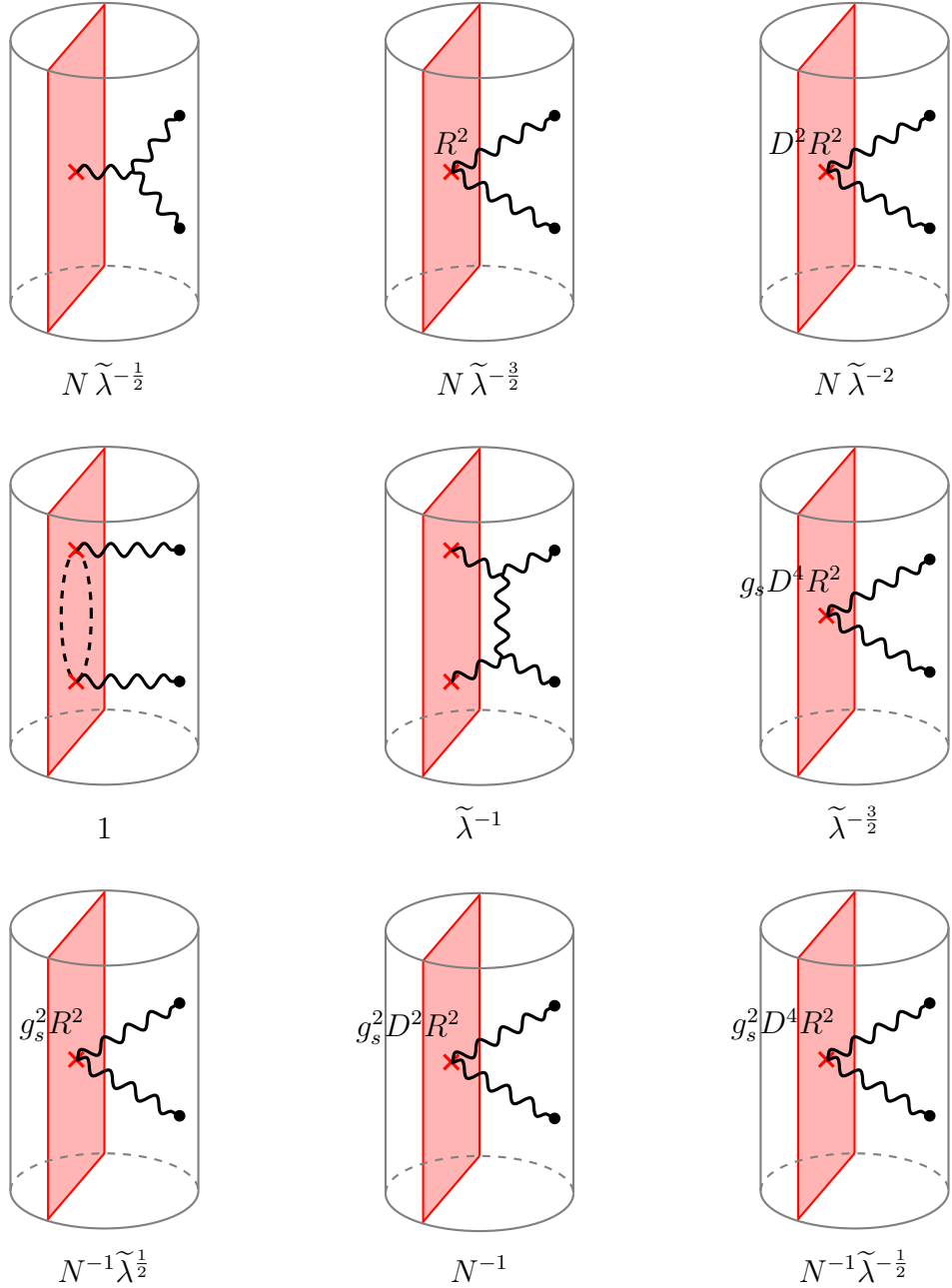

Using these rules, in Figure~\ref{fig:sampleF1diagrams} we give a few examples of diagrams that contribute to the first few terms in the $1/N$ and $1/\lambda$ expansion given in \eqref{Expansions}.\footnote{For the sake of brevity, we only present one diagram at each order in Figure~\ref{fig:sampleF1diagrams} (also in Figure~\ref{fig:sampleAdSdiagramsD1}), even though there are multiple diagrams contributing at any given order.}

{\bf D1-string.}  The S-dual case of the D1-brane (or more general $(p,q)$-strings with $q\neq 0$), whose holographic integrated correlator corresponds to that of the 't Hooft line (\ref{ZTLtwoderiv}) (or more general dyonic lines), also admits an 't Hooft expansion in which the  higher order terms in $1/N$ can be understood in terms of effective field theory diagrams in AdS\@. 

In this case, the boundary-boundary and bulk-boundary interactions are given by the low energy effective action of the massless degrees of freedom on the D1-brane:
\ie
S_{\rm D1} = - T_{\rm D1} \int d^{2}\xi \vphantom{e^{-\Phi}} \sqrt{- \det( G_{ab} + B_{ab} +2\pi \ell_s^2 F_{ab} )} + \cdots\,, \qquad T_{\rm D1} = \frac{1}{2 \pi \ell_s^2 g_s} \,.
\label{eq:DBI}
\fe
Here, the embedding of the D1-brane in spacetime is described by the fields $X^\mu(\xi)$ with coordinates $\xi^{a}$, $a=1,2$, on the D1-brane, $G_{ab}(\xi) = G_{\mu\nu}(X(\xi)) \partial_a X^\mu(\xi) \partial_b X^\mu(\xi)$ and $B_{ab} = B_{\mu\nu}(X(\xi)) \partial_a X^\mu(\xi) \partial_b X^\mu(\xi)$ are the induced metric and antisymmetric tensor on the brane, $F_{ab}(\xi)$ is the field strength of the gauge fields living on the brane, and the $\cdots$ include terms involving fermionic degrees of freedom and RR gauge fields as well as higher derivative interactions on top of those implied by the expansion of the square root in the DBI action \eqref{eq:DBI}.  Interaction vertices of degrees of freedom on the D1 brane scale as $T_{\rm D1}$, whereas their propagators scale as $T_{\rm D1}^{-1}$.  Expanding the action (\ref{eq:DBI}) in static gauge and around a background $G_{\mu\nu}^{0}$, $G_{\mu\nu} =G^{(0)}_{\mu\nu} + h_{\mu\nu}$, we can deduce the scaling of various interaction vertices between bulk and boundary degrees of freedom.   See the last two rows of Figure~\ref{fig:Feymanrules}.  For diagrams in AdS, $\ell_s$ and $T_{\rm D1}$ will always appear in the dimensionless combinations
 \es{TD1Dimless}{
  T_{\rm D1} L^2 \propto  \frac{N}{\sqrt{\widetilde{\lambda}}}  \,, \qquad 
   \frac{\ell_s}{L} \propto \widetilde{\lambda}^{-\frac 14} \,.
}
(As in the previous sections, we use $\widetilde{g}$ and $\widetilde{\lambda} = \widetilde{g}^2 N$ for the gauge coupling and the 't Hooft coupling in the case of a 't Hooft line / D1-brane.)

We can then obtain the scaling with $N$ and $\widetilde \lambda$ in the 't Hooft expansion for several classes of diagrams with $l$ loop order on the brane worldsheet and $L$ loop order in the bulk as follows:
\ie
\begin{tikzpicture}[scale=0.75,baseline={([yshift=-.5ex]current bounding box.center)},vertex/.style={anchor=base,
    circle,fill=black!25,minimum size=18pt,inner sep=2pt}]
\draw [color=white, thick] (-2,0) -- (2,0);
\filldraw [color=candyapplered, thick, fill=candyapplered!30] (0,-3) -- (0,0.475) -- (-0.75,-0.4) -- (-0.75,-3.875) -- (0,-3);
\draw [color=gray, thick] (0,0) ellipse (1.25 and 0.5);
\draw [color=gray, thick] (-1.25,0) -- (-1.25,-3.5);
\draw [color=gray, thick] (-1.25,-3.5) arc (180:360:1.25 and 0.5);
\draw [color=gray, thick, dashed] (-1.25,-3.5) arc (180:360:1.25 and -0.5);
\draw [color=gray, thick] (1.25,-3.5) -- (1.25,0);
\draw [very thick, dashed] (-0.375,-1.75) ellipse (0.15 and 0.85);
\node[cross=4pt, color=candyapplered, very thick] at (-0.375,-0.875) {};
\node[cross=4pt, color=candyapplered, very thick] at (-0.375,-2.625) {};
\filldraw [very thick, fill = black!40] (-0.375,-1.75) ellipse (0.25 and 0.3);
\node[left] at (-0.65,-1.75) {$l$};
\draw [very thick, snake it] (-0.375,-0.875) -- (1,-0.875);
\filldraw (1,-0.875) circle (2pt);
\draw [very thick, snake it] (-0.375,-2.625) -- (1,-2.625);
\filldraw (1,-2.625) circle (2pt);
\end{tikzpicture}
&\sim \, \frac{\lambdad^{\frac{l-1}{2}}}{N^{l-1}},
&~~~~~~
&\begin{tikzpicture}[scale=0.75,baseline={([yshift=-.5ex]current bounding box.center)},vertex/.style={anchor=base,
    circle,fill=black!25,minimum size=18pt,inner sep=2pt}]
\draw [color=white, thick] (-2,0) -- (2,0);
\filldraw [color=candyapplered, thick, fill=candyapplered!30] (0,-3) -- (0,0.475) -- (-0.75,-0.4) -- (-0.75,-3.875) -- (0,-3);
\draw [color=gray, thick] (0,0) ellipse (1.25 and 0.5);
\draw [color=gray, thick] (-1.25,0) -- (-1.25,-3.5);
\draw [color=gray, thick] (-1.25,-3.5) arc (180:360:1.25 and 0.5);
\draw [color=gray, thick, dashed] (-1.25,-3.5) arc (180:360:1.25 and -0.5);
\draw [color=gray, thick] (1.25,-3.5) -- (1.25,0);
\node[cross=4pt, color=candyapplered, very thick] at (-0.375,-1.75) {};
\draw [very thick, snake it] (-0.375,-1.75) -- (0.375,-1.75);
\draw [very thick, snake it] (0.375,-1.75) -- (1,-1);
\filldraw (1,-1) circle (2pt);
\draw [very thick, snake it] (0.375,-1.75) -- (1,-2.5);
\filldraw (1,-2.5) circle (2pt);
\filldraw[very thick, fill = black!40] (0.375,-1.75) circle (7pt);
\node[above] at (0.375,-1.5) {$L$};
\end{tikzpicture}
&\sim \, \frac{1}{N^{2L-1}\,\lambdad^{1\over 2}}, & \\
\begin{tikzpicture}[scale=0.75,baseline={([yshift=-.5ex]current bounding box.center)},vertex/.style={anchor=base,
    circle,fill=black!25,minimum size=18pt,inner sep=2pt}]
\draw [color=white, thick] (-2,0) -- (2,0);
\filldraw [color=candyapplered, thick, fill=candyapplered!30] (0,-3) -- (0,0.475) -- (-0.75,-0.4) -- (-0.75,-3.875) -- (0,-3);
\draw [color=gray, thick] (0,0) ellipse (1.25 and 0.5);
\draw [color=gray, thick] (-1.25,0) -- (-1.25,-3.5);
\draw [color=gray, thick] (-1.25,-3.5) arc (180:360:1.25 and 0.5);
\draw [color=gray, thick, dashed] (-1.25,-3.5) arc (180:360:1.25 and -0.5);
\draw [color=gray, thick] (1.25,-3.5) -- (1.25,0);
\node[cross=4pt, color=candyapplered, very thick] at (-0.375,-0.875) {};
\node[cross=4pt, color=candyapplered, very thick] at (-0.375,-2.625) {};
\draw [very thick, snake it] (-0.375,-0.875) -- (0.3,-1.75);
\draw [very thick, snake it] (0.3,-1.75) -- (1,-0.875);
\filldraw (1,-0.875) circle (2pt);
\draw [very thick, snake it] (-0.375,-2.625) -- (0.3,-1.75);
\draw [very thick, snake it] (0.3,-1.75) -- (1,-2.625);
\filldraw (1,-2.625) circle (2pt);
\filldraw[very thick, fill = black!40] (0.375,-1.75) circle (7pt);
\node[above] at (0.375,-1.5) {$L$};
\end{tikzpicture}
&\sim\, \frac{1}{N^{2L} \, \lambdad},
&~~~~~~
&\begin{tikzpicture}[scale=0.75,baseline={([yshift=-.5ex]current bounding box.center)},vertex/.style={anchor=base,
    circle,fill=black!25,minimum size=18pt,inner sep=2pt}]
\draw [color=white, thick] (-2,0) -- (2,0);
\filldraw [color=candyapplered, thick, fill=candyapplered!30] (0,-3) -- (0,0.475) -- (-0.75,-0.4) -- (-0.75,-3.875) -- (0,-3);
\draw [color=gray, thick] (0,0) ellipse (1.25 and 0.5);
\draw [color=gray, thick] (-1.25,0) -- (-1.25,-3.5);
\draw [color=gray, thick] (-1.25,-3.5) arc (180:360:1.25 and 0.5);
\draw [color=gray, thick, dashed] (-1.25,-3.5) arc (180:360:1.25 and -0.5);
\draw [color=gray, thick] (1.25,-3.5) -- (1.25,0);
\draw [very thick, dashed] (-0.375,-1.75) ellipse (0.15 and 0.85);
\node[cross=4pt, color=candyapplered, very thick] at (-0.375,-0.875) {};
\node[cross=4pt, color=candyapplered, very thick] at (-0.375,-2.625) {};
\filldraw [very thick, fill = black!40] (-0.375,-1.75) ellipse (0.25 and 0.3);
\node[left] at (-0.65,-1.75) {$l$};
\draw [very thick, snake it] (-0.375,-0.875) -- (0.3,-1.75);
\draw [very thick, snake it] (0.3,-1.75) -- (1,-0.875);
\filldraw (1,-0.875) circle (2pt);
\draw [very thick, snake it] (-0.375,-2.625) -- (0.3,-1.75);
\draw [very thick, snake it] (0.3,-1.75) -- (1,-2.625);
\filldraw (1,-2.625) circle (2pt);
\filldraw[very thick, fill = black!40] (0.375,-1.75) circle (7pt);
\node[above] at (0.33,-1.5) {$L$};
\end{tikzpicture}
&\sim\, \frac{\lambdad^{\frac{l-1}{2}}}{N^{2L+l+1}},
& ~~~~ \ldots
\label{eq:loopdiagramsD1}
\fe

In addition, higher derivative contact interactions in (\ref{eq:DBI}) between bulk supergravitons and the D1-brane induced from an open string amplitude at $l$ loop order, schematically denoted by $g_s^l D^{2m} R^2$, with $m=0$,$1$, \ldots, give rise to a contribution to the two-point supergraviton amplitude that scales as
\ie
\begin{tikzpicture}[scale=0.75,baseline={([yshift=-.5ex]current bounding box.center)},vertex/.style={anchor=base,
    circle,fill=black!25,minimum size=18pt,inner sep=2pt}]
\draw [color=white, thick] (-2,0) -- (2,0);
\filldraw [color=candyapplered, thick, fill=candyapplered!30] (0,-3) -- (0,0.475) -- (-0.75,-0.4) -- (-0.75,-3.875) -- (0,-3);
\draw [color=gray, thick] (0,0) ellipse (1.25 and 0.5);
\draw [color=gray, thick] (-1.25,0) -- (-1.25,-3.5);
\draw [color=gray, thick] (-1.25,-3.5) arc (180:360:1.25 and 0.5);
\draw [color=gray, thick, dashed] (-1.25,-3.5) arc (180:360:1.25 and -0.5);
\draw [color=gray, thick] (1.25,-3.5) -- (1.25,0);
\node[cross=4pt, color=candyapplered, very thick] at (-0.375,-1.75) {};
\node[above] at (-0.95,-1.65) {$g_s^{l} D^{2m} R^2$};
\draw [very thick, snake it] (-0.375,-1.75) -- (1,-1);
\filldraw (1,-1) circle (2pt);
\draw [very thick, snake it ] (-0.375,-1.75) -- (1,-2.5);
\filldraw (1,-2.5) circle (2pt);
\end{tikzpicture}
\sim \, \frac{1}{N^{l-1}}\, \lambdad^{\frac{1}{2}(2l-3-m)}.
\label{eq:higherderivD1}
\fe
Using (\ref{eq:loopdiagramsD1}) and (\ref{eq:higherderivD1}) we can identify sample perturbative AdS diagrams, as shown in Figure~\ref{fig:sampleAdSdiagramsD1}, that are expected to contribute to the integrated two-point correlator and give rise to the result (\ref{DualTthooftexpansion}) of the holographic computation of the integrated two-point correlator in the presence of the 't Hooft line.

It is also interesting to consider the case of small but nonzero $\tilde{\theta}$, whose leading order contribution to the integrated two-point correlator in the presence of the 't Hooft line is given by (\ref{eq:linearthetat2}). In this case, the nonzero $\tilde{\theta}$ corresponds to a nonzero expectation value for the RR axion field $C_0 = \tau_1 + \delta C_0$, where $\tau_1 = \tilde{\theta}/(2\pi)$ and $\delta C_0$ denotes the fluctuation of the axion field about its expectation value. A linear term in $\tilde\theta$ may come from the Wess-Zumino coupling
\ie
S_{\rm WZ} = T_{\rm D1} \, g_s \int_{\rm D1} C_0 \wedge (B_2 + 2\pi\ap F_2),
\label{eq:WZcoupling}
\fe
where the integral is over the worldsheet of the D1-brane. 
For example, using the Wess-Zumino D1-brane coupling (\ref{eq:WZcoupling}) we could consider the following diagrams,\footnote{The bulk action for the RR fields,
$$
S_R = -\frac{1}{4\kappa_{10}^2}\int d^{10}x (-G)^{1\over 2} ( |F_1|^2 + |\tilde{F_3}|^2 + \cdots ) \,,
$$
where $\tilde F_3 = F_3 - C_0\wedge H_3$, $F_{p+1} = dC_p$, $H_3 = dB_2$, and where $\kappa_{10}\sim \lambdad^2$, implies that the $C_0$ propagator scales as $\lambdad^2$ whereas the $C_0$-$C_2$-$B_2$ bulk coupling scales as $\lambdad^{-2}$. Similarly, from the WZ coupling (\ref{eq:WZcoupling}), after noting that the field strength $F_2$ obtains a nontrivial expectation value proportional to $\tilde \theta$ as well, we see that the contact interaction between $\delta C_0$ and the D1-brane scales as $\tilde\theta\lambdad^{3\over 2}N^{-1}$.
}
\ie
\begin{tikzpicture}[scale=0.75,baseline={([yshift=-.5ex]current bounding box.center)},vertex/.style={anchor=base,
    circle,fill=black!25,minimum size=18pt,inner sep=2pt}]
\draw [color=white, thick] (-2,0) -- (2,0);
\filldraw [color=candyapplered, thick, fill=candyapplered!30] (0,-3) -- (0,0.475) -- (-0.75,-0.4) -- (-0.75,-3.875) -- (0,-3);
\draw [color=gray, thick] (0,0) ellipse (1.25 and 0.5);
\draw [color=gray, thick] (-1.25,0) -- (-1.25,-3.5);
\draw [color=gray, thick] (-1.25,-3.5) arc (180:360:1.25 and 0.5);
\draw [color=gray, thick, dashed] (-1.25,-3.5) arc (180:360:1.25 and -0.5);
\draw [color=gray, thick] (1.25,-3.5) -- (1.25,0);
\node[cross=4pt, color=candyapplered, very thick] at (-0.375,-1.75) {};
\node[above left] at (-0.375,-1.75) {${}_{\rm WZ}$};
\draw [very thick] (-0.375,-1.75) -- (0.375,-1.75);
\node[below] at (0,-1.75) {$\delta C_0$};
\draw [very thick] (0.375,-1.75) -- (1,-1);
\filldraw (1,-1) circle (2pt);
\node[above right] at (1,-1) {$C_2$};
\draw [very thick] (0.375,-1.75) -- (1,-2.5);
\filldraw (1,-2.5) circle (2pt);
\node[below right] at (1,-2.5) {$B_2$};
\end{tikzpicture}
\sim \, \tilde{\theta}\, \frac{\lambdad^{3\over 2}}{N}, ~~~~~~ 
\begin{tikzpicture}[scale=0.75,baseline={([yshift=-.5ex]current bounding box.center)},vertex/.style={anchor=base,
    circle,fill=black!25,minimum size=18pt,inner sep=2pt}]
\draw [color=white, thick] (-2,0) -- (2,0);
\filldraw [color=candyapplered, thick, fill=candyapplered!30] (0,-3) -- (0,0.475) -- (-0.75,-0.4) -- (-0.75,-3.875) -- (0,-3);
\draw [color=gray, thick] (0,0) ellipse (1.25 and 0.5);
\draw [color=gray, thick] (-1.25,0) -- (-1.25,-3.5);
\draw [color=gray, thick] (-1.25,-3.5) arc (180:360:1.25 and 0.5);
\draw [color=gray, thick, dashed] (-1.25,-3.5) arc (180:360:1.25 and -0.5);
\draw [color=gray, thick] (1.25,-3.5) -- (1.25,0);
\node[cross=4pt, color=candyapplered, very thick] at (-0.375,-0.875) {};
\node[above left] at (-0.375,-0.875) {${}_{\rm WZ}$};
\node[cross=4pt, color=candyapplered, very thick] at (-0.375,-2.625) {};
\draw [very thick] (-0.375,-0.875) -- (0.3,-1.75);
\node[above] at (0.1,-1.25) {$B_2$};
\draw [very thick, snake it] (0.3,-1.75) -- (1,-0.875);
\filldraw (1,-0.875) circle (2pt);
\draw [very thick] (-0.375,-2.625) -- (0.3,-1.75);
\node[below] at (0.1,-2.25) {$B_2$};
\draw [very thick, snake it] (0.3,-1.75) -- (1,-2.625);
\filldraw (1,-2.625) circle (2pt);
\end{tikzpicture}
\sim \, \frac{\tilde{\theta}}{N}
\fe
which, however, are subleading in the $1\over N$ expansion with respect to (\ref{eq:linearthetat2}). 
Furthermore, diagrams with loops on the D1-brane, in the bulk, or both, are further suppressed by higher powers of $1\over N$ and cannot account for the leading order contribution (\ref{eq:linearthetat2}). 

Thus, we find that the linear in $\tilde\theta$ leading order term (\ref{eq:linearthetat2}) must come from a higher-derivative contact interaction, with eight more derivatives with respect to the leading order interaction proportional to $T_{\rm D1}$, that gives rise to the following diagram
\ie
\begin{tikzpicture}[scale=0.75,baseline={([yshift=-.5ex]current bounding box.center)},vertex/.style={anchor=base,
    circle,fill=black!25,minimum size=18pt,inner sep=2pt}]
\draw [color=white, thick] (-2,0) -- (2,0);
\filldraw [color=candyapplered, thick, fill=candyapplered!30] (0,-3) -- (0,0.475) -- (-0.75,-0.4) -- (-0.75,-3.875) -- (0,-3);
\draw [color=gray, thick] (0,0) ellipse (1.25 and 0.5);
\draw [color=gray, thick] (-1.25,0) -- (-1.25,-3.5);
\draw [color=gray, thick] (-1.25,-3.5) arc (180:360:1.25 and 0.5);
\draw [color=gray, thick, dashed] (-1.25,-3.5) arc (180:360:1.25 and -0.5);
\draw [color=gray, thick] (1.25,-3.5) -- (1.25,0);
\node[cross=4pt, color=candyapplered, very thick] at (-0.375,-1.75) {};
\node[above] at (-0.95,-1.65) {$D^{4} R^2$};
\draw [very thick, snake it] (-0.375,-1.75) -- (1,-1);
\filldraw (1,-1) circle (2pt);
\draw [very thick, snake it ] (-0.375,-1.75) -- (1,-2.5);
\filldraw (1,-2.5) circle (2pt);
\draw [very thick, dashed] (-0.375,-1.75) -- (0.45,-3.3);
\node[below] at (1.1,-3.2) {$C_0|_{p\to 0}$};
\end{tikzpicture}
\sim \, \tilde{\theta} \, \frac{N}{\lambdad^{5 \over 2}}.
\label{eq:higherderivTheta}
\fe
This higher derivative contact interaction would be induced from the string worldsheet amplitude on a disk with two graviton vertex operators and one $C_0$ vertex operator that is taken to be soft, i.e.~such that the momentum of the $C_0$ boson is taken to zero.

A class of string theory amplitudes of this type, namely the disk three-point amplitude between two massless NSNS states and one massless RR state was considered in \cite{Becker:2011bw,Becker:2011ar,Becker:2016bzb} and computed up to quartic order in the external momenta. 
A useful selection rule was noted \cite{Becker:2016bzb} in order to identify the nonzero amplitudes as follows. The D1-brane boundary state respects the $\bZ_2$ symmetry\footnote{Note that gauging this $\bZ_2$ symmetry gives rise to the orientifold plane O1.} generated by $\Omega R_{\perp}$, where $\Omega$ denotes worldsheet parity, and $R_\perp$ denotes a spacetime reflection perpendicular to the D1-brane worldsheet.  Under $\Omega$, the worldsheet bosonic fields transform as $X^{\mu}_L(z)\leftrightarrow X^{\mu}_R(z)$, which in particular implies that $B_2$ is odd and $C_2$ is the only even massless RR state.  Under $R_\text{perp}$, $X^i(z,\zbar) \to - X^i(z,\zbar)$ for $i=2,\ldots,9$ in the case where the D1-brane is placed along the $X^1$ direction.   
The total $\bZ_2$ charge of the vertex operator insertions on the disk with D1-brane boundary condition must be even in order to have a nonzero amplitude. 
In particular, this argument implies that the disk three-point amplitude between $C_0$ and two gravitons vanishes since $C_0$ is odd whereas the graviton vertex operator is even under this $\bZ_2$. 

While it is difficult to write down a possible $\bZ_2$-invariant three-point coupling between $C_0$ and two gravitons, it is not hard to write down higher-point couplings at eight derivative order that are $\bZ_2$-invariant. For example,
\ie
\Delta S = \int_{\rm D1} C_0 \wedge \star ( a\, \tr(R^4) +b\,\tr(R^2)\wedge\tr(R^2)) \,,
\label{eq:8derivcoupling}
\fe
where $R$ denotes the bulk curvature two-form and $a$ and $b$ are constants. Notice that in this case the ten-dimensional Hodge star makes the coupling (\ref{eq:8derivcoupling}) $\bZ_2$-even, and thus satisfies the selection rule of \cite{Becker:2016bzb}. 
Although (\ref{eq:8derivcoupling}) is a higher-point contact interaction, it is possible that when expanded around the ${\rm AdS}_5\times {\rm S}^5$ background it leads to a three-point interaction between two gravitons and $C_0$ at eight derivative order as in (\ref{eq:higherderivTheta}). In fact, one may view the result (\ref{eq:linearthetat2}) as a \emph{prediction} for this type of higher derivative  coupling to the D1-brane.

Although the analysis of this section offers a strong consistency check on the results (\ref{IExplicit}) in the 't Hooft limit of large $N$ and fixed 't Hooft coupling $\lambda$ (or $\lambdad$), it is worth emphasizing that (\ref{IExplicit}) is a much stronger result beyond this limit. Specifically, it provides exact expressions for higher derivative couplings of supergravitons scattering off long $(p,q)$-strings at finite gauge coupling $\gym$ (or string coupling $g_s$) in the very strong coupling limit.

For simplicity, we have focused on the case of D1-brane in the above discussion. The generalization to $(p,q)=(p,1)$-strings is immediate.
The general $(p,q)$-strings are bound states of $p$ F1-strings and $q$ D1-branes \cite{
Harvey:1995rn,Schwarz:1995dk,Witten:1995im,Callan:1995xx} with the following tension,
\ie 
T_{(p,q)}={|p+q\tau|\over 2\pi \ell_s^2}={1\over 2\pi \ell_s^2}\sqrt{(p+q\chi)^2+{q^2\over g_s^2}}\,.
\label{pqtension}
\fe
Note that in Planck units, the tension depends on $\tau$ via $\frac{ |p+q\tau|}{\Im \tau}$ and is invariant under  $SL(2,\mZ)$ transformation that acts simultaneously on $\tau$ \eqref{tauTransf} and on the dyonic charges \eqref{SL2dyonic}.

The worldvolume action for $(p,1)$-strings take the same form as in \eqref{eq:DBI} for the D1-brane but the ground state involves quantized worldvolume electric flux which sources the NS-NS B-field with $p$ units of F1 charge \cite{Witten:1995im,Callan:1995xx}. Equivalently, this can be seen from a shift of the type IIB axion $C_0\to C_0+p$, which induces the same electric flux on the D1-brane and thus describe the $(p,1)$-string \cite{Witten:1995im}, which is the type IIB string dual of the Witten effect \cite{Witten:1978mh} in the $\cN=4$ SYM (see for example \cite{Harvey:1996ur}) that relates the 't Hooft and dyonic loops \eqref{Tdual2pt}.  
Consequently, the worldvolume effective action for  $(p,1)$-string follows from that of the D1-brane by the shift $C_0\to C_0+p$.
The Feynman rules from the type IIB supergravity coupled to the DBI action for $(p,1)$-string are identical to that of the D1-brane in the last two rows of Figure~\ref{fig:Feymanrules}, with the modification  that $T_{\rm D1}$ is replaced by $T_{(p,1)}$ in \eqref{pqtension}. The structure of higher derivative interactions on the $(p,1)$-string are more nontrivial, but can be readily determined once we restore the full $\tilde\theta$ dependence in the effective action on the D1-brane. In particular, the expansion \eqref{WTexpand} of the integrated correlator with the dyonic loop in the 't Hooft limit follows from the bulk effective field theory in the same way as for the 't Hooft loop case \eqref{DualTthooftexpansion}, where the coefficients in the expansion will now depend on $p$. We leave the detailed analysis of the effective action on $(p,1)$-string to future work.\footnote{The $(p,q)$-string with $q>1$ contains an $\cN=8$ supersymmetric-Yang-Mills with $U(q)$ gauge group on its worldvolume (and an non-abelian DBI extension) with $p$ probe quarks in the fundamental representation of $U(q)$ that sources a electric field \cite{Witten:1995im}. It would be interesting to extend our analysis to learn about the non-abelian DBI action and its higher derivative corrections in this case.
In particular, a good starting point would be the $(1,q)$-string which is S-dual to the $(q,1)$-string discussed in the main text (see \cite{Gukov:1997tt,Verlinde:1997fx,Klebanov:2000pp} for related works).
}

\subsection{Reflection Amplitudes off Long Strings and D1-branes in flat space}
\label{sec:flatspacescattering}

As mentioned in the Introduction, the ``very strong coupling limit" of large $N$ and finite $\gym$ corresponds to the limit in which the radius of AdS become very large. In this limit, we expect to recover flat space scattering amplitudes.  In what follows we will compare the structure of our result for the integrated two-point correlator in the presence of a $1\over 2$-BPS 't Hooft line (\ref{DualTthooftexpansion}) with the structure of the derivative expansion of the two-point reflection amplitude of two gravitons off a D1-brane, order by order in the open string genus expansion, in type IIB string theory in flat space.  We focus on the 't Hooft line / D1-brane case first because it is easier to understand than the Wilson line / fundamental string case which we come to near the end of the section. Along the way, we also comment on generalizations to the dyonic Wilson-'t Hooft lines corresponding to $(p,1)$-strings. 

It is most convenient to express the scattering amplitude of massless particles in type IIB string theory in the presence of a D1-brane in the spinor helicity formalism \cite{Elvang:2015rqa, Boels:2012ie, Lin:2015ixa}. The null momentum $p_i$ of a given external massless particle $i$ propagating in 10D is expressed in terms of spinor helicity variables $\zeta_{iA}^{\alpha}$ as
\ie
p_i^{\mu} (\gamma_\mu)^{\alpha\beta} = \zeta_{iA}^\alpha \zeta_{i}^{\beta A} \,,
\fe
where $\alpha=1,\ldots,16$ is a chiral spinor index of $SO(1,9)$, and $A=1,\ldots,8$ is an $SO(8)$ little group index. The one-particle supercharges are defined as $q_{i+}^{\alpha} = \zeta_{iA}^{\alpha} \eta_i^A$ and $q_{i-}^{\alpha} = \zeta_{i}^{\alpha A} \frac{\partial}{\partial\eta_i^A}$, where $\eta_i^A$ is a Grassmann odd variable and $\{\eta_i^A,\frac{\partial}{\partial\eta_j^B}\} = \delta^A_B \delta_{ij}$. The total supercharges
\ie
Q^\alpha_{+} = \sum_i q^{\alpha}_{i+}\,, \qquad Q^\alpha_{-} = \sum_i q^{\alpha}_{i-} 
\label{IIBsupercharges}
\fe
then satisfy  $\{Q^\alpha_{+},Q^\beta_{-}\} = -(\gamma_\mu)^{\alpha\beta}P^\mu$, where $P^\mu=\sum_i p_i^\mu$ is the total momentum.

In the presence of a D1-brane, the spinor helicity variables $\zeta_{iA}^{\alpha}$ as well as the supercharges can be decomposed with respect to the preserved $SO(1,1)\times SO(8)_\perp$ symmetry as $\zeta_{iA}^{\alpha} = (\xi_{iA}^a,\widetilde{\xi}_{iA}^{\dot a})$, where $a=1,\ldots,8$ and ${\dot a}=1,\ldots,8$ are indices for the spinor representations ${\bm 8}_s$ and ${\bm 8}_c$ of $SO(8)_\perp$ respectively.  Similarly, the one-particle supercharges split as
\ie
q_{i+}^{\alpha} = ( q_{i+}^{a} , \widetilde{q}_{i+}^{\dot a} )\,, \qquad  q_{i-}^{\alpha} = ( q_{i-}^{a} , \widetilde{q}_{i-}^{\dot a} ) \,.
\fe
The D1-brane preserves 16 out of the 32 supercharges, which we take to be\footnote{Since we are only interested in the scattering of 10D supergravitons off a D1-brane, we will not consider the supercharges of (external) two-dimensional particles living on the worldvolume of the D1-brane.}
\ie
Q_{+}^a &= \sum_i q_{i+}^a = \sum_i \xi_{iA}^a\eta_{i}^A\,,  \\
 Q_{-}^{ a} &= \sum_i   q_{i-}^{  a} = \sum_i {\xi}_{i}^{a A} \frac{\partial}{\partial\eta_i^A} \,.
 \label{eq:supercharges1}
\fe

A typical $n$-point superamplitude in the presence of a D1-brane takes the form
 \es{Amplitude}{
   \cA =  \delta^2(P^{||}_\mu) \delta^8(Q_{+}^{||}) \cF(\zeta_i,\eta_i) \,,
 }
where $P^{||}_\mu \equiv \sum_i^n p^{||}_{i,\mu}$ is the total momentum along the directions parallel to the worldvolume of the D1-brane and $\delta^8(Q_{+}^{||}) \equiv \prod_a Q_{+}^a$. The expression \eqref{Amplitude} automatically obeys $Q_+^a \cA = 0$, but one still has to impose the condition $Q_-^a {\cal A} = 0$.  This condition implies that the function $\cF$ must satisfy the supersymmetry Ward identities $\delta^2(P^{||}_\mu) \delta^8(Q_{+}^{||})   Q_{-}^{  a} \cF = 0$.   For a $1\to 1$ superamplitude, the solution of this equation is that ${\cal F}$ is independent of the $\eta_i^A$ and depends only on the $\zeta_{i A}^\alpha$ through the kinematic invariants \cite{Lin:2015ixa}
\ie
s = 2p^2_{1||} = 2p^2_{2||} \,, \qquad t = p_1\cdot p_2  \,.
\label{eq:defsandt}
\fe
So, ${\cal F} = {\cal F}(s, t)$.  This is of course the case for the tree-level superamplitude that one can calculate using only interaction vertices from supergravity and the DBI action.  For instance, for the case of external gravitons, this diagram is
\begin{center}
\begin{tikzpicture}
\draw [color=white, thick] (-2,0) -- (2,0);
\filldraw [color=candyapplered, thick, fill=candyapplered!30] (0,-3) -- (0,0.475) -- (-0.75,-0.4) -- (-0.75,-3.875) -- (0,-3);
\node[cross=4pt, color=candyapplered, very thick] at (-0.375,-1.75) {};
\draw [very thick, snake it] (-0.375,-1.75) -- (1,-1.75);
\draw [very thick, snake it] (1,-1.75) -- (3,-0.5);
\draw [very thick, snake it] (1,-1.75) -- (3,-3);
\end{tikzpicture} 
\end{center}
The superamplitude is \cite{Lin:2015ixa} 
 \es{ASGDBI}{
\cA_{\rm SGDBI\,tree}=-i(2\pi)^2\delta^2(P^{||}_\mu) \delta^8(Q_+^{||}) {T_{\rm D1}\over 2st}\,.
 }

\noindent The most general supersymmetry-invariant $1 \to 1$ amplitude is then
\ie
\cA_{1\to 1} =  
\cA_{\rm SGDBI\,tree} f_{\mathbb{T}}(s,t) \,,
\label{eq:defFst}
\fe
for some function $f_{\mathbb{T}}(s,t)$.  

For simplicity, let us focus on the case where the expectation value of the axion vanishes, $\chi= 0$.  In this case, the function $f_{\mathbb{T}}(s,t)$ admits a perturbative genus expansion 
\ie
f_{\mathbb{T}}( s, t) = \sum_{l=0}^\infty g_s^{l} f_{\mathbb{T},l}( s, t)  + \text{(non-perturbative)} \,,
\label{Fgenusexp}
\fe
where $f_{\mathbb{T},l}(  s,  t)$ is proportional\footnote{The explicit polarization factor associated with the specific states in the supergraviton multiplet being scattered are conveniently packaged in the term $\delta^8(Q_{+}^{||})$ of the superamplitude.} to the perturbative string amplitude at open string loop level $l$.  For instance, at tree level ($l=0$) in string perturbation theory, the flat space scattering amplitude of two gravitons with momenta $p_i$ and polarizations $e_{i,\mu\nu}$, with $i=1,2$, is computed by the disk two-point diagram with D1-brane boundary conditions on the disk, where the two bulk graviton insertions are represented by the following worldsheet vertex operators \cite{Garousi:1996ad}:
\ie
\label{wsvops}
\cV_i^{(-1,-1)} &=   c\widetilde{c} e_{i,\mu\nu} e^{-\phi}\psi^{\mu} e^{-\widetilde{\phi}}\widetilde{\psi}^{\nu} e^{i p_i \cdot X} \,,  \\
\cV_i^{(0,0)} &=   c\widetilde{c} e_{i,\mu\nu} \left( i\partial X^{\mu} + \frac{\ap}{2} p_i\cdot \psi \psi^\mu \right) \left( i\overline{\partial} X^{\nu} + \frac{\ap}{2} p_i\cdot \widetilde{\psi} \widetilde{\psi}^\mu \right) e^{i p_i \cdot X}\,, 
\fe
in the $(-1,-1)$ and $(0,0)$ picture,  respectively (see also \cite{Polchinski:1998rr} for details on the worldsheet formulation including the notation in \eqref{wsvops}).  This disk string amplitude for graviton-brane scattering (which is a component of the full superamplitude (\ref{eq:defFst})) evaluates to \cite{Hashimoto:1996bf, Garousi:1996ad,Bachas:1999um},\footnote{Note that our graviton polarizations are normalized as $G_{\m\n}=\eta_{\m\n}+2 e_{\m\n} e^{i p\cdot X}$ which differs from the choice in \cite{Garousi:1996ad} by a factor of $\kappa$.}
\ie \label{HKamp}
\cA^{\rm disk}_{{\rm grav}\to {\rm grav}}(p_1,e_1;p_2,e_2) =-{i\over 8}  T_{\rm D1} \ap^2\,(2\pi)^2\delta^2(P^{||}_\mu)\,\frac{\Gamma({\ap\over 2 } s)\Gamma({\ap\over 2 } t)}{\Gamma(1 + {\ap\over 2 } s + {\ap\over 2 } t)} K(p_1,e_1;p_2,e_2)\,,
\fe
where $K(p_1,e_1;p_2,e_2) = s a_1 - t a_2$ is a kinematical structure with 
\ie \label{a1a2}
a_1 =~& \Tr(e_1\cdot D) p_1\cdot e_2 \cdot p_1 - p_1\cdot e_2 \cdot D \cdot e_1 \cdot p_2 - 2 p_1 \cdot e_2 \cdot e_1 \cdot D \cdot p_1 - p_1 \cdot e_2 \cdot e_1 \cdot p_2 \\
& + \frac{s}{2}\Tr(e_1\cdot e_2) + (1\leftrightarrow 2)\,, \\
a_2 =~& \Tr(e_1\cdot D)\left( p_1\cdot e_2 \cdot D \cdot p_2 + p_2\cdot D \cdot e_2 \cdot p_1 + p_2\cdot D \cdot e_2 \cdot D \cdot p_2 \right) \\
& + p_1\cdot D \cdot e_1 \cdot D \cdot e_2 \cdot D \cdot p_2 - p_2 \cdot D \cdot e_2 \cdot e_1 \cdot D \cdot p_1 + \frac{s}{2}\Tr(e_1\cdot D \cdot e_2 \cdot D) \\
& - \frac{s}{2}\Tr(e_1\cdot e_2) - \frac{s+t}{2}\Tr(e_1\cdot D)\Tr(e_2\cdot D) + (1\leftrightarrow 2) \,,
\fe
and ${D^\mu}_\nu = \text{diag}(1,1,-1,\ldots,-1)$ for the case in which the D1-brane lies parallel to the $X^1$ direction. In \eqref{a1a2}, $\cdot$ indicates the contraction between adjacent 10D spacetime indices.

We normalize $\delta^8(Q_+^{||})$ such that the component $K(p_1,e_1;p_2,e_2)$ for the two-graviton scattering appears with unit coefficient. Then, $f_{\mathbb{T},0}( s, t)$ in \eqref{Fgenusexp} takes the explicit form
\ie
{f_{\mathbb{T},0}( s, t) \over st} &= { \ell_s^4 \over 4}\,\frac{\Gamma({\ell_s^2 \over 2 } s)\Gamma({\ell_s^2 \over 2 } t)}{\Gamma(1 + {\ell_s^2 \over 2 } s + {\ell_s^2 \over 2 } t)}\,,
\fe 
and its low energy expansion is 
\ie 
{f_{\mathbb{T},0}( s, t) \over st}
&= {1\over st}- \frac{\pi^2}{24} \ell_s^4 + {\zeta(3) \over 8}\ell_s^6(s + t) - \frac{\pi^4}{5760}\ell_s^8(4s^2+st+4t^2) + \cdots  \,.
\label{eq:disc2ptexpansion}
\fe
In \eqref{eq:disc2ptexpansion}, the leading term captures the two-supergraviton tree amplitudes from the D1-brane DBI action, and the subleading terms capture the corresponding terms that arise from higher derivative contact interactions of the types $R^2$, $D^2R^2$, and $D^4R^2$ on the D1-brane.
This has precisely the same structure as the leading order in $N$ contribution to ${\cal{I}}_{\mathbb{T},\rm pert}$ in (\ref{DualTthooftexpansion}).

Although in this paper we have not explicitly derived the relation between the integrated correlator (\ref{eq:IL2massderivs}) and the two-point correlation function in the presence of the Wilson (or 't Hooft) line (\ref{2pfL}), nor the precise flat space limit to extract the type IIB scattering amplitude, it is encouraging that the low-energy expansion, and in particular the $\zeta(3)$ coefficient of the $\ap^3=\ell_s^6$ term in (\ref{eq:disc2ptexpansion}), are already apparent from the integrated correlator (\ref{DualTthooftexpansion}).

The disk amplitude \eqref{HKamp} for the D1-brane has a simple generalization for $(p,1)$-strings \cite{Gukov:1997tt,Garousi:1998bj} after taking into account the modification to the boundary state due to the worldvolume electric flux \cite{Callan:1995xx}. In our normalization, this amounts to simply replacing the D1-brane tension $T_{\rm D1}$ that appears as a prefactor in \eqref{HKamp} by the tension of the $(p,1)$-string (see \eqref{pqtension}). Consequently the above discussion for the momentum expansion on the D1-brane simply carries over for the $(p,1)$-string, and so is the comparison to the large $N$ expansion of the integrated correlator with the dyonic loop ${\cal{I}}_{\mathbb{W}_p\mathbb{T},\rm pert}$ in \eqref{WTexpand}.
In particular, at $\tau=i/g_s$,
since the tension $T_{(p,1)}=T_{\rm D1}(1+\cO(p^2 g_s^2))=T_{\rm D1}(1+\cO({p^2\tilde\lambda^2\over N^2}))$ in the 't Hooft large $N$ limit, this is consistent with the statement below \eqref{WTexpand} that the order $N$ and order $N^0$ contributions to the integrated correlator are $p$-independent.

Going beyond the disk diagram in the flat space open-closed string amplitudes, from the result (\ref{DualTthooftexpansion}), we can also extract the structure of the low-energy expansion at the next order in $g_s$, computed from the two-punctured annulus diagram with D1-brane boundary conditions on both boundaries. Namely, the order $\cO(N^0)$ contributions in (\ref{DualTthooftexpansion}) imply a low energy expansion for the annulus amplitude of the following form,
\ie
\cA^{\rm annulus}_{{\rm grav}\to {\rm grav}}(p_1,e_1;p_2,e_2)= &\,\ap^{-1}\delta^2(P_\mu^{||})K(p_1,e_1;p_2,e_2) 
\\
&\times 
\left(  g_{0}(s,t) + \ap^2 g_{2}(s,t) + \vphantom{\zeta(3)}\ap^3 g_{3}(s,t) + \cdots \right) \,,
\fe

This amplitude, however, is somewhat subtle in string perturbation theory. Explicitly, it takes the following form \cite{Pasquinucci:1997di,Lee:1997gwa},
\ie\label{eq:annulus2pt}  
\cA&^{\rm annulus}_{{\rm grav}\to {\rm grav}}(p_1,e_1;p_2,e_2) \\
&= C'\ap^{-1}\delta^2(P_\mu^{||})K(p_1,e_1;p_2,e_2) \int_{0}^{\infty}\frac{dT}{T^2} \int_0^{\pi}dx_1 \int_0^{2\pi T}dy_1 \int_0^{\pi}dx_2 \int_0^{2\pi T}dy_2 \\
&~~~ \times\left| \frac{\theta_1(\frac{w_1-w_2}{2\pi}|iT)\theta_1(\frac{\overline{w}_1-\overline{w}_2}{2\pi}|iT)}{\theta_1(\frac{w_1+\overline{w}_2}{2\pi}|iT)\theta_1(\frac{\overline{w}_1+w_2}{2\pi}|iT)} \right|^{{\ap\over 2 } t} \left| \frac{\theta_1(\frac{w_1+\overline{w}_1}{2\pi}|iT)\theta_1(\frac{w_2+\overline{w}_2}{2\pi}|iT)}{\theta_1(\frac{w_1+\overline{w}_2}{2\pi}|iT)\theta_1(\frac{\overline{w}_1+w_2}{2\pi}|iT)} \right|^{{\ap\over 2 } s} e^{\frac{\ap s}{4\pi T}(y_1-y_2)^2} \,,
\fe
where we have parametrized the cylinder by $0\leq \Re w \leq \pi$ and $0\leq \Im w \leq 2\pi T$, $C'$ is a numerical constant, and where $w_j = x_j + i y_j$ with $j=1,2$ denote the positions of the two graviton vertex operators.  This amplitude, however, suffers from an IR logarithmic divergence at $T\to \infty$, for which the amplitude (\ref{eq:annulus2pt}) behaves as
\ie\label{eq:logdiv}
\cA&^{\rm annulus}_{{\rm grav}\to {\rm grav}}(p_1,e_1;p_2,e_2) \\ 
&\sim C'\ap^{-1} \delta^2(P_\mu^{||})K(p_1,e_1;p_2,e_2)\,  \int^{\infty}\frac{dT}{T^2}\times T \frac{(2\pi)^3}{2} {\ap\over 2 } t \frac{\Gamma({\ap\over 2 } s)\Gamma({\ap\over 2 } t)}{\Gamma(1+{\ap\over 2 } s + {\ap\over 2 } t)} \,.
\fe
On the other hand, there is no such divergence in the limit $T\to 0$ of (\ref{eq:annulus2pt}), as can be seen by changing variables to $U=1/T$ and performing a modular transformation of the theta functions to pass to the closed string channel expansion; the amplitude (\ref{eq:annulus2pt}) behaves in this case as $\sim\int^\infty\frac{dU}{U^3}$.
The logarithmic divergence (\ref{eq:logdiv}) has been proposed in \cite{Periwal:1996pw} and in \cite{Fischler:1996ja} to be canceled via a Fischler-Susskind mechanism \cite{Fischler:1986ci,Fischler:1986tb} that incorporates the recoil of the D1-brane. Of course, a meaningful comparison with the results implied by our integrated correlator computation would require adding both the annulus amplitude (\ref{eq:annulus2pt}) and the recoil contribution. 
Furthermore, in order to obtain an unambiguous result after the cancellation of this logarithmic divergence, a string field theory analysis similar to that of \cite{Sen:2019qqg,Sen:2020cef,Sen:2020eck,Eniceicu:2022xvk} appears to be necessary. 
While this computation lies outside the scope of this paper, we hope that our holographic computation will eventually help clarify the effects of the D1-brane recoil.

Although the analysis presented in this section serves as a valuable qualitative consistency check on the results for ${\cal I}_{\mathbb{W}}(\tau, \overline{\tau})$ in \eqref{IExplicit}, or equivalently for those for ${\cal I}_{\mathbb{T}}(\tau, \overline{\tau}) = {\cal I}_{\mathbb{W}}(-1/\tau, -1/\overline{\tau})$ in Section~\ref{sec:verystrongtH}, it is important to note that these results include information about {\em all orders} in the genus expansion, albeit only at the first few orders in the derivative expansion.  Away from the small $g_s$ limit, these results cannot be checked using current string worldsheet perturbation theory.  Instead, they can be used to make the following prediction.   The loop diagrams of the form (\ref{eq:loopdiagramsD1}), re-expanded in the very strong coupling limit of large $N$ with fixed $\tau=i/g_s$ (i.e.~$\chi = 0$, or fixed $\tilde{g}$ and $\tilde \theta=0$ in Section~\ref{sec:verystrongtH}), lead to the following conjecture for the derivative expansion of $f_{\mathbb{T}}(s,t)$:
\ie
\chi = 0: \qquad \frac{f_{\mathbb{T}}(s,t)}{st} & = \frac{1}{st} + \ell_s^2\, g_s {\rm NA}_{1}(s,t)  + \ell_s^4\left( g_s^2{\rm NA}_2(s,t)  + c_1 \right) \\
& + \ell_s^6 \left(g_s^3 {\rm {\rm NA}_{3}}(s,t) +g_s {\rm NA}_{4}(s,t)  +  c_2  g_s^2 {\cal I}_{\mathbb{T},1}(\tau,\overline{\tau})(s+t) \vphantom{\sum} \right) + \cdots,
\label{eq:FD1conjecture}
\fe
where ${\rm NA}_{i}(s,t)$ are $g_s$-independent functions with non-analytic dependence on $s$ and $t$, and $c_1$ and $c_2$ are $g_s$-independent constants whose precise values are yet to be determined. 

In \eqref{eq:FD1conjecture}, the coefficients of the various powers of $\ell_s$ are as follows:
 \begin{itemize}
     \item The coefficient of $\ell_s^0$ is the same as the leading term in \eqref{eq:disc2ptexpansion}, and it corresponds to the tree-level SUGRA/DBI diagram \eqref{ASGDBI}.  This term is independent of $g_s$.
     \item The coefficient of $\ell_s^2$ is non-analytic and it corresponds to brane one-loop diagrams, such as the diagram in the top-left diagram of \eqref{eq:loopdiagramsD1} with $l = 1$ and the AdS space removed.  This term is proportional to $g_s$.
     \item The coefficient of $\ell_s^4$ has a non-analytic contribution proportional to $g_s^2$ that comes from brane two-loop diagrams, such as the diagram in the top-left diagram of \eqref{eq:loopdiagramsD1} with $l = 2$ and the AdS space removed.  In addition, there is a $g_s$-independent constant that comes from an $R^2$ contact interaction on the brane.\footnote{An efficient way to classify these higher derivative interactions is by on-shell supervertices (superamplitudes without poles) that solve the supersymmetry Ward identity for the preserved 16 supercharges by the D1 brane (a direct generalization of the analysis in \cite{Lin:2015ixa} for the D3-brane). At the derivative order of $R^2$, there is a unique supervertex of the form $\cA_{R^2}=\D^8(Q^{||}_+)\delta^2(P_\m^\parallel)$. At the order of $D^2R^2$, there are two independent supervertices of the form $s \cA_{R^2}$ and $t \cA_{R^2}$ respectively, and one linear combination is expected to appear in the open-closed string amplitude (for the case of D3-brane, this follows from the $SL(2,\mZ)$ duality \cite{Lin:2015ixa}).\label{footInvariants}}
     \item At order $\ell_s^6$, there are two distinct types of non-analytic contributions, one proportional to $g_s^3$ that comes from brane three-loop diagrams such as the diagram in the top-left diagram of \eqref{eq:loopdiagramsD1} with $l = 3$ and the AdS space removed, and one proportional to $g_s$ that comes from a bulk contact interaction (see the bottom-left diagram in \eqref{eq:loopdiagramsD1} with $L=0$ and the AdS space removed.  The last term in the coefficient of $\ell_s^6$ comes from a $D^2 R^2$ contact interaction and, through ${\cal I}_{\mathbb{T},1}(\tau,\overline{\tau})$,  it contains highly non-trivial contribution from D-instantons.   
 \end{itemize}

Note that there are two different SUSY invariants of the schematic form $D^2 R^2$ (see Footnote~\ref{footInvariants}), one proportional to $s$ and one proportional to $t$. The arguments presented in this paper do not show that the entire contribution proportional to ${\cal I}_{\mathbb{T}, 1}$ should be proportional to $s+t$.  We conjecture this dependence as the simplest option consistent with the perturbative contributions in \eqref{eq:disc2ptexpansion}.\footnote{This is also the case for the $D^2R^2$ interaction on the D3-brane \cite{Basu:2008gt,Lin:2015ixa}.}

The same analysis of the $1\to 1$ superamplitude in equations \eqref{eq:supercharges1}--\eqref{eq:defFst} can be performed for the case of two supergravitons scattering off a long F1 string. 
In this case, the supercharges ${Q'}_+^{||}$ preserved by the long F1 string form an algebra that is isomorphic to that of the supercharges ${Q}_+^{||}$ preserved by the D1-brane considered in this section, transformed into one another by an S-transformation.\footnote{Under the $SL(2,\mZ)$ transformation \eqref{tauTransf}, the type IIB supercharges \eqref{IIBsupercharges} (which are complex combinations of the two 10d chiral Majorana-Weyl spinors) transform as $Q_\pm \to e^{\pm{1\over 2} i\varphi_{c,d}(\tau)}$, with $e^{i\varphi_{c,d}(\tau)}={c \tau +d \over |c \tau+d|}$.  The F1-string extended in the $01$ directions preserves the supercharges satisfying $Q_-=\Gamma^{01} Q_+$ (see \cite{Polchinski:1998rr}). Then, \eqref{SL2dyonic} and the $SL(2, \mZ)$ transformation of the supercharges implies that the $(p,q)$-string along the $01$ directions  preserves the supercharges that satisfy $Q_-=e^{i\varphi_{q,p}(\tau)}\Gamma^{01} Q_+$. In particular, the solutions to the linear constraints for the $(p,q)=(0,1)$ case (D1-brane) correspond to the supercharges in \eqref{eq:supercharges1} (denoted collectively by $Q_\pm^\parallel$).
}

Thus, the $1\to 1$ superamplitude off a long F1 string takes the same form
\ie
\cA^{(\rm{F1})}_{1\to 1} = \cA^{(\rm{F1})}_{\text{tree}}f_{\mathbb{W}}(s,t) \,,
\label{eq:deffW}
\fe
for some function $f_{\mathbb{W}}(s,t)$, where $s$ and $t$ are the same kinematic invariants \eqref{eq:defsandt} and the tree-level superamplitude is
\ie
\cA^{(\rm{F1})}_{\text{tree}} = -i(2\pi)^2\delta^2(P_\mu^{||})\delta({Q'}_+^{||})\frac{T_{\rm F1}}{2st} \,.
\fe

One can analyze the diagrammatics of the type \eqref{eq:loopdiagramsF1} in the very strong coupling limit of large $N$ and fixed gauge coupling $g$, as was done in the paragraph following equation \eqref{eq:FD1conjecture}, in order to deduce the derivative expansion \eqref{FF1Conjecture} of $f_{\mathbb{W}}(s,t)$:
 \es{FF1Conjecture}{
  \frac{f_{\mathbb{W}}(s, t)}{st}
   &= \frac{1}{st}
    + \ell_s^2 \text{NA}_1(s, t)
     + \ell_s^4 (\text{NA}_2(s, t) + c_1 g_s^2) \\
  &{}+ \ell_s^6 \left(\text{NA}_3(s, t) + g_s^2 \text{NA}_4(s, t)
   + c_2 g_s {\cal I}_{\mathbb{W}, 1}(\tau, \bar \tau) (s + t) \right) + \cdots \,,
 }
where ${\cal I}_{\mathbb{W}, 1}(\tau, \bar \tau)$, whose explicit form at finite $g$ and $\theta$ is given in  \eqref{IExplicit}, contains highly nontrivial instanton contributions.  Alternatively, for $\chi = 0$, one can perform a direct S-transformation of \eqref{eq:FD1conjecture}, under which\footnote{Or equivalently, sending $g_s\to g_s^{-1}$ and then $\ell_s \to \ell_s g_s^{1\over 2}$.} $g_s\to g_s^{-1}$ when \eqref{eq:FD1conjecture} is written in units of the Planck length $\ell_{\rm P}\sim\kappa^{1\over 4}$, obtaining the same expression for $f_{\mathbb{W}}(s,t)$.

By implementing the bootstrap program outlined in Section~\ref{sec:unintegrated}, in the future we hope to make Eqs.~\eqref{eq:FD1conjecture} and \eqref{FF1Conjecture} precise, and more generally, provide analogous formulas for scattering of gravitons from  $(p,q)$-strings.

\section{Discussion}
\label{sec:discussion}

In this paper, we studied the partition function of the $\cN = 2$ supersymmetry-preserving mass deformation of the $\cN = 4$ super Yang-Mills theory on $S^4$ in the presence of a $1\over 2$-BPS fundamental Wilson loop placed along the equator of the $S^4$, further differentiated twice in the mass deformation parameter $m$. 
On the one hand, this observable can be readily computed via a particular matrix model due to supersymmetric localization. On the other hand, this quantity has the interpretation as the two-point correlation function in the presence of the Wilson line of two $1\over 2$-BPS local operators residing in the stress tensor multiplet of the $\cN=4$ SYM, and further integrated against a measure that is yet to be determined. 

The main technical results of this work are the exact expressions for the integrated two-point correlator in the presence of the $1\over 2$-BPS Wilson line $\cal{I}_{\mathbb{W}}(\tau,\bar{\tau})$ in 
 (\ref{IExplicit}) and correspondingly $\cal{I}_\mathbb{L}(\tau,\bar{\tau})$ for general $1\over 2$-BPS Wilson-'t Hooft lines $\mathbb{L}$ in the same $SL(2,\mathbb{Z})$ orbit.
 For a fixed $SL(2,\mathbb{Z})$ duality frame, we also spell out the structure of perturbative contributions to $\cal{I}_\mathbb{L}(\tau,\bar{\tau})$.  While we have not computed the full un-integrated two-point correlators in this work,
we were able to identify examples of Witten diagrams in ${\rm AdS}_5$ that we expect to contribute to the integrated correlators, order by order in the large $N$ expansion. 
These include higher-derivative contact interactions with a D1-brane, induced by type IIB string theory amplitudes, for a vanishing as well as non-vanishing but small expectation value of the axion (i.e.~a small but nonzero theta angle in the dual SYM), the latter of which to the best of our knowledge has not been studied in the literature. 
Furthermore, assuming the existence of a well-defined flat space limit of AdS correlators in the presence of a line defect analogous to that of \cite{Polchinski:1999ry,Susskind:1998vk,Giddings:1999jq,Penedones:2010ue}, we were able to qualitatively compare our results for the integrated correlator in the presence of the 't Hooft line ${\cal{I}}_{\mathbb{T}}(\tau,\overline{\tau})$ against the low-energy (or $\ap$) expansion of the two-graviton scattering amplitude off a D1-brane in type IIB string theory at leading order in the string coupling $g_s$, as computed by the twice-punctured disk diagram \cite{Hashimoto:1996bf}, and predict nontrivial D-instanton contributions starting at the fourth derivative order.

This work constitutes one step in the broader program (summarized in Figure~\ref{fig:2pfsummary}) of bootstrapping the full two-point correlator of $1\over 2$-BPS local operators in the presence of a $1\over 2$-BPS line defect in the  $\cN=4$ super Yang-Mills theory, at least to the first few orders in the large $N$ limit. 
Under the AdS/CFT duality, this corresponds to bootstrapping scattering amplitudes of, for instance, supergravitons bouncing off an extended string or brane in type IIB string theory in ${\rm AdS}_5\times {\rm S}^5$ as well as in flat space, via a flat space limit that is yet to be spelled out. 
The recent success of \cite{Binder:2019jwn, Chester:2019jas,Chester:2020vyz, Chester:2020dja, Chester:2021aun} of incorporating the techniques of supersymmetric localization, analytic bootstrap, and the flat space limit, in the study of local operator four-point functions makes us hopeful that this program can be extended to include higher-dimensional objects such as D-branes. 
Our results will serve as a crucial input for the analytic bootstrap of supersymmetry-protected interactions in the low-energy expansion of type IIB string theory in the presence of a long F1-string or D1-brane (and the entire family of $(p,q)$-strings), and hopefully as important input to the numerical bootstrap of unprotected interactions as well. 

In addition to pursuing the bootstrap program outlined in Figure~\ref{fig:2pfsummary} for the un-integrated two-point function 
with a fundamental Wilson loop from the results in this work, there are a number of immediate generalizations worth investigating already at the level of the integrated two-point functions. The first is the possibility of an independent integrated correlator from derivatives with respect to the squashing parameter $b$. Secondly, thus far we have focused on a special class of line defects in the SYM theory coming from the $SL(2,\mZ)$ orbit of the Wilson loop in the fundamental representation of $SU(N)$. It would be interesting to pursue the matrix model analysis for higher representations (see \cite{Zarembo:2016bbk} for a recent review). In particular, Wilson loops in the symmetric representations at large rank (of order $\cO(N)$) correspond to D3-branes that extend entirely on ${\rm AdS}_5$ \cite{Drukker:2005kx},\footnote{Since the D3-brane is self-dual under $SL(2,\mZ)$, its worldvolume higher-derivative interactions come with modular invariant functions of $(\tau, \bar \tau)$. Together with nontrivial constraints from supersymmetry, this has lead to indirect arguments for the exact $(\tau, \bar \tau)$ dependence of the leading higher-derivative interactions \cite{Bachas:1999um,Green:2000ke,Basu:2008gt,Lin:2015ixa}. It would be interesting to derive these modular functions directly from the integrated correlator for the Wilson loop in the symmetric representation. } while Wilson loops in the anti-symmetric representations are described by D5-branes that wrap ${\rm S}^4\subset {\rm S}^5$ \cite{Yamaguchi:2006tq}.\footnote{In particular, the Wilson loop in the anti-fundamental representation is described a degeneration limit where the D5-brane shrinks to a point on the ${\rm S}^5$.} Consequently, they provide a promising pathway to study higher-derivative interactions on D3- and D5-branes in both ${\rm AdS}_5\times {\rm S}^5$ and on flat space. Furthermore, it would be interesting to understand the origin of the differential equations we found in Section~\ref{sec:diffeqns}. In light of the analysis in \cite{Lin:2015ixa}, they are expected to arise from supersymmetry Ward identities in the open-closed type II string theory. Finally, it would be interesting to identify a Laplace-difference equation for these defect integrated correlators that would hold at finite $N$, in the same spirit as for such equations for the integrated four-point functions studied in \cite{
Dorigoni:2021bvj,Dorigoni:2021guq,Dorigoni:2022cua} and derived in \cite{Brown:2023cpz}, which has lead to exact results at finite $N$ for these integrated correlators in these works and also in \cite{Collier:2022emf,Paul:2022piq,Dorigoni:2022cua,Brown:2023cpz,Paul:2023rka}. Such an equation has the potential of determining a fully non-perturbative defect CFT observable. 
We hope to come back to some of these questions in the near future.

\section*{Acknowledgments} 

We thank Lorenz Eberhardt, Aleix Gimenez-Grau, Simone Giombi, Igor Klebanov, Juan Maldacena, Ashoke Sen, and Pierre Vanhove for interesting discussions. SSP was supported in part by the US National Science Foundation under Grant No.~PHY-2111977 and by the Simons Foundation Grant No.~488653.   VAR was supported in part by the Simons Foundation Grant No.~488653  and by the Future Faculty in the Physical Sciences Fellowship at Princeton University.  The work of YW was supported in part by NSF grant PHY-2210420 and by the Simons Junior Faculty Fellows program.

\appendix

\section{More details on resolvents}
\label{RESOLVENTS}

\subsection{Eynard's topological recursion}
\label{EYNARD}

Let us briefly review the recursion relations for the ``genus" $m$, $n$-body resolvents $W^n_m$ for the case of interest of the Gaussian Hermitian matrix model. 
The seed term is the one-body resolvent at genus zero,
\ie
W^{1}_{0}(y_1) = 2a^2 y_1\left( 1 -\sqrt{1-(ay_1)^{-2}} \right) \,, 
\label{eq:W1m0}
\fe
where, as in the main text (see \eqref{ScaleWnm}), we have defined $a = \frac{2\pi}{\sqrt{\lambda}}$. The genus zero $n$-body resolvent is determined by
\ie
W^n_0(y_1,\ldots, y_n) =\, & \frac{1}{4a^2y_1\sqrt{1-(ay_1)^{-2}}} \left[ \sum_{l=1}^{n-2} \sum_{I\in R^n_l} W^{l+1}_0(y_1, y_I)W^{n-l}_0(y_1, y_{R^n - I}) \right. \\
& + \left. \sum_{l=2}^n \partial_{y_l} \frac{W^{n-1}_0(y_2, \ldots, y_l, \ldots, y_n ) - W^{n-1}_0(y_2, \ldots, y_1, \ldots, y_n )}{y_l-y_1}\right] \,,
\label{eq:recgenus0}
\fe
where $R^n=\{2,\ldots,n\}$, and $R^n_l$ are subsets of $R^n$ of length $l$.
Finally, the higher-genus resolvents are then computed using
\ie
W^1_m(y_1) = \frac{1}{4a^2y_1\sqrt{1-(ay_1)^{-2}}} \left[ W^2_{m-1}(y_1,y_1) + \sum_{r=1}^{m-1}W^1_{m-r}(y_1)W^1_r(y_1) \right] \,,
\label{eq:recgenusall1}
\fe
for $m\geq 1$, and
\ie
W^n_m(y_1,\ldots,y_n) =\,& \frac{1}{4a^2y_1\sqrt{1-(ay_1)^{-2}}} \left[ W^{n+1}_{m-1}(y_1,y_1,\ldots, y_n) + 2\sum_{r=0}^{m-1}W^1_{m-r}(y_1)W^{n}_{r}(y_1,\ldots, y_n) \right. \\
& + \sum_{r=0}^m \sum_{l=1}^{n-2} \sum_{I\in R^n_l} W_{r}^{l+1}(y_1,y_I)W_{m-r}^{n-l}(y_1,y_{R^n - I}) \\
& + \left. \sum_{l=2}^n \partial_{y_l} \frac{W^{n-1}_m(y_2, \ldots, y_l, \ldots, y_n ) - W^{n-1}_m(y_2, \ldots, y_1, \ldots, y_n )}{y_l-y_1}\right] \,,
\label{eq:recgenusall2}
\fe
with $m\geq 1$ and $n\geq 2$.

\subsection{Inverse Laplace transforms}
\label{LAPLACE}

Performing an inverse Laplace transform of the resulting resolvents is straightforward since most resolvents have a factorized form in the variables $y_i$. A useful formula to compute the relevant inverse Laplace transforms is 
 \es{LT}{
\cL\left[\frac{a^{p-1}\sqrt{\pi}}{(-2i)^k\Gamma(k+\frac{1}{2})}\,\partial_z^p\left( \left(\frac{z}{a}\right)^k J_k\left(-i\frac{z}{a}\right) \right)\right](z) = \frac{(ay)^p}{\left((ay)^2-1\right)^{k+\frac{1}{2}}} \,, \qquad  p\leq 2k \,.
 }
The only nontrivial case is the two-body resolvent at genus zero,
\ie
W^{2}_{0}(y_1,y_2) = \frac{ay_1\,ay_2-1-ay_1\,ay_2\sqrt{1-(ay_1)^{-2}}\sqrt{1-(ay_2)^{-2}}}{2(y_1-y_2)^2\,ay_1\,ay_2\sqrt{1-(ay_1)^{-2}}\sqrt{1-(ay_2)^{-2}}} \,,
\label{eq:W2m0}
\fe
which is not of factorized form. However, taking a derivative with respect to $a$ does in fact factorize,
\ie
\frac{\partial W^2_0}{\partial a}  = -\frac{a}{2}\frac{1+ay_1\,ay_2}{(ay_1)^3(1-(ay_1)^{-2})^{3\over 2}(ay_2)^3(1-(ay_2)^{-2})^{3\over 2}} \,, 
\fe
and hence its inverse Laplace transform is easily evaluated to
 \es{LapDerivative}{
  {\cal L}^{-1}\left[\frac{\partial W^2_0}{\partial a} \right](z_1, z_2)
   = - \frac{z_1 z_2}{2 a^3}
    \left( I_0\left[\frac{z_1}{a}\right]I_0\left[\frac{z_2}{a}\right] + I_1\left[\frac{z_1}{a}\right]I_1\left[\frac{z_2}{a}\right] \right) \,.
 }  
Integrating in $a$, we obtain that
\ie
\cL^{-1}[W^2_0](z_1,z_2) = \frac{1}{2} \frac{\frac{z_1}{a}\,\frac{z_2}{a}}{\frac{z_1}{a}+\frac{z_2}{a}} \left( I_0\left[\frac{z_1}{a}\right]I_1\left[\frac{z_2}{a}\right] + I_1\left[\frac{z_1}{a}\right]I_0\left[\frac{z_2}{a}\right] \right) + f(z_1, z_2)\,.
\label{eq:LinvW2m0TMP}
\fe
where $f(z_1, z_2)$ is an integration constant.  This integration constant can be fixed from the $a \to \infty$ limit, where $W^2_0(y_1, y_2) = \frac{1}{4 a^2 y_1^2 y_2^2} + O(a^{-4})$.  The inverse Laplace transform of this expression is $\frac{z_1 z_2}{4 a^2} + O(a^{-4})$.  Comparing with the large $a$ asymptotic behavior of \eqref{eq:LinvW2m0TMP}, we find that $f(z_1, z_2) = 0$, so
\ie
\cL^{-1}[W^2_0](z_1,z_2) = \frac{1}{2} \frac{\frac{z_1}{a}\,\frac{z_2}{a}}{\frac{z_1}{a}+\frac{z_2}{a}} \left( I_0\left[\frac{z_1}{a}\right]I_1\left[\frac{z_2}{a}\right] + I_1\left[\frac{z_1}{a}\right]I_0\left[\frac{z_2}{a}\right] \right) \,.
\label{eq:LinvW2m0}
\fe

\section{Perturbative contributions to higher orders}
\label{PERTURBATIVE}

At next-to-leading order in $1/N^2$ we have
 \es{IPert1Explicit}{
 {\cal I}_{\mathbb{W},\text{pert}, 1} &= -\frac{1}{2^4 3 \pi^2}\int_0^\infty d\omega \frac{\omega}{(\omega^2 + \pi^2) \sinh^2 \omega} \left\{\vphantom{\left(\frac{\sqrt{1}}{\sqrt{1}}\right)^2} J_0\left(\frac{\omega}{\pi}\sqrt{\lambda}\right)^2 \lambda^2\omega^2 (6\pi^2+5\omega^2) \right. \\
&{}+ J_0\left(\frac{\omega}{\pi}\sqrt{\lambda}\right)J_1\left(\frac{\omega}{\pi}\sqrt{\lambda}\right) \left[ -\frac{I_0(\sqrt{\lambda})}{I_1(\sqrt{\lambda})} 2 \lambda^2 \pi^3 \omega + 4\lambda^{3\over 2}\pi\omega(\omega^2 + \pi^2) \right] \\
&{}+ \left. J_1\left(\frac{\omega}{\pi}\sqrt{\lambda}\right)^2 \left[ -\left(\frac{I_0(\sqrt{\lambda})}{I_1(\sqrt{\lambda})}\right)^2 \lambda^2 \pi^4 +\lambda(\omega^2+\pi^2) (7\omega^2\lambda - \pi^2(4+\lambda)) \right] \right\},
 }
and at next-to-next-to-leading order,
 \es{IPert2Explicit}{
  {\cal I}_{\mathbb{W},\text{pert}, 2} &=  -\frac{1}{2^9 3^2 5 \pi^4}\int_0^\infty d\omega \frac{\omega}{(\omega^2 + \pi^2) \sinh^2 \omega} \\
&{}\times \left\{\vphantom{\left(\frac{\sqrt{1}}{\sqrt{1}}\right)^2} J_0\left(\frac{\omega}{\pi}\sqrt{\lambda}\right)^2 \left[ -\frac{I_0(\sqrt{\lambda})}{I_1(\sqrt{\lambda})} 5\omega^4\lambda^{7\over 2}(10\pi^2 + 11\omega^2) -2\omega^4\lambda^2(11\omega^2\lambda + 2\pi^2(5\lambda - 12)) \right] \right. \\
&{}+ J_0\left(\frac{\omega}{\pi}\sqrt{\lambda}\right)J_1\left(\frac{\omega}{\pi}\sqrt{\lambda}\right) \left[ \left(\frac{I_0(\sqrt{\lambda})}{I_1(\sqrt{\lambda})}\right)^2 10\pi^5\omega\lambda^{7\over 2} \right. \\
&{}+ \frac{I_0(\sqrt{\lambda})}{I_1(\sqrt{\lambda})} 4\pi\omega\lambda^2(48\pi^4 + 24\pi^2\omega^2 + 2\pi^4\lambda - 10\pi^2\omega^2\lambda - 11\omega^4\lambda^2) \\
&{}- \left. \frac{2}{\pi}\omega(\omega^2 + \pi^2) \lambda^{3\over 2} (192\pi^4 + 52\pi^4\lambda -188\pi^2\omega^2\lambda + 5\pi^4\lambda^2 - 11\pi^2\omega^2\lambda^2 + 5\omega^4\lambda^2) \vphantom{\left(\frac{\sqrt{1}}{\sqrt{1}}\right)^2}\right] \\
&{}+ J_1\left(\frac{\omega}{\pi}\sqrt{\lambda}\right)^2 \left[ -\left(\frac{I_0(\sqrt{\lambda})}{I_1(\sqrt{\lambda})}\right)^3 5\pi^6\lambda^{7\over 2} - \left(\frac{I_0(\sqrt{\lambda})}{I_1(\sqrt{\lambda})}\right)^2 2\pi^6\lambda^2(\lambda + 24) \right.\\
&{}+ \frac{I_0(\sqrt{\lambda})}{I_1(\sqrt{\lambda})} \lambda^{3\over 2}(\omega^2+\pi^2)(-192\pi^4 - 4\pi^4\lambda + 332\pi^2\omega^2\lambda + 5\pi^4\lambda^2 + 7\pi^2\omega^2\lambda^2 - 65\omega^4\lambda^2) \\
&{}+ \left.\left. (\omega^2+\pi^2)\lambda (87\omega^4\lambda^2 - \pi^2\omega^2\lambda(712 + 127\lambda) + \pi^4(576 + \lambda(152 + 7\lambda))) \vphantom{\left(\frac{\sqrt{1}}{\sqrt{1}}\right)^2}\right] \right\} \,.
 }
Lastly, ${\cal I}_{\mathbb{W},\text{pert}, 3}$ is given by\\
\tiny
\ie
{\cal I}&_{\mathbb{W},\text{pert}, 3} = -\int_0^\infty d\omega \frac{2\omega}{\sinh^2(\omega)} \left\{ J_0\left(\frac{\sqrt{\lambda } \omega }{\pi }\right)^2 \left[\frac{\lambda ^5 \omega ^2
   I_2(\sqrt{\lambda }){}^2}{147456 \pi ^2 I_1(\sqrt{\lambda }){}^2} \right.\right.\\
&-\frac{\lambda ^3
   \left(96 \lambda  \omega ^8+\pi ^2 (\lambda  (\lambda +288)+768) \omega ^6+96 \pi ^4 (\lambda +24) \omega
   ^4-96 \pi ^6 (\lambda -8) \omega ^2-768 \pi ^8\right) I_0(\sqrt{\lambda }){}^2}{884736 \pi ^4
   \omega ^2 \left(\omega ^2+\pi ^2\right) I_1(\sqrt{\lambda }){}^2}\\
&-\frac{1}{1105920 \pi ^4 \omega
   ^2 \left(\omega ^2+\pi ^2\right) I_1(\sqrt{\lambda })}\left( \lambda ^{5/2}
   \left(-66 \lambda ^2 \omega ^8-65 \pi ^2 \lambda ^2 \omega ^6-144 \lambda  \omega ^8-1080 \pi ^2 \lambda 
   \omega ^6 \right.\right. \\
& \left.\left. -480 \pi ^4 \lambda  \omega ^4+480 \pi ^6 \lambda  \omega ^2-3840 \pi ^2 \omega ^6-11520 \pi ^4
   \omega ^4-3840 \pi ^6 \omega ^2+3840 \pi ^8\right) I_0(\sqrt{\lambda }) \right) \\
&+\frac{1}{92897280
   \pi ^8 \omega ^2 \left(\omega ^2+\pi ^2\right)} \left( \lambda ^2 \left(70 \lambda ^3
   \omega ^{12}+12 \pi ^2 \lambda ^2 (105 \lambda -604) \omega ^{10}-3 \pi ^4 \lambda  (7 \lambda  (19
   \lambda +504)-17280) \omega ^8 \right.\right.\\
&\left.\left.\left. -6 \pi ^6 (7 \lambda  (\lambda  (37 \lambda +82)+864)+61440) \omega ^6-40320
   \pi ^8 (\lambda +24) \omega ^4+40320 \pi ^{10} (\lambda -8) \omega ^2+322560 \pi ^{12}\right) \right) \vphantom{ \frac{I_0\left(\sqrt{\lambda}\right)}{I_0\left(\sqrt{\lambda}\right)} } \right] \\
&+J_0\left(\frac{\sqrt{\lambda } \omega }{\pi }\right) J_1\left(\frac{\sqrt{\lambda } \omega }{\pi }\right)
   \left[-\frac{\pi  \lambda ^5 \omega  I_0\left(\sqrt{\lambda }\right){}^3}{442368 \left(\omega ^2+\pi
   ^2\right) I_1\left(\sqrt{\lambda }\right){}^3}\right.\\
&-\frac{1}{1105920 \pi ^3 \omega ^3 \left(\omega ^2+\pi
   ^2\right) I_1\left(\sqrt{\lambda }\right){}^2} \left( \lambda ^{5/2} \left(54 \lambda ^2 \omega ^8+25 \pi
   ^2 \lambda ^2 \omega ^6-27 \pi ^4 \lambda ^2 \omega ^4-456 \pi ^2 \lambda  \omega ^6 \right.\right.\\
& \left.\left. -168 \pi ^4 \lambda 
   \omega ^4+240 \pi ^6 \lambda  \omega ^2-3840 \pi ^2 \omega ^6-11520 \pi ^4 \omega ^4-3840 \pi ^6 \omega
   ^2+3840 \pi ^8\right) I_0\left(\sqrt{\lambda }\right){}^2 \right) \\
&+\frac{1}{46448640 \pi ^7 \omega ^3 \left(\omega ^2+\pi ^2\right) I_1\left(\sqrt{\lambda
   }\right)} \left( \lambda ^2 \left(140 \lambda ^3 \omega ^{12}-150 \pi
   ^2 \lambda ^3 \omega ^{10}-252 \pi ^4 \lambda ^3 \omega ^8+108 \pi ^6 \lambda ^3 \omega ^6+105 \pi ^8
   \lambda ^3 \omega ^4 \right.\right.\\ 
&-14496 \pi ^2 \lambda ^2 \omega ^{10}+8208 \pi ^4 \lambda ^2 \omega ^8+17448 \pi ^6
   \lambda ^2 \omega ^6-5364 \pi ^8 \lambda ^2 \omega ^4+127872 \pi ^4 \lambda  \omega ^8+32256 \pi ^6
   \lambda  \omega ^6 \\
&\left.\left. -62784 \pi ^8 \lambda  \omega ^4+40320 \pi ^{10} \lambda  \omega ^2-737280 \pi ^6 \omega
   ^6-2073600 \pi ^8 \omega ^4-645120 \pi ^{10} \omega ^2+645120 \pi ^{12}\right) I_0\left(\sqrt{\lambda}\right) \right) \\
&-\frac{1}{15482880 \pi ^7 \omega ^3} \left( \lambda ^{3/2} \left(413 \lambda ^3 \omega ^{10}-1191 \pi ^2 \lambda ^3 \omega ^8+817 \pi
   ^4 \lambda ^3 \omega ^6-77 \pi ^6 \lambda ^3 \omega ^4-14832 \pi ^2 \lambda ^2 \omega ^8+24528 \pi ^4
   \lambda ^2 \omega ^6 \right.\right.\\
&\left.\left.\left. -4744 \pi ^6 \lambda ^2 \omega ^4+100992 \pi ^4 \lambda  \omega ^6-57408 \pi ^6
   \lambda  \omega ^4+13440 \pi ^8 \lambda  \omega ^2-307200 \pi ^6 \omega ^4-430080 \pi ^8 \omega ^2+215040
   \pi ^{10} \right) \right)  \vphantom{ \frac{I_0\left(\sqrt{\lambda}\right)}{I_0\left(\sqrt{\lambda}\right)} } \right] \\
&+J_1\left(\frac{\sqrt{\lambda } \omega }{\pi }\right)^2 \left[\frac{\pi ^2 (\lambda +24) \lambda ^{7/2}
   I_0\left(\sqrt{\lambda }\right){}^3}{1105920 \left(\omega ^2+\pi ^2\right) I_1\left(\sqrt{\lambda
   }\right){}^3}-\frac{\lambda ^5 \left(\pi ^2-7 \omega ^2\right) I_2\left(\sqrt{\lambda }\right){}^2}{884736
   \pi ^2 I_1\left(\sqrt{\lambda }\right){}^2}
	+\frac{\pi ^2 \lambda ^5 I_0\left(\sqrt{\lambda
   }\right){}^4}{884736 \left(\omega ^2+\pi ^2\right) I_1\left(\sqrt{\lambda }\right){}^4} \right.\\
&+\frac{1}{92897280 \pi ^4 \omega ^4 \left(\omega ^2+\pi ^2\right)
   I_1\left(\sqrt{\lambda }\right){}^2} \left( \lambda ^2
   \left(105 (\lambda -48) \lambda ^2 \omega ^{10}-21 \pi ^2 \lambda  (\lambda  (7 \lambda +72)-1920) \omega
   ^8 \right.\right. \\
&-84 \pi ^4 (\lambda  (\lambda  (3 \lambda +17)-528)+3840) \omega ^6-\pi ^6 (\lambda  (\lambda  (35
   \lambda +4848)-11520)+921600) \omega ^4-322560 \pi ^8 \omega ^2\\
&\left.\left. +322560 \pi ^{10}\right)
   I_0\left(\sqrt{\lambda }\right){}^2 \right) \\
&-\frac{1}{46448640 \pi ^6 \omega ^4 I_1\left(\sqrt{\lambda
   }\right)} \left( \lambda ^{3/2} \left(2338 \lambda ^3 \omega ^{10}-6343 \pi ^2
   \lambda ^3 \omega ^8+73 \pi ^4 \lambda ^3 \omega ^6+287 \pi ^6 \lambda ^3 \omega ^4-55968 \pi ^2 \lambda
   ^2 \omega ^8 \right.\right.\\
&+51648 \pi ^4 \lambda ^2 \omega ^6-9288 \pi ^6 \lambda ^2 \omega ^4+302976 \pi ^4 \lambda 
   \omega ^6-22464 \pi ^6 \lambda  \omega ^4-829440 \pi ^6 \omega ^4-1290240 \pi ^8 \omega ^2\\
&\left.\left. +645120 \pi
   ^{10}\right) I_0\left(\sqrt{\lambda }\right) \right) \\
&+\frac{1}{92897280 \pi
   ^8 \omega ^4} \left( \lambda  \left(-70 \lambda ^4 \omega ^{12}+2 \pi ^2 \lambda ^3 (917 \lambda +5962) \omega
   ^{10}-\pi ^4 \lambda ^2 (\lambda  (2371 \lambda +52196)+221760) \omega ^8 \right.\right.\\
&+\pi ^6 \lambda  (\lambda 
   (\lambda  (217 \lambda +20276)+324864)+1096704) \omega ^6+\pi ^8 (\lambda  (\lambda  (\lambda  (35 \lambda
   +4)-46032)-290304)-2211840) \omega ^4 \\
&\left.\left.\left. -2580480 \pi ^{10} \omega ^2+1290240 \pi ^{12}\right) \right) \vphantom{ \frac{I_0\left(\sqrt{\lambda}\right)}{I_0\left(\sqrt{\lambda}\right)} } \right] \\
& +\frac{I_2\left(\sqrt{\lambda}\right)^2}{I_1\left(\sqrt{\lambda }\right)^2} \left[-\frac{\lambda ^4 \left(-2 (\lambda +4) \omega ^4+\pi ^2
   (\lambda -8) \omega ^2+4 \pi ^4\right) J_3\left(\frac{\sqrt{\lambda } \omega }{\pi }\right){}^2}{147456
   \pi ^4} \right. \\
&\left.\left. -\frac{\lambda ^3 \left(6 \pi ^2 \lambda ^2 \omega ^4-\lambda  (11 \lambda +96) \omega ^6-768 \pi
   ^4 \omega ^2+384 \pi ^6\right) J_2\left(\frac{\sqrt{\lambda } \omega }{\pi }\right){}^2}{884736 \pi ^4
   \omega ^2}\right] \right\}.
\fe
\normalsize

These expressions as well as the expression for ${\cal I}_{\mathbb{W},\text{pert}, 0}$ given in \eqref{eq:MMpertLO} can be expanded in $1/\lambda$ as follows.  We can use the Mellin representation of the product of two Bessel functions 
 \es{JJId}{
  J_m(x) J_n(x) = \int_{\gamma - i \infty}^{\gamma + i \infty} \frac{ds}{2 \pi i} 
   \frac{ \left( \frac x2 \right)^{m + n + 2s} \Gamma(-s) \Gamma(2 s + m + n + 1) }{\Gamma(s + m + 1) \Gamma(s + n + 1) \Gamma(s + m + n + 1) } \,,
 } 
with $-\frac 12 < \gamma  < 0$, to write each ${\cal I}_{\mathbb{W},\text{pert}, n}$ as
 \es{IPertMellin}{
  {\cal I}_{\mathbb{W},\text{pert}, n} 
   = \int_{\gamma - i \infty}^{\gamma + i \infty} \frac{ds}{2 \pi i} \sum_p C_{n, p}(s, \lambda) \int_0^\infty d\omega\, \frac{\omega^{2p + 3 + 2s} \lambda^s }
    {(\omega^2 +\pi^2) \sinh^2 \omega} \,,
 }
where $p$ runs over a finite range of integer values, and $C_{n, p}(s, \lambda)$ are coefficients.   Closing the contour to the left and picking up poles at $s = - 1/2, -1, -3/2, -2, -5/2, \ldots$ gives an expansion of the result in powers of $1/\sqrt{\lambda}$.
 
 To make this more explicit, let us define the function
  \es{Js}{
   J(s) = \int_0^\infty d\omega\, \frac{\omega^{3 + 2s}}{(\omega^2 + \pi^2) \sinh^2 \omega} \,.
  }
This function is defined for $\Re s > -1$, but can be extended meromorphically to the entire complex plane.  Indeed, for $\Re s > 0$, it is not hard to see that it obeys
 \es{Relation}{
  J(s) + \pi^2 J(s-1) = \int_0^\infty d\omega\, \frac{\omega^{1 + 2s}}{ \sinh^2 \omega}
   = 4^{-s} \Gamma(2 + 2s) \zeta(2s + 1) \,.
 }
This expression can be used repeatedly to extend $J(s)$ beyond the original range $\Re s > -1$.  The analytically extended function is meromorphic with simple poles at $s = -1,  -2, -3, \ldots$.
 
In terms of $J(s)$ we have
  \es{IPertMellinAgain}{
  {\cal I}_{\mathbb{W},\text{pert}, n} 
   = \int_{\gamma - i \infty}^{\gamma + i \infty} \frac{ds}{2 \pi i} \sum_p C_{n, p}(s, \lambda) J(s + p) \lambda^s \,.
 } 
When closing the contour, we pick up the poles at $s = -1, -2, -3, \ldots$ from $J(s + p)$, and poles at $s = -1/2, -3/2, -5/2, \ldots$ from the coefficients $C_{n, p}(s, \lambda)$.  Thus, 
   \es{IPertMellinAgain2}{
  {\cal I}_{\mathbb{W},\text{pert}, n} 
   &= \sum_{m=0}^\infty  \sum_p  \text{Res}_{s = - \frac 12 - m} \left[ C_{n, p}(s, \lambda)\right]  J(p-m-1/2) \lambda^{-\frac 12 - m} \\
    &{}+ \sum_{m=0}^\infty  \sum_p \left[ C_{n, p}(-m, \lambda)\right]   \text{Res}_{s = - m + p} \left[ J(s)\right]  \lambda^{ - m}  \,.
 } 
Thus, we only need the recursion formula to compute the values of $J(s)$ at half-integers and the residues of $J(s)$ at negative integers.  For the former, we can use $J(1/2) = - \frac{\pi^2}{6} (\pi^2 - 10)$, while for the second we can use the recursion formula for residues
 \es{ResRecursion}{
  \text{Res}_{s = -m} [J(s)] = - \frac{1}{\pi^2} \text{Res}_{s = -m + 1} [J(s)] - \frac{(-1)^m (2m-3) \zeta(2 (m-1)}{\pi^{2m}}  \,,
 }
with the boundary condition that $\text{Res}_{s = 0} [J(s)] = 0$.

This procedure produces the result \eqref{Expansions}.

\section{Details on the other integrated correlator $\cE_{\mathbb{L}}(\tau,\bar\tau)$}
\label{app:otherintegrated}

Here we provide more details on
the other integrated two-point function $\cE_{\mathbb{L}}(\tau,\bar\tau) =  
\pa_\tau \pa_{\bar\tau} \left.\log \la \mathbb{L} \ra  \right|_{m=0}$ defined in \eqref{eq:IL2tauderivs}.
This integrated correlator has been studied in the $\cN=4$ SYM when $\mathbb{L}$ is the half-BPS Wilson loop \cite{Giombi:2009ds,Giombi:2012ep,Beccaria:2020ykg} (see also \cite{Giombi:2020mhz,Beccaria:2021alk,Beccaria:2021ksw}). In particular, the Wilson loop expectation value $\la \ra|_{m=0}=\llangle \mathbb{W}(a_i)\rrangle$ is computed by the Gaussian matrix model as in \eqref{fExp}.  For this integrated correlator, we have
\ie 
\cE_{W}(\tau,\bar\tau)= \pi^2 \left( 
{\llangle   (\sum_i a_i^2 )^2  \mathbb{W}(a_i) \rrangle   \over \llangle    \mathbb{W}(a_i) \rrangle} -
{\llangle    \sum_i a_i^2   \mathbb{W}(a_i) \rrangle^2   \over \llangle    \mathbb{W}(a_i) \rrangle^2} \right)\,.
\fe
In fact, the supersymmetric integrated insertions brought down by $(\tau,\bar \tau)$ derivatives are equivalent to completely localized insertions of the superconformal primary operators of the stress tensor multiplet at the North and South poles of the $S^4$,  up to total supersymmetry variations with respect to a supercharge $\D_\ve$ in the massive $\cN=2$ subalgebra  $\mf{osp}(2|4)$ 
preserved by the supergravity background on $S^4$
\cite{Gomis:2014woa,Gerchkovitz:2016gxx},
\ie 
\int d^4 x \sqrt{g} M(x)  = \cO(x_N,Y)+\D_\ve(\dots) \,,
\quad 
\int d^4 x \sqrt{g} \overline{M}(x)= \cO(x_S,\overline{Y})+\D_\ve(\dots)\,.
\label{NSinsertions}
\fe 
Here, $M(x)$ is the exactly marginal superconformal descendant of $\cO(\vec x,Y)$ (similarly for its complex conjugate). 
The R-symmetry polarization $Y$ (and its complex conjugate $\overline{Y}$) depends on the choice of embedding $\mf{osp}(2|4)\subset \mf{psu}(2,2|4)$. The corresponding components $\cO(\vec x,Y)$ and $\cO(\vec x,\overline{Y})$  of the $\cN=4$ primary are often referred to as the chiral and the anti-chiral primaries respectively. An immediate consequence of \eqref{NSinsertions} is that for any observable that is $\D_\ve$-invariant, the $(\tau,\bar\tau)$ derivatives are equivalent to  additional localized insertions of $\cO(x_N,{Y})$ and $\cO(x_S,\overline{Y})$ at the poles of the $S^4$. This obviously applies to correlation functions of these insertions \eqref{NSinsertions} themselves, and also to the $\cN=2$ preserving mass deformations, which all preserve the full $\mf{osp}(2|4)$ algebra. As we explain below, the same simplification also applies to insertions of half-BPS line operators.

For general $\cN=2$ theories, the $\mf{osp}(2|4)$ subalgebra is the maximal superalgebra in the $\cN=2$ superconformal algebra $\mf{su}(2,2|2)$.  The bosonic parts of these algebras are $\mf{u}(1)_R \times \mf{so}(3, 2)$ and $\mf{su}(2)_R \times \mf{u}(1)_r \times \mf{so}(4, 2)$, respectively.  The $\mf{u}(1)_R$ generator of $\mf{osp}(2|4)$ is a Cartan generator of the $\mf{su}(2)_R$. With respect to the $\cN=4$ algebra $\mf{psu}(2,2|4)$, we fix the embedding 
\ie 
\mf{osp}(2|4)\subset \mf{su}(2,2|2) \times \mf{su}(2)_F \subset \mf{psu}(2,2|4)\,,
\fe 
 by requiring the six adjoint scalars in the $\cN=4$ SYM to split into $(\Phi_1,\Phi_2,\Phi_3,\Phi_4)$   and $(\Phi_5,\Phi_6)$, which transform, respectively, as $(\bm 2,\bm 2)_0$ and $(\bm 1,\bm 1)_{\pm 1}$ under $\mf{su}(2)_R\times \mf{su}(2)_F \times \mf{u}(1)_r \subset \mf{so}(6)_R$. In this case, the $\mf{so}(6)_R$ polarization for the localized insertions in \eqref{NSinsertions} can be fixed to be 
 \ie 
 Y=(0,0,0,0,-i,1)\,.
 \label{Ychoice}
 \fe
To relate to the correlation functions on $\mR^4$ discussed in the beginning of Section~\ref{sec:unint2pf}, it is convenient to  
work with a set of stereographic coordinates on $S^4$ with metric 
\ie 
ds^2=e^{2\Omega} d\vec{x}^2 \,,\qquad e^{\Omega}={1\over 1+{\vec{x}^2\over 4 }}\,,
\fe 
such that the North and South poles are located at $\vec{x}_N=0$ and $\vec{x}_S$, with $\abs{\vec{x}_S} \to \infty$, respectively. We would like to include a half-BPS line operator in this setup. We will work with the $\cN=2$ subalgebra since the conclusion will apply to general $\cN=2$ SCFTs.

In these stereographic coordinates, the $\cN=2$ superconformal symmetries are parametrized by conformal Killing spinors on $S^4$ (related to those on $\mR^4$ by a Weyl rescaling),
\ie 
e^{\Omega\over 2}(\ep_a+ x_\m \C^\m \eta_a)\,,
\label{N2csp}
\fe
where $\ep_a,\eta_a$ are constant Dirac spinors with $\mathfrak{su}(2)_R$ doublet indices $a=1,2$  and $\C^\m$ denotes the usual constant gamma matrices. The projection to the massive $\mf{osp}(2|4)$ subalgebra is implemented by 
\ie 
\eta_a=(\sigma_3)_a{}^b \ep_b\,.
\label{projtomassiveN2}
\fe
The supercharge $\ve$ (acting on fields as $\D_\ve$) for which \eqref{NSinsertions} holds is the same chiral supercharge used in the supersymmetric localization of \cite{Pestun:2007rz} and comes from imposing further the following conditions \cite{Gomis:2014woa} (up to discrete choices),
\ie 
\ep_{a}=\C_5 \ep_a\,,\qquad 
\ep_{a}= i \C_{12} (\sigma_3)_a{}^b \ep_b\,.
\label{projtopestunN2}
\fe
In particular, it squares to the following combination of rotations on $S^4$ (preserving the poles) and a $\mf{u}(1)_R$ R-symmetry rotation
\ie 
\D^2_\ve = J_{12}+J_{34}+R\,.
\fe
A general half-BPS line operator $\mathbb{L}$ in an $\cN=2$ SCFT can preserve a maximal $\mf{osp}(4^*|2)$ subalgebra when wrapping a great circle on $S^4$. The choice  compatible with the supercharge $\D_\ve$ is when $\mathbb{L}$ wraps the equator at $x^3=x^4=0$ and $(x^1)^2+(x^2)^2=4$, in which case the $\mf{osp}(4^*|2)$ subalgebra is generated by conformal killing spinors \eqref{N2csp} satisfying (up to discrete choices) \cite{Pestun:2007rz},
\ie 
\eta_a=i \C_{12} \ep_a\,,
\fe
which is clearly compatible with \eqref{projtomassiveN2} and \eqref{projtopestunN2}. 

Specializing to the $\cN=4$ SYM on $S^4$ with an half-BPS line operator $\mathbb{L}$ on the equator as above, we conclude that the integrated correlator \eqref{eq:IL2tauderivs} from $\tau$ and $\bar\tau$ derivatives is in fact equal to the following un-integrated \textit{connected} correlator,
\ie 
\cE_{\mathbb{L}}(\tau,\bar\tau)={2c \over \tau_2^2}\la \cO(x_N ,Y)\cO(x_S,\overline{Y})\ra_{S^4}^{\mathbb{L},{\rm conn}}\,,
\label{chiral3pfunintS4}
\fe
where $c$ is the central charge (i.e. $c={N^2-1\over 4}$ for the $SU(N)$ SYM) and the identity channel on the defect is subtracted off,
\ie 
&\la \cO(x_N ,Y)\cO(x_S,\overline{Y})\ra_{S^4}^{\mathbb{L},{\rm conn}}
\\&\equiv {\la \cO(x_N ,Y)\cO(x_S,\overline{Y}) \mathbb{L}(Y_3)\ra_{S^4} \over 
\la   \mathbb{L}(Y_3)\ra_{S^4}} -{\la  \cO(x_N,Y) \mathbb{L}(Y_3)\ra_{S^4}\la  \cO(x_S,\overline{Y}) \mathbb{L}(Y_3)\ra_{S^4}\over 
\la   \mathbb{L}(Y_3)\ra_{S^4}^2}\,.
\label{defconn}
\fe 
Here, in the $\cN=4$ SYM theory, we choose the R-symmetry polarization of $\mathbb{L}$ to be\footnote{There are  discrete choices of $Y_3$ that are compatible with the supersymmetry we want to preserve. We have picked \eqref{Y3choice} to be specific but other choices are equivalent.}
\ie 
Y_3=(0,0,0,0,0,1)\,.
\label{Y3choice}
\fe 
 By a Weyl transformation,
\ie 
\la \cO(x_N ,Y)\cO(x_S,\overline{Y})\ra_{S^4}^{\mathbb{L}}
=4^{-\Delta_\cO}
\la \cO(x_N ,Y)\cO(x_S,\overline{Y})\ra_{\mR^4}^{\mathbb{L},{\rm conn}} \,.
\fe 
Thus, the correlator on $S^4$ is related to the flat space correlator (defined in the similar way as in \eqref{defconn}):\footnote{For general BPS operators, the normalized flat space operators (as in \eqref{2pfnorm}) undergo mixing on $S^4$ due to the nontrivial supergravity background and, correspondingly, the nontrivial conformal anomalies \cite{Gomis:2015yaa}. To recover the correlation functions of the normalized flat space operators, one needs to implement the Gram-Schmidt procedure on the two-point functions of the $S^4$ operators \cite{Gerchkovitz:2016gxx}. See \cite{Rodriguez-Gomez:2016cem} for explicit formulas that implement the Gram-Schmidt procedure in the $U(N)$ SYM theory.} 
\ie 
\cE_{\mathbb{L}}(\tau,\bar\tau)={c\over 8 \tau_2^2}
\la \cO(\vec x_1 ,Y)\cO(\vec x_2,\overline{Y})\ra_{\mR^4}^{\mathbb{L},{\rm conn}}
\label{chiral3pfunintR4}
\fe
with $x_1=(1,0,0,0)$, $x_2=(-1,0,0,0)$, and  $\mathbb{L}$ along the $x^4$ direction at $x^1=x^2=x^3=0$, as discussed in the beginning of Section~\ref{sec:unint2pf}.

The same conclusion above can be reached from the $S^2$ topological sector of the $\cN=4$ SYM theory, which computes topological correlation functions of twisted local operators and defects (including the half-BPS line operators) represented by certain insertions in the (multi-matrix) Gaussian matrix model \cite{Drukker:2007yx,Pestun:2009nn,Giombi:2009ds,Giombi:2009ek,Wang:2020seq}. On $S^4$, this topological sector resides on a great $S^2$ which we can take to be at $x^2=0$ and $(x^1)^2+(x^3)^2+(x^4)^2=4$. For special configurations of the local operators and the defects, such as in \eqref{chiral3pfunintS4}, the topological correlator coincides with the physical correlator.
Indeed, at the level of the defect two-point function \eqref{2pfL}, this topological sector corresponds to the special configuration \eqref{topconstraint}
and the topological correlator is given in \eqref{topF}.

From the configuration in \eqref{chiral3pfunintR4} and the choice of R-symmetry polarizations in \eqref{Ychoice} and \eqref{Y3choice}, we have $z=\bar z=w=-1$, and, consequently, we obtain 
\ie 
\cE_{\mathbb{L}}(\tau,\bar\tau)={c\over 8\tau_2^2}(c^-_\mathbb{L}(\tau,\bar\tau)-a_{\cO,\mathbb{L}}(\tau,\bar\tau)^2)\,,
\fe
as given in \eqref{Eintfinal}.

The relations \eqref{chiral3pfunintS4} and \eqref{chiral3pfunintR4} hold identically for general $\cN=2$ SCFTs with half-BPS line defects $\mathbb{L}$. Furthermore, a simple generalization is when 
the chiral primary operator $\cO(\vec{x}, Y)$ of dimension $\Delta_\cO=2$ is replaced by another generator of the Coulomb branch chiral ring, namely an operator $\cO_p$ with dimension $\Delta_{\cO_p}=p$ that couples to a background chiral multiplet with chiral coupling $\tau_p$. Similarly, $\cO(\vec{x}, \overline{Y})$ is replaced by an anti-chiral primary generator $\bar \cO_p$ with anti-chiral coupling $\bar\tau_p$.  These operator insertions can be related to an $\mf{osp}(2|4)$-preserving deformation on $S^4$ that generalizes \eqref{NSinsertions} (see \cite{Gerchkovitz:2016gxx} for details).  In this case, \eqref{eq:IL2tauderivs} and \eqref{chiral3pfunintS4} becomes
\ie 
\cE_{\mathbb{L}}^{p,q,a,b}(\tau,\bar\tau)= \pa_{\tau_p}^a \pa_{\bar\tau_q}^b \left.\log \la \mathbb{L} \ra  \right|_{m=0,\tau_{p\neq 2}=0}
 =\la \cO_p^a(x_N  )\bar\cO_q^b (x_S) \ra_{S^4}^{\mathbb{L},{\rm conn}}
\,,
\label{eq:IL2tauderivsp}
\fe
for the defect two-point function of chiral and anti-chiral primaries generated by products of $\cO_p$ and $\bar\cO_q$. The generalization to more general operators in the chiral ring is immediate. 
In particular this includes the defect one-point function of a general (anti)chiral primary. For example, the one-point function coefficient for $\cO$ (see \eqref{O1pf}) is  
\ie 
a_{\cO,\mathbb{L}}={2\sqrt{2}\tau_2\over \sqrt{c}}\pa_\tau \log \la \mathbb{L}\ra \,,
\fe
as given in \eqref{O1pffinal}.

Concretely when $\mathbb{L}=\mathbb{W}$ is the  half-BPS Wilson loop in the fundamental representation in the $SU(N)$ SYM theory, its expectation value was determined exactly in \cite{Drukker:2007yx}
\ie 
\la \mathbb{W}\ra={1\over N}e^{{\pi (N-1)\over 2N\tau_2}} L_{N-1}^1\left(-{\pi\over \tau_2}\right)\,,
\fe 
where $L_n^k$ is the associated Laguerre polynomial defined by 
\ie 
L_n^k(x)={e^x \over x^k n!} {d^n\over dx^n} e^{-x} x^{n+k}=(-1)^k {d^k\over dx^k} L_{n+k}(x)\,,
\fe
in terms of the usual Laguerre polynomial $L_{n}(x)\equiv L_{n}^0(x)$.  The quantities $a_{\cO,W}(\tau,\bar\tau)$ and $c^-_W(\tau,\bar\tau)$ follow immediately by taking derivatives of $\log \la \mathbb{W}\ra$ with respect to $(\tau,\bar\tau)$. In the large $N$ limit (fixed $\tau$), we have
\ie 
\log \la \mathbb{W} \ra 
=\sqrt{4\pi N\over \tau_2}
-{3\over 4} \log {N\over \tau_2}-\log 2-{5\over 4}\log \pi 
+\cO(N^{-{1\over 2}})\,.
\fe
In particular, all dependence on $(\tau, \bar\tau )$ is perturbative.

\bibliographystyle{JHEP}
\bibliography{WL.bib}

\end{document}